\documentclass{article}
\usepackage{amsfonts,amssymb,amsmath,graphicx}
\usepackage{theorem, caption, subcaption}
\usepackage{vmargin}
\usepackage{url}
\usepackage{tikz}
\usetikzlibrary{arrows}
\tikzstyle{block}=[draw opacity=0.7,line width=1.4cm]
\usetikzlibrary{positioning,arrows,chains,matrix,scopes,fit,decorations.markings,decorations,decorations.pathreplacing,decorations.pathmorphing,patterns}
\tikzset{
  on each segment/.style={
    decorate,
    decoration={
      show path construction,
      moveto code={},
      lineto code={
        \path [#1]
        (\tikzinputsegmentfirst) -- (\tikzinputsegmentlast);
      },
      curveto code={
        \path [#1] (\tikzinputsegmentfirst)
        .. controls
        (\tikzinputsegmentsupporta) and (\tikzinputsegmentsupportb)
        ..
        (\tikzinputsegmentlast);
      },
      closepath code={
        \path [#1]
        (\tikzinputsegmentfirst) -- (\tikzinputsegmentlast);
      },
    },
  },
  mid arrow/.style={postaction={decorate,decoration={
        markings,
        mark=at position .9 with {\arrow[#1]{stealth}}
      }}},
}

\newtheorem{definition}{Definition} 

\newtheorem{theorem}[definition]{Theorem}

\newtheorem{lemma}[definition]{Lemma}
\newtheorem{rmk}[definition]{Remark}
\numberwithin{equation}{section}

\newcommand{\beq}{\begin{equation}}
\newcommand{\eeq}{\end{equation}}
\newcommand{\bea}{\begin{eqnarray}}
\newcommand{\eea}{\end{eqnarray}}
\newcommand{\beano}{\begin{eqnarray*}}
\newcommand{\eeano}{\end{eqnarray*}}
\newcommand{\bma}{\begin{pmatrix}}
\newcommand{\ema}{\end{pmatrix}}

\def\mathbi#1{\textbf{\em #1}}
\def\a{\alpha}
\def\b{\beta{}} 
            
\def\cD{{\cal D}}

\def\cM{{\cal M}}

\def\fb{{\mathfrak b}}

\def\fm{{\mathfrak m}}

\def\fq{{\mathfrak q}}
\def\fp{{\mathfrak p}}

\newcommand{\CC}{{\mathbb C}}

\newcommand{\PP}{\mbox{${\mathbb P}$}}

\newcommand{\ZZ}{{\mathbb Z}}

\newcommand{\prf}{\underline{Proof:}\ }
\newcommand{\finprf}{\null \hfill {\rule{5pt}{5pt}}\\ \medskip}
\newcommand{\ie}{{\it i.e.}\ }
\newcommand{\cf}{{\it c.f.}\ }

\title{Classification of integrable boundary equations \\ for integrable  quad-graph systems\footnote{The authors are supported by National Natural Science Foundation of China (No.~11875040, 12171306).}}

\date{\empty}
\author{ Pengyu Sun\,,~~~~  Cheng Zhang\footnote{Correspondence:  ch.zhang.maths@gmail.com} \\ \\
  \sc \small Department of Mathematics \\  \sc \small Shanghai University \\ \sc \small Shanghai, 200444, China
 }

\begin{document}
\maketitle

\begin{abstract}
In the context of integrable systems on quad-graphs, the boundary consistency around a half of a rhombic dodecahedron, as a companion notion to the three-dimensional consistency around a cube, was introduced as a criterion for defining integrable boundary conditions for quad-graph systems with a boundary. In this paper, we formalize the notions of boundary equations as boundary conditions for quad-graph systems, and provide a systematic method for solving the boundary consistency, which results in a classification of integrable boundary equations for quad-graph equations in the Adler-Bobenko-Suris classification. This relies on factorizing, first the quad-graph equations into  pairs of dual boundary equations, and then the consistency on a rhombic dodecahedron into two equivalent boundary consistencies. Generalizations of the method to  rhombic-symmetric equations are also considered.
\end{abstract}

\section{Introduction}

In the context of integrable partial difference equations, also known as fully discrete integrable equations in contrast to some well-known continuous integrable equations in the soliton theory, the notion of {\em three-dimensional consistency} has emerged as the key feature of defining integrability for fully discrete systems on quad-graphs \cite{nijhoff2002lax, BS}. A major result was the classification of scalar equations on quad-graphs led by Adler, Bobenko and Suris \cite{adler2003classification, ABSII}, also known as the ABS classification.
One of the particular aspects of fully discrete integrable equations is that classes of equations are formulated as polynomial or rational functions,  the classification techniques are naturally connected to languages and methods in some classical aspects of geometry and invariant theories. For instance, quad-graph equations in the ABS classification are expressed in terms of affine-linear polynomials of four variables, and the classification was based on analysis of  canonical forms of multivariate polynomials and their invariants with respect to common M\"obius transformations. Also in the classification of Yang-Baxter maps \cite{adler2003geometry, papageorgiou2010quadrirational}, the maps are birational functions, and the Yang-Baxter properties were described as consistency conditions among classical geometric objects such as lines and pencils of quadratic curves.
This is in an apparent contrast to classifications of continuous integrable systems, where notions in differential geometry and related group/algebraic structures are of prominent importance.

This paper aims to classify one particular type of discrete equations, known as integrable boundary equations, which, together with the $3$D-consistent equations, satisfy the {\em boundary consistency condition}. The latter condition was introduced as the integrability criterion for integrable boundary conditions for quad-graph systems with boundary \cite{CCZ}. In this introductory section, necessary backgrounds on quad-graph systems with boundary are provided. 
\subsection{Quad-graph systems with a boundary}
\begin{figure}[h]
    \centering

  \begin{tikzpicture}[scale=1.8]
    \def\d{1}%
    \def\le{0.2}%
    \def\r{0.06}
    \def\ld{1.5}%
    \tikzstyle{nod}= [circle, inner sep=0pt, fill=white, minimum size=5pt, draw]
    \tikzstyle{nod1}= [circle, inner sep=0pt, fill=black, minimum size=5pt, draw]

    \coordinate (u00) at (0,0);
    \coordinate (u10) at (\d,0);
    \coordinate (u01) at (0,\d);
    \coordinate (u11) at (\d,\d);
    \draw[-] (u00)   -- node [below]{ $\a$} (u10)  -- (u11) -- (u01) -- node [left]{ $\b$}(u00);

    \node[nod] (u00) at (0,0) [label=below: $x$] {};
    \node[nod1] (u10) at (\d,0) [label=below: $u$] {};
    \node[nod1] (u01) at (0,\d) [label=above:$v$] {};
    \node[nod] (u11) at (\d,\d) [label=above:$y$] {};
    
  \end{tikzpicture}\hspace{2cm}
  \begin{tikzpicture}[scale=2]
	\def\d{.8}%
	\def\le{0.2}%
	\def\r{0.06}
	\def\ld{1.5}%
	\tikzstyle{nod}= [circle, inner sep=0pt, fill=white, minimum size=5pt, draw]
	\tikzstyle{nod1}= [circle, inner sep=0pt, fill=black, minimum size=5pt, draw]

	\coordinate (w00) at (\le+\ld+\d,0);
	\coordinate (w10) at (\le+\ld+\d+\d,0);
	\coordinate (w01) at (\le+\ld+\d+\d,\d);
	\draw (w00)  -- node [below]{ $\a$} (w10)  ;
	\draw[ultra thick] (w01) --  (w00);
	\draw (w10) -- node [right]{ $\b$} (w01);
	\node[nod1] (w10)at (\le+\ld+\d+\d,0) [label=below: $y$] {};
	\node[nod] (w01) at (\le+\ld+\d+\d,\d)[label=above: $z$]{};
	\node[nod] (w00) at (\le+\ld+\d,0) [label=below:$x$] {};
	\end{tikzpicture}
	\caption{Elementary configurations for a quad-graph with  boundary: a quadrilateral supporting a bulk quad equation (left); and a triangle supporting a boundary equation (right) where the thick line represents the boundary connected by boundary values.}
	\label{fig:tri}
      \end{figure}
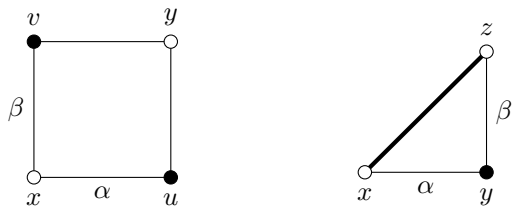

The notion of integrable systems on quad-graph was introduced by Bobenko and Suris \cite{BS} (partly inspired by earlier work of Mercat \cite{M1} on discretizing Riemann surfaces). A quad-graph means a cellular decomposition of a surface with all faces being quadrilateral. It can be obtained from an arbitrary planar graph: first, set up the dual graph of an arbitrary cellular decomposition of a surface called {\em original graph},  then connect the adjacent vertices of original graph and its {\em dual graph}. A quad-graph system is determined by equations of elementary type that are quad equations in the form (possibly depending on the lattice parameters, see Figure \ref{fig:tri})
\begin{equation}
\label{eq:lqeleQ}
Q(x,u,v,y;\alpha,\beta) =0\,. 
\end{equation}

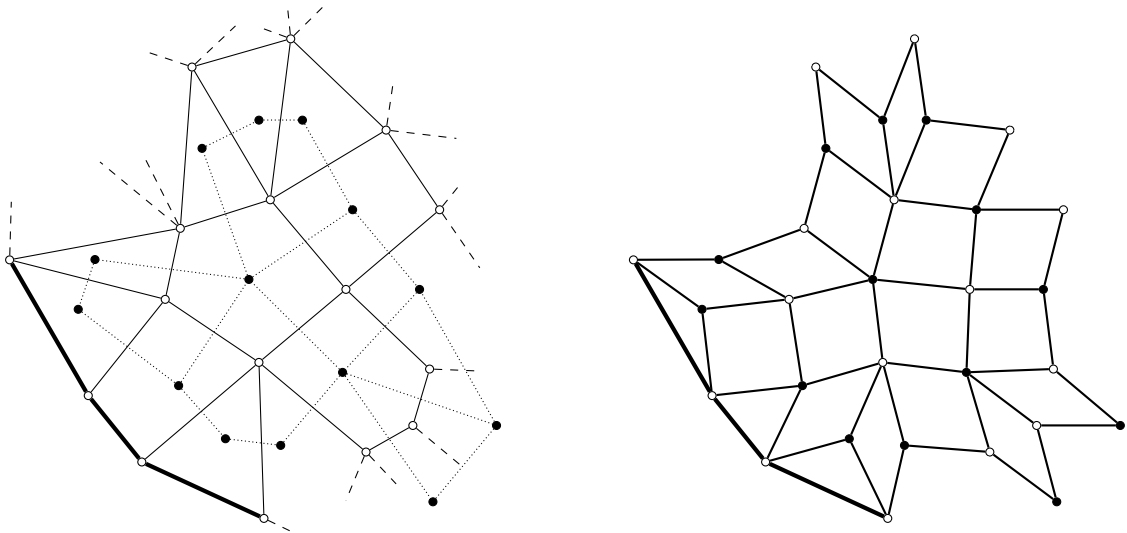
\begin{figure}
 \begin{subfigure}[bt]{0.4\textwidth}
  \begin{tikzpicture}[scale=2.2]
    \tikzstyle{nod}= [circle, inner sep=0pt, minimum size=3pt, draw]
    \tikzstyle{nod1}= [circle, inner sep=0pt, fill=black, minimum size=3pt, draw]
    \tikzstyle{vertex}=[circle,minimum size=20pt,inner sep=0pt]
    \tikzstyle{selected vertex} = [vertex, fill=red!24]
    \tikzstyle{selected edge} = [draw,line width=2pt,-]
    \tikzstyle{edge} = [draw, thin,-,black] 
    \tikzstyle{ddedge} = [draw, densely dotted,-,black] 
    \tikzstyle{dedge} = [draw, dashed,-,black] 
    \tikzstyle{eedge} = [draw, thick,-,black] 

    \pgfmathsetmacro \l {.02}
    \node[nod] (v1) at (4.5*\l,182.9*\l) {};
    \draw[dashed] (v1) -- (5*\l,200.4*\l);
    \node[nod1] (v2) at (30*\l,183*\l) {};
    \node[nod] (v3) at (55.5*\l,192.4*\l) {};
    \node[nod1] (v4) at (62*\l,216.5*\l) {};
    \draw[dashed] (v3) -- (44.5*\l,214.4*\l);
    \draw[dashed] (v3) -- (31.5*\l,212.4*\l);
    \node[nod] (v5) at (59*\l,241*\l) {};
    \draw[dashed] (v5) -- (46*\l,245.4*\l);
    \draw[dashed] (v5) -- (72*\l,253.4*\l);

    \node[nod1] (u1) at (25*\l,168*\l) {};
    \node[nod] (u2) at (51*\l,171*\l) {};
    \node[nod1] (u3) at (76*\l,177*\l) {};
    \node[nod] (u4) at (82.4*\l,201*\l) {};
    \node[nod1] (u5) at (79*\l,225*\l) {};

    \node[nod] (w1) at (28*\l,142*\l) {};
    \node[nod1] (w2) at (55*\l,145*\l) {};
    \node[nod] (w3) at (79*\l,152*\l) {};
    \node[nod1] (w4) at (104*\l,149*\l) {};
    \node[nod] (w5) at (130*\l,150*\l) {};
    \draw[dashed] (w5) -- (145*\l,149.4*\l);
    
    \node[nod] (x1) at (44*\l,122*\l) {};
    \node[nod1] (x2) at (69*\l,129*\l) {};
    \node[nod] (x3) at (80.5*\l,105*\l) {};
    \draw[dashed] (x3) -- (90*\l,100.4*\l);
    \node[nod1] (x4) at (85.5*\l,127*\l) {};
    \node[nod] (x5) at (111*\l,125*\l) {};
    \draw[dashed] (x5) -- (105*\l,110.4*\l);
    \draw[dashed] (x5) -- (120*\l,115.4*\l);
    \node[nod1] (x6) at (131*\l,110*\l) {};
    \node[nod] (x7) at (125*\l,133*\l) {};
    \draw[dashed] (x7) -- (140*\l,120.4*\l);
    \node[nod1] (x8) at (150*\l,133*\l) {};

    \node[nod] (y1) at (88.5*\l,249.5*\l) {};
    \draw[dashed] (y1) -- (80.5*\l,252.5*\l);
    \draw[dashed] (y1) -- (87.5*\l,259.5*\l);
    \draw[dashed] (y1) -- (99.5*\l,260.5*\l);
    \node[nod1] (y2) at (92*\l,225*\l) {};
    \node[nod] (y3) at (117*\l,222*\l) {};
    \draw[dashed] (y3) -- (119*\l,235.5*\l);
    \draw[dashed] (y3) -- (138*\l,219.5*\l);
    \node[nod1] (y4) at (107*\l,198*\l) {};
    \node[nod] (y5) at (133*\l,198*\l) {};
    \draw[dashed] (y5) -- (139*\l,205.5*\l);
    \draw[dashed] (y5) -- (145*\l,180.5*\l);
    \node[nod1] (y6) at (127*\l,174*\l) {};
    \node[nod] (y7) at (105*\l,174*\l) {};

    \draw[thin] (v5) 
    --   (u4) -- (y1) ;
    \draw[thin] (u4) -- (v3);
    \draw[thin] (u4) -- (y7);
    \draw[thin] (u4) -- (y3);
    \draw[thin] (v3) -- (u2);
    \draw[thin] (u2) -- (w3);
    \draw[thin] (u2) -- (v1);
    \draw[thin] (u2) -- (w1);
    \draw[thin] (w3) -- (y7);
    \draw[thin] (y7) -- (y5);
    \draw[thin] (y7) -- (w5);
    \draw[thin] (x7) -- (w5);
    \draw[thin] (x7) -- (x5);
    \draw[thin] (w3) -- (x5);
    \draw[thin] (w3) -- (x1);
    \draw[thin] (w3) -- (x3);
    \draw[thin] (y3) -- (y5);
    \draw[thin] (v1) --(v3) -- (v5) -- (y1)-- (y3);
    \draw[densely dotted] (u3) -- (v2) --(u1) -- (w2) -- (u3)-- (y4)-- (y6)-- (w4) -- (u3);
    \draw[densely dotted] (w2) -- (x2) --(x4) -- (w4);
    \draw[densely dotted] (u3) -- (v4) -- (u5) --(y2) -- (y4);
    \draw[densely dotted] (w4) -- (x6) -- (x8) --(y6);
    \draw[densely dotted] (w4) --  (x8);
    \draw[ultra thick] (v1) -- (w1) -- (x1) -- (x3);
    
  \end{tikzpicture}

\end{subfigure}
\begin{subfigure}[bt]{0.4\textwidth}
\hspace{2cm}  \begin{tikzpicture}[scale=2.2]
    \tikzstyle{nod}= [circle, inner sep=0pt, minimum size=3pt, draw]
    \tikzstyle{nod1}= [circle, inner sep=0pt, fill=black, minimum size=3pt, draw]
    \tikzstyle{vertex}=[circle,minimum size=20pt,inner sep=0pt]
    \tikzstyle{selected vertex} = [vertex, fill=red!24]
    \tikzstyle{selected edge} = [draw,line width=2pt,-]
    \tikzstyle{edge} = [draw, thin,-,black] 
    \tikzstyle{ddedge} = [draw, densely dotted,-,black] 
    \tikzstyle{dedge} = [draw, dashed,-,black] 
    \tikzstyle{eedge} = [draw, thick,-,black] 

    \pgfmathsetmacro \l {.02}
    \node[nod] (v1) at (4.5*\l,182.9*\l) {};

    \node[nod1] (v2) at (30*\l,183*\l) {};
    \node[nod] (v3) at (55.5*\l,192.4*\l) {};
    \node[nod1] (v4) at (62*\l,216.5*\l) {};
    \draw[dashed, white] (v3) -- (44.5*\l,214.4*\l);
    \draw[dashed, white] (v3) -- (31.5*\l,212.4*\l);
    \node[nod] (v5) at (59*\l,241*\l) {};
    \draw[dashed, white] (v5) -- (46*\l,245.4*\l);
    \draw[dashed, white] (v5) -- (72*\l,253.4*\l);

    \node[nod1] (u1) at (25*\l,168*\l) {};
    \node[nod] (u2) at (51*\l,171*\l) {};
    \node[nod1] (u3) at (76*\l,177*\l) {};
    \node[nod] (u4) at (82.4*\l,201*\l) {};
    \node[nod1] (u5) at (79*\l,225*\l) {};

    \node[nod] (w1) at (28*\l,142*\l) {};
    \node[nod1] (w2) at (55*\l,145*\l) {};
    \node[nod] (w3) at (79*\l,152*\l) {};
    \node[nod1] (w4) at (104*\l,149*\l) {};
    \node[nod] (w5) at (130*\l,150*\l) {};
    \draw[dashed, white] (w5) -- (145*\l,149.4*\l);
    
    \node[nod] (x1) at (44*\l,122*\l) {};
    \node[nod1] (x2) at (69*\l,129*\l) {};
    \node[nod] (x3) at (80.5*\l,105*\l) {};
    \draw[dashed, white] (x3) -- (90*\l,100.4*\l);
    \node[nod1] (x4) at (85.5*\l,127*\l) {};
    \node[nod] (x5) at (111*\l,125*\l) {};
    \draw[dashed, white] (x5) -- (105*\l,110.4*\l);
    \draw[dashed, white] (x5) -- (120*\l,115.4*\l);
    \node[nod1] (x6) at (131*\l,110*\l) {};
    \node[nod] (x7) at (125*\l,133*\l) {};
    \draw[dashed, white] (x7) -- (140*\l,120.4*\l);
    \node[nod1] (x8) at (150*\l,133*\l) {};

    \node[nod] (y1) at (88.5*\l,249.5*\l) {};
    \draw[dashed, white] (y1) -- (80.5*\l,252.5*\l);
    \draw[dashed, white] (y1) -- (87.5*\l,259.5*\l);
    \draw[dashed, white] (y1) -- (99.5*\l,260.5*\l);
    \node[nod1] (y2) at (92*\l,225*\l) {};
    \node[nod] (y3) at (117*\l,222*\l) {};
    \draw[dashed, white] (y3) -- (119*\l,235.5*\l);
    \draw[dashed, white] (y3) -- (138*\l,219.5*\l);
    \node[nod1] (y4) at (107*\l,198*\l) {};
    \node[nod] (y5) at (133*\l,198*\l) {};
    \draw[dashed, white] (y5) -- (139*\l,205.5*\l);
    \draw[dashed, white] (y5) -- (145*\l,180.5*\l);
    \node[nod1] (y6) at (127*\l,174*\l) {};
    \node[nod] (y7) at (105*\l,174*\l) {};

    \draw[thick] (v1) -- (v2) -- (v3)    --   (v4) -- (v5) ;
    \draw[thick] (u1) -- (u2) -- (u3)    --   (u4) -- (u5) ;
    \draw[thick] (w1) -- (w2) -- (w3)    --   (w4) -- (w5) ;
    \draw[thick] (x1) -- (x2) -- (x3)    --   (x4) -- (x5) --(x6) --(x7)  --(x8)   ;
    \draw[thick] (y1) -- (y2) -- (y3)    --   (y4) -- (y5) --(y6) --(y7)   ;
    \draw[thick] (v1) -- (u1) -- (w1)    --   (v1) ;
    \draw[thick] (v2) -- (u2) -- (w2);
    \draw[thick] (v3) -- (u3) -- (w3) --(x2);
    \draw[thick] (v4) -- (u4) -- (y4);
    \draw[thick] (v5) -- (u5) -- (y1);
    \draw[thick] (w3) --(x4);
    \draw[thick] (w1) -- (x1) -- (w2) ;
    \draw[thick] (x1) -- (x3);
    \draw[thick] (u4) -- (y2);
    \draw[thick] (u3) -- (y7) -- (y4);
    \draw[thick] (y7) -- (w4) -- (x5);
    \draw[thick] (w4) -- (x7);
    \draw[thick] (y6) -- (w5) --(x8);

        \draw[ultra thick] (v1) -- (w1) -- (x1) -- (x3);
  \end{tikzpicture}
\end{subfigure}
\caption{A quad-graph with boundary: the left figure shows the original graph (white dots) with a boundary (thick line), and its dual graph (black dots); the right figure shows the resulting quad-graph with a boundary by connecting the adjacent vertices of different  types (black or white dots). }
\label{fig:qgwb}
\end{figure}

Quad-graphs were constructed for original graphs  {\bf without boundary}. When taking an original graph with boundary, it was realized in \cite{CCZ} that the boundary should be represented by triangles as illustrated by  Figure \ref{fig:qgwb}. 
Then, a quad-graph system with  boundary is determined by two  types of elementary equations: the quad equations representing the {\em bulk} dynamics, and the triangle equations, called {\em boundary equations} in this context,  representing the {\em boundary conditions}. A boundary equation with dependence on the lattice  parameters is expressed as (see Figure \ref{fig:tri})
\begin{equation}
\label{eq:lqeleq}
 q(x, y,z; \a, \b)=0\,.
\end{equation}
By convention, the first and third arguments in $q$, \ie $x,z$ in \eqref{eq:lqeleq},  are the values at the boundary vertices, and $\alpha,\beta$ are the lattice parameters attached to edges
connecting the value $y$ to the boundary values $x$ and $z$, respectively.

Cauchy problems for generic (integrable or nonintegrable) quad-graph systems without boundary were investigated in \cite{AV} with the help of characteristic lines connecting the lattice parameters. A constructive approach was provided in \cite{VdK2}. In a broader geometric setting, the quad-graph systems appeared as a natural approach to discretizing Riemann surfaces \cite{M1, BG1} (see also the monograph \cite{BS11}). The question of connecting quad-graph systems with concrete examples of discrete Riemann surfaces and discrete complex analysis has its own significance that is well beyond the original motivations of solving initial value problems for quad-graph systems, \cf \cite{BIts1}. 

Similar questions could be asked for quad-graph systems with boundary. Some basic examples of well-posed initial-boundary value problems were given in \cite{CCZ, CVZ1}. However, a general criterion for the well-posedness of initial-boundary problems for quad-graphs with boundary still remains to be investigated. And the connections of quad-graphs with boundary to discrete Riemann surfaces and discrete complex analysis are  completely open. The present work is  partly motivated by these tantalizing perspectives. 
     
\subsection{Consistency as integrability criterion}
\begin{figure}[th]
	\centering
	\begin{tikzpicture}[scale=.55, decoration={markings,mark=at position 0.55 with {\arrow{latex}}}]
	\tikzstyle{nod1}= [circle, inner sep=0pt, fill=white, minimum size=5pt, draw]
	\tikzstyle{nod}= [circle, inner sep=0pt, fill=black, minimum size=5pt, draw]
	\def\lx{3}%
	\def\ly{1.22}%
	\def\lz{ (sqrt(\x*\x+\y*\y))}%
	\def\l{4}%
	\def\d{4}%
	\coordinate (u00) at (0,0);
	\coordinate (u10) at (\l,0);
	\coordinate (u01) at (\lx,\ly);
	\coordinate (u11) at (\l+\lx,\ly);
	\coordinate (v00) at (0,\d);
	\coordinate (v10) at (\l,\d);
	\coordinate (v01) at (\lx,\d+\ly);
	\coordinate (v11) at (\l+\lx,\d+\ly);
	\draw[-]  (u10)  
	-- (u11) ;
	\draw [dashed]  (u01)  -- (u11);
	\draw [dashed]  (u00)--node [above]{$\alpha_2$}(u01) ;
	\draw[-] (v00) --  (v10) -- (v11) -- (v01) -- (v00);
	\draw[-] (u11) --  (v11);
	\draw[-] (u00) -- node [left]{$\alpha_3$}(v00);
	\draw[-] (u10) -- (v10);
	\draw[dashed] (u01) -- (v01);
	\coordinate (u011) at (1.5*\lx,1.5* \ly);
	\draw[-] (u00) 
	-- node [below]{$\alpha_1$} (u10);
	\node[nod] (v00) at (0,\d) [label=above: $x_3$] {};
	\node[nod1] (v10) at (\l,\d) [label=above: $x_{13}$] {};
	\node[nod1] (v01) at (\lx,\d+\ly) [label=above: $x_{23}$] {};
	\node[nod] (v11) at (\l+\lx,\d+\ly) [label=above: $x_{123}$] {};
	\node[nod1] (u00) at (0,0) [label=below: $x$] {};
	\node[nod] (u10) at (\l,0) [label=below: $x_1$] {};
	\node[nod] (u01) at  (\lx,\ly) [label=below: $x_2$] {};
	\node[nod1] (u11) at (\l+\lx,\ly) [label=below: $x_{12}$] {};
	\end{tikzpicture}
	\caption{$3$D consistency:  $x$ can be seen as a discrete field living on a $3$D lattice with $x:=x(n,m,l)$, $x_{1}:=x(n+1,m,l)$, $x_{2}:=x(n,m+1,l)$ \dots The parameters $\alpha_1,\alpha_2,\alpha_3$ remain the same on the opposite edges. 
} \label{fig:3d}
\end{figure}
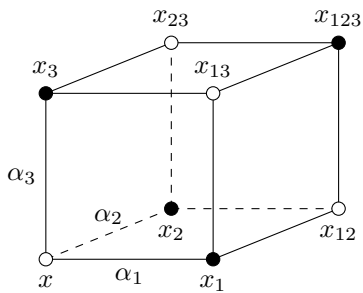

We take the integrability of quad equations as synonymous with the $3$D consistency. By $3$D consistency, it means that a quad equation $Q=0$ can be consistently imposed on a cube \cite{nijhoff2002lax, BS}. When taking $Q=0$ as a two-dimensional lattice equation depending on the lattice parameters,  the $3$D consistency implies that a collective shift, or {\em covariance}, of $Q=0$   in a multi-dimensional lattice is an auto-B\"acklund transformation  (see Figure \ref{fig:3d}).  For a scalar quad equation,  this interpretation, together with some extra assumptions of $Q$ such as irreducibility and  affine-linearity, allows us  to express $Q=0$ in terms of the {\em discrete zero curvature conditions}. Namely, the equation $Q(x,x_1,x_2,x_{12};\alpha_1,\alpha_2)=0$ is a result of the constraint 
\begin{equation}
  \label{eq:3mllm1}
  \mathbi{g}_{(x_2, x_{12};\alpha_1)}(\alpha_3)\, \mathbi{g}_{(x, x_2;\alpha_2)}(\alpha_3) =  \mathbi{g}_{(x_1, x_{12};\alpha_2)}(\alpha_3)\, \mathbi{g}_{(x, x_1;\alpha_1)}(\alpha_3)\,,  
\end{equation} where $ \mathbi{g}$ is certain transition matrix in forms of M\"obius transformation acting on the multi-dimensional lattice. For instance, $x_{13} =  \mathbi{g}_{(x, x_{1};\alpha_1)}(\alpha_3)[ x_3]$, 
where $\mathbi{g}_{(x, x_{1};\alpha_1)}(\alpha_3)$ is attached to the edge $(x, x_1;\alpha_1)$ with $\alpha_3$ being the B\"acklund (spectral) parameter. 

Based on the $3$D consistency condition, the ABS classification \cite{adler2003classification}, up to few extra assumptions such as the $\cD_4$-symmetry, affine-linearity and tetrahedron property (these properties are also discussed in Section \ref{sec:31}),  exhausted all scalar integrable quad equations.  The classification contains nine equations (their canonical forms are provided in Appendix \ref{sec:app1}) called respectively Q4, Q3($\delta$), Q2, Q1($\delta$) as {\bf Q-type} equations, A2, A1($\delta$) as {\bf A-type} equations, and H3($\delta$), H2, H1 as {\bf H-type} equations. Thees equations are discrete analogs of the (continuous) Korteweg-de Vries (KdV)-type equations. For instance, the Q4 equation \cite{Q4, AS11} as the ``master'' equation in the classificaiton is the lattice version of the famous Krichever-Novikov equation \cite{KN11};  Q1($0$), H3($0$) and H1 are in connection to the Schwarzian KdV, modified KdV, and KdV equations respectively \cite{NC1}. A follow-up study of the classification, by relaxing the $\cD_4$-symmetry and tetrahedron property, was taken following a more in-depth analysis of multi-affine polynomials and their invariants \cite{ABSII}. In particular, an extra list of integrable quad equations was provided, and named {\bf H$^\epsilon$-type} (see Section \ref{sec:cd4}). The list contains three rhombic-symmetric equations, and can be seen as an extension of the H-type list.  These integrable quad equations are the main objects of study of the paper.

\begin{figure}[th]
  \centering
  \begin{subfigure}[b]{0.4\textwidth}
  \begin{tikzpicture}[scale=2.2]
    \tikzstyle{nod1}= [circle, inner sep=0pt, minimum size=5pt, draw]
    \tikzstyle{nod}= [circle, inner sep=0pt, fill=black, minimum size=5pt, draw]
    \tikzstyle{vertex}=[circle,minimum size=20pt,inner sep=0pt]
    \tikzstyle{selected vertex} = [vertex, fill=red!24]
    \tikzstyle{selected edge} = [draw,line width=1.5pt,-]
    \tikzstyle{edge} = [draw, thick,-,black] 
    \tikzstyle{ddedge} = [draw,  densely dotted,-,black] 
    \tikzstyle{dedge} = [draw, dashed,-,black] 
    \tikzstyle{eedge} = [draw,line width=1.5pt,-,black] 

    \pgfmathsetmacro \ll {6}
    \pgfmathsetmacro \rl {7}
    \pgfmathsetmacro \h {2}
    \pgfmathsetmacro \an {32}
    \pgfmathsetmacro \bn {16}
    \pgfmathsetmacro \cn {18}
    \pgfmathsetmacro \ra {.2}
    \pgfmathsetmacro \s {.4}

    \pgfmathsetmacro \llx {{\ra*\ll*cos(\an)}}
    \pgfmathsetmacro \lly {{\ra*\ll*sin(\an)}}
    \pgfmathsetmacro \rlx {{\ra*\rl*cos(\bn)}}
    \pgfmathsetmacro \rly {{\ra*\rl*sin(\bn)}}
    \pgfmathsetmacro \dh {{\h*tan(\cn)}}
    \node[nod1] (v) at (0,0) [label=right:$x$] {};
    \node[nod] (v1) at (-\llx,\lly) [label=left:$y$] {};
    \node[nod1] (v14) at (\rlx-\llx,\rly+\lly) [label=right:$z$] {};
    \draw[dedge] (v1) --node [above] {$\beta$} (v14)  ;
    \draw[edge] (v1) -- node [below] {$ \alpha$} (v) ;
    \draw[eedge] (v) --  (v14)  ;
    \node[nod1] (w) at (0,\h) [label=right: $r$] {};
    \node[nod] (w1) at (-\llx,\lly+\h) [label=left: $s$] {};
    \node[nod1] (w14) at (\rlx-\llx,\rly+\lly+\h) [label=above:$t$] {};
    \draw[edge] (w1) -- node [above]{$\beta$}(w14);
    \draw[edge] (w1) -- node [below]{$\alpha$}(w) ;
    \draw[eedge] (w) -- (w14) ;
    \node[nod] (u1) at (-\dh,\h/2) [label=above:$u$] {};
    \draw[edge] (v) --  node [left] {$\lambda$} (u1) -- node [left] {$\eta$} (w);
    \draw[eedge] (v) -- (w);
    \node[nod1] (u5) at (-\dh -\llx,\h/2+\lly) [label=left:$v$] {};
    \node[nod] (u6) at (-\dh+\rlx-\llx,\rly+\lly+\h/2) [label= left :$w$] {};
    \draw[edge] (u1) -- node [below]{$\alpha$} (u5);
    \draw[dedge] (u5) --node [above]{$\beta$} (u6);
    \draw[edge] (v1) -- node [left]{$\lambda$}   (u5) --  node [left]{$\eta$} (w1);
    \draw[dedge] (v14) --  node [right]{$\lambda$}  (u6) -- node [left]{$\eta$}  (w14) ;
    \draw[eedge] (v14)  --  (w14) ;
    \draw [dotted, opacity =.4] (0,0) --(0,\h) -- (\rlx-\llx,\rly+\lly+\h)                  --  (\rlx-\llx,\rly+\lly) --cycle;
  \end{tikzpicture}
\end{subfigure}
\begin{subfigure}[b]{0.4\textwidth}
  \begin{tikzpicture}[scale=1.1]
    \tikzstyle{nod1}= [circle, inner sep=0pt, minimum size=5pt, draw]
    \tikzstyle{nod}= [circle, inner sep=0pt, fill=black, minimum size=5pt, draw]
    \tikzstyle{vertex}=[circle,minimum size=20pt,inner sep=0pt]
    \tikzstyle{selected vertex} = [vertex, fill=red!24]
    \tikzstyle{selected edge} = [draw,line width=1.5pt,-]
    \tikzstyle{edge} = [draw, thick,-,black] 
    \tikzstyle{ddedge} = [draw, densely dotted,-,black] 
    \tikzstyle{dedge} = [draw, dashed,-,black] 
    \tikzstyle{eedge} = [draw, line width=1.5pt,-,black] 
    \pgfmathsetmacro \lv {1.1}
    \pgfmathsetmacro \lh {0.8}
    \pgfmathsetmacro \ld {2}
    \pgfmathsetmacro \an {38}

    \pgfmathsetmacro \llx {{\ld*tan(\an)}}
    \pgfmathsetmacro \lly {{\ld/sin(\an)}}
    \node (em) at (0,0) [label=right:${}$] {};
    \node[nod1] (z) at (0+\lh,\lv) [label=below:$z$] {};
    \node[nod1] (x) at (2*\ld+\lh,0+\lv) [label=below:$x$] {};
    \node[nod1] (bz) at (0+\lh,2*\ld+\lv) [label=above:$t$] {};
    \node[nod1] (bx) at (2*\ld+\lh,2*\ld+\lv) [label=above:$r$] {};
    \draw[eedge] (z) -- node[below]{\footnotesize $\alpha = \sigma(\beta)       $}   (x)  -- node[right]{\footnotesize $\eta = \sigma(\lambda)       $}   (bx) --node[above]{\footnotesize $\beta = \sigma(\alpha)       $}   (bz) --node[left]{\footnotesize $\lambda = \sigma(\eta)       $}   (z);

    \node[nod] (by) at (0+\lh+\ld,\lv+\lly) [label=below right:$s$] {};
    \node[nod] (w) at (2*\ld+\lh-\lly,0+\lv+\ld) [label=left:$w$] {};
    \node[nod] (u) at (0+\lh+\lly,\lv+\ld) [label=right:$u$] {};
    \node[nod] (y) at (0+\lh+\ld,2*\ld+\lv-\lly) [label=above right :${y}$] {};
    \draw[edge] (bz) -- node [below left] {$\beta$} (by)  -- node [below right] {$\alpha$}(bx) -- node [left] {$\eta$} (u) --node [left] {$\lambda$} (x) --node [above right] {$\alpha$} (y) --node [above left] {$\beta$} (z) -- node [right] {$\lambda$} (w) -- node [right] {$\eta$}(bz);
    \node[nod1] (v) at (0+\lh+\ld,\lv+\ld) [label=below right:$v$] {};
    \draw[edge] (w) --  
    (v) --  
    (u);
    \draw[edge] (by) --  
    (v) -- 
    (y);

    \coordinate (Y0) at (2*\ld+\lh,\ld+\lv);
    \coordinate (Y1) at (2*\ld/2+\lh/2+0+\lh/2+\lly/2,2*\ld/2+\lv/2+\lv/2+\ld/2);
    \coordinate (Y2) at ( \lh/2+\ld/2+\lh/2+\ld/2,\lv/2+\lly/2+\lv/2+\ld/2);
    \coordinate (Y3) at ( 2*\ld/2+\lh/2-\lly/2+\lh/2,2*\ld/2+\lv/2+\lv/2+\ld/2);
    \coordinate (Y4) at (0+\lh,2*\ld/2+\lv);
    \coordinate (Y5) at ( 2*\ld/2+\lh/2-\lly/2+\lh/2,\lv/2+\lv/2+\ld/2);
    \coordinate (Y6) at ( \lh/2+\ld/2+\lh/2+\ld/2,2*\ld/2+\lv/2-\lly/2+\lv/2+\ld/2);
    \coordinate (Y7) at  (2*\ld/2+\lh/2+0+\lh/2+\lly/2,\lv/2+\lv/2+\ld/2);
 \path [draw=black,  thick, densely dotted, rounded corners=6pt, postaction={on each segment={mid arrow=black}}] (Y0) -- (Y1) --(Y2) -- (Y3) -- (Y4) --  (Y5) -- (Y6)-- (Y7) -- (Y0);

    \coordinate (X0)  at (\ld+\lh,0+\lv);
    \coordinate (X1) at (\lh/2+\ld/2+2*\ld/2+\lh/2, 2*\ld/2+\lv/2-\lly/2+\lv/2);
  \coordinate (X2) at (\lh/2+\ld/2+\lh/2+\lly/2, \lv/2+\ld/2+\lv/2+\ld/2);
  \coordinate (X3) at ( 2*\ld/2+\lh/2+\lh/2+\ld/2,2*\ld/2+\lv/2+\lv/2+\lly/2 );
  \coordinate (X4) at  (0+\lh/2+0+\lh/2+\ld/2,2*\ld/2+\lv/2+\lv/2+\lly/2);
  \coordinate (X41) at (\ld+\lh,0+2*\ld+\lv);
  \coordinate (X5) at  (0+\lh/2+\ld/2+2*\ld/2+\lh/2-\lly/2,\lv/2+\ld/2+0+\lv/2+\ld/2);
  \coordinate (X6) at (\lh/2+\ld/2+\lh/2,\lv/2+2*\ld/2+\lv/2-\lly/2);
 \path [draw=black,  thick, densely dotted, rounded corners=6pt, postaction={on each segment={mid arrow=black}}] (X0) -- (X1) --(X2) -- (X3) -- (X41) --  (X4) -- (X5)-- (X6) -- (X0);

  \end{tikzpicture}
\end{subfigure}
  \caption{Boundary consistency around a half of a rhombic dodecahedron (left) and its planar projection (right), where $Q=0$ is imposed on  four quadrilaterals and $q=0$ is imposed on four triangles. The (closed) characteristic lines (the dotted lines) are reflected at the boundary edges. } 
    \label{fig:bccy}
\end{figure}
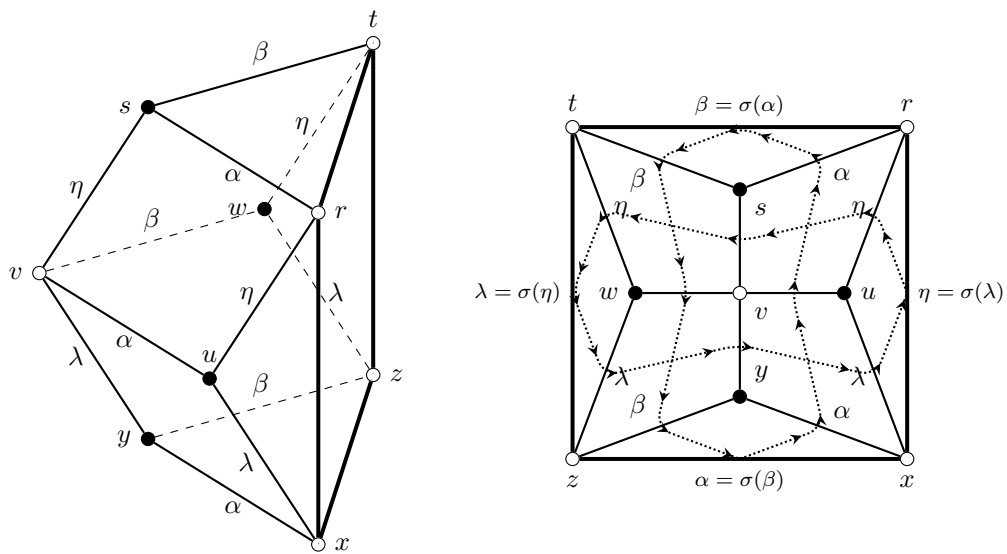

As to quad-graph systems with boundary, the integrability criterion for boundary equations, known as the boundary consistency condition,  was introduced in \cite{CCZ}.  It emerged from  the notions of {\em set-theoretical reflection equation} and its solutions as {\em reflection maps} \cite{CHE, sklyanin1988boundary, CZ} 
as companions to the  {\em set-theoretical Yang–Baxter equation} \cite{Drinfeld1} and its solutions as  {\em Yang-Baxter maps} \cite{veselov2003yang}. In a similar way to connecting integrable quad equations with Yang-Baxter maps \cite{papageorgiou2006yang}, integrable boundary equations were connected with reflection maps, and the integrability condition inherits from that of the {\em set-theoretical reflection equation}, which can be imposed on a half of a rhombic dodecahedron  (see Figure \ref{fig:bccy}) \cite{CCZ1}. Here, we adapt the definition given in \cite{CCZ} to our context. 

\begin{definition}\label{def:bc1}
  A {\bf nondegenerate} boundary equation $q=0$ is {\bf   boundary consistent} with an integrable quad equation $Q=0$ if there is an involution relation $\sigma$ between the parameters  such that the initial value problem on the half rhombic dodecahedron in Figure \ref{fig:bccy} with $\beta=\sigma(\alpha)$ and $\eta =\sigma(\lambda)$ is well-posed, \ie the three ways of computing $t$ from initial values $x, y, u$ and $\alpha,\lambda$ yield the same result. 
  A boundary equation which is boundary consistent is called {\bf integrable}.
\end{definition}

The nuance between the above definition and that give in \cite{CCZ} is the requirement that $q=0$ is {\em nondegenerate}, which will clarified in Definition \ref{def:nondeg} below. Therefore, the above definition of boundary consistency based on the involution relation $\sigma$ is reserved to nondegenerate boundary equations.  For nondegenerate $q=0$, the involution relation $\sigma$, as consistency between the parameters (see Figure \ref{fig:bccy}), is taken as part of an integrable boundary equation \cite{CCZ}. Comparing to \eqref{eq:lqeleq}, an integrable boundary equation is simply put as
\begin{equation}
  q(x,y,z;\alpha) = 0\,,
\end{equation}
with $\beta = \sigma(\alpha)$. 
In the picture of classical soliton dynamics, $\sigma$ can be understood as the change of velocities when solitons interacting with an integrable boundary. We refer readers to \cite{CVZ1} for the origin and significance of $\sigma$ in this respect. Despite of its importance, $\sigma$ obtained in some known examples  was largely based on guesswork \cite{CCZ}.  %
In this paper, a systematic derivation of $\sigma$ is provided.

As to degenerate $q=0$,  a class of boundary equations, known as   degenerate  integrable boundary equations, is shown to be boundary consistent with  $Q=0$ in this paper as well. However, it turns out that  the involution relation $\sigma$  is no longer needed for degenerate  integrable boundary equations (see Section \ref{sec:degbc}). 

\subsection{Preliminary results}\label{sec:prri}
Various aspects of integrable boundary equations were obtained in \cite{CCZ, CVZ1}. These include the notions of {\em dual boundary equations} and {\em discrete boundary zero curvature conditions}. A partial list of integrable boundary equations was also given for quad equations in ABS classification.  We illustrate these results by taking  Q1($0$) as our primary example.

 Q1($0$), also known as the lattice Schwarzian KdV,  or cross-ratio  equation, reads $Q = 0$ with
\begin{equation}
 Q:= Q(x,u,v,y;\alpha,\beta) = \alpha (x-v)(u-y)- \beta(x-u)(v-y)\,. 
\end{equation}
It has a pair of integrable boundary equations $p =0$ and $ q=0$ with  $\sigma(\alpha) =  \mu^2/\alpha$
\begin{equation}\label{eq:prr1}
  q(x, y , z;\alpha) =  \mu( x-y) + \alpha (y-z)\,, \quad
  p(x,y,z;\alpha) =  \alpha (y-z  ) + \mu(y -z)\,,
\end{equation}
where $\mu\neq 0$ is an arbitrary constant. They have the following  properties.
\begin{description}
\item[Duality property:]   $q=0$ and $p=0$ are dual in the following sense: taking the compatibility of
  \begin{subequations}
  \begin{equation}
   Q(x,u,v,y;\alpha,\sigma(\alpha)) = 0\,,\quad  q(x,u,y;\alpha) = 0\,. 
  \end{equation}
Here, the compatibility means to put $Q =0$ and $q=0$  on the same quadrilateral. By eliminating, for instance,  $y$, the remaining equation is proportional to $p(u,x,v;\alpha) = 0$. Similarly, taking the compatibility of
  \begin{equation}
   Q(x,u,v,y;\alpha,\sigma(\alpha)) = 0\,,\quad  p(u,x,v;\alpha) = 0\,,
  \end{equation}
  by eliminating $v$, the remaining equation is proportional to $q(x,u,y;\alpha) = 0$. For a given integrable quad equation $Q=0$ and an integrable boundary equation $q=0$, one could obtain a dual $p = 0$ which is also an integrable boundary equation, and {\em vice verse}.
  \end{subequations}

\item[Discrete boundary zero curvature conditions:] consider the pair of dual integrable boundary equations   $p = 0$ and $ q = 0$ given in \eqref{eq:prr1}. Taking, for instance, $p(x,y,z;\alpha) =0$, one can express $z$ as
\begin{equation}
  \label{eq:4k}
z = \mathbi{k}_{y}(\alpha)[x]\,,
  \end{equation}
where $\mathbi{k}_{y}(\alpha)$ is called {\em boundary matrix} that is attached to the value $y$ with $\alpha$ being a parameter. Clearly, it can be set up as a M\"obius transformation.  One requires that the action of a boundary matrix is  accompanied by a change of parameter following $\alpha \to \sigma(\alpha)$. Then, the integrable boundary equation $q=0$,  dual to $p =0$,  can be cast into 
\begin{equation}
  \label{eq:3mllm}
      \mathbi{g}_{(u,y;\sigma(\alpha))}(\sigma(\lambda))\,\mathbi{g}_{(x,u;\alpha)}(\sigma(\lambda))\,
 \mathbi{k}_{x}(\lambda) =
 \mathbi{k}_{y}(\lambda)      \, \mathbi{g}_{(u,y;\sigma(\alpha))}(\lambda)\,\mathbi{g}_{(x,u;\alpha)}(\lambda)\,,
\end{equation}
which holds precisely when $q(x,u,y;\alpha)=0$ \cite{CVZ1}. The above equation is referred to as the discrete boundary zero curvature condition for $q=0$, where the boundary matrix is obtained using its dual boundary equation $p=0$. 

\end{description}

In \cite{CVZ1}, the notion of duality was formulated in a more general setting, and the  discrete boundary zero curvature condition was constructed following a {\em dual boundary consistency}.  Moreover, a special type of initial-boundary value problems for quad-graph systems on a trip with two parallel boundaries was investigated,  which led to to the {\em open boundary reduction} technique, as an alternative to the {\em periodic reduction}. Some nontrivial examples of integrable mappings   \cite{CVZ1, vdk3} were accordingly obtained.

In this paper, based on a systematic characterization of boundary equations in terms of multivariate polynomials, the duality property is formulated as the factorization property of quad equations. The role of the dual boundary equations played in the boundary consistency is also clarified. 

\subsection{Plan of the  paper}
The aim of the paper is to classify integrable boundary equations for quad equations in the ABS classification. 
In Section \ref{sec:2fact}, after giving necessary ingredients to the ABS classification (detailed expressions are given in Appendix \ref{sec:app1}), we provide the notions of boundary polynomials and boundary equations, and formulate the duality property as factorization of quad equations. This leads to pairs of boundary equations dual to each. We classify all such pairs for quad equations in the ABS classification. The results are given in Appendix \ref{app:cbp} and \ref{app:hp}.
In Section \ref{sec:fcrd}, for a given quad equation, we provide a criterion of selecting the integrable ones among the list of dual boundary equations. This relies on the factorization of the consistent system around a rhombic dodecahedron into two equivalent halves. The construction gives rise to the involution relation $\sigma$ needed in Definition \ref{def:bc1}. The degenerate boundary equations are also considered, and contribute to degenerate integrable boundary equations as solutions to the boundary consistency.
In Section \ref{sec:cd4}, we derive integrable boundary equations for the  H$^\epsilon_1$ equation. Since  H$^\epsilon_1$  is rhombic-symmetric only, 
one can obtain two sets of boundary consistency conditions depending on the {\em patterns} of the quadrilaterals and  triangles. This allows us to design some interesting initial-boundary value problems on quad-graphs on a strip with two parallel boundaries that consist of different patterns of triangles. Section \ref{sec:conc} contains concluding remarks.

\section{Factorization of quad equations}
\label{sec:2fact}
In this section, we provide the notions of {\em boundary polynomials} and {\em boundary equations} that are the natural objects to characterize discrete boundary conditions for quad-graph systems with  boundary.  In the context of quad equations in the ABS classification, we provide systematic methods for factorizing the quad equations, which leads to  pairs of boundary equations dual to each other.  Factorization of quad equations plays a crucial role in the derivation of integrable boundary equations. The exhausted list of boundary equations satisfying the factorization property is obtained and provided in Appendix \ref{app:cbp} (for Q-type and A-type equations) and \ref{app:hp} (for H-type equations).

\subsection{ABS classification}\label{sec:31}
Let us first collect some useful ingredients of the ABS classification. Their canonical forms (with respect to simultaneous M\"obius transformations) of types Q, H and A are given in Appendix \ref{sec:app1}. Details of the classification results  can be found in the original papers  \cite{adler2003classification, ABSII}.

A quad equation $Q=0$ in the ABS classification   is in the form \eqref{eq:lqeleQ} (see Figure~\ref{fig:tri}).
The arguments $x,u,v,y$ are understood as discrete fields, and $\alpha, \beta$ are the lattice parameters. Each argument of $Q$ is living in the complex projective space $\CC \PP^1$,    and $Q$ is an irreducible  polynomial affine-linear with respect to each of its arguments.  In our context of quad-graph systems with  boundary, we refer to such $Q$ as {\bf bulk polynomial} and to $Q=0$ as the associated {\bf bulk equation}. Due to the irreducibility of $Q$, one  requires  $\alpha \neq \beta$\footnote{One can check by direct computations that all bulk polynomials listed in Appendix \ref{sec:app1} can be  reduced to
  \[
    Q\vert_{\alpha=\beta} \propto (u - v) (x - y)\,. 
  \]
}.       
Here, we briefly present a series of properties of $Q$ that is of crucial importance in the rest of the paper.
\begin{description}

\item[${\cal D}_4$-symmetry:] a bulk polynomial $Q$ obeys
  \begin{subequations}\label{eq:QQpm}
  \begin{align}
\label{eq:QQs}    Q(x,u,v,y;\alpha,\beta) &=   Q(u,x,y,v;\alpha,\beta)\,,   \\
\label{eq:QQs1}    Q(x,u,v,y;\alpha,\beta) &=  - Q(x,v,u,y;\beta,\alpha)\,. 
\end{align}
    
  \end{subequations}

\item[Biquadratic polynomials:] let a bulk polynomial $Q:=Q(x,u,v,y;\alpha,\beta)$   
  be given,  one can associate the edges and diagonals of the underlying quadrilateral with biquadratic polynomials that are obtained using a discriminant-type operator $ \delta_{\cdot,\cdot}  $ which eliminates two of the four arguments. For instance, one has (with the subscripts denoting partial derivatives)
  \begin{equation}\label{eq:dvy}
\delta_{v,y}Q: = Q Q_{vy} -Q_vQ_y\,,    
  \end{equation}
as a biquadratic polynomial in $x,u$  attached to the edge connected by $ x$ and $u$, and
  \begin{equation}\label{eq:duv}
\delta_{u,v}Q: = QQ_{uv} -Q_uQ_v\,,    
  \end{equation}  as a biquadratic polynomial in $x,y$  attached to the diagonal connected by $ x$ and $y$. 
  The biquadratic polynomials attached to other edges ({\em resp}. to the other diagonal)  can be similarly derived, and have similar forms as \eqref{eq:dvy} ({\em resp}. as \eqref{eq:duv}) due to the $\cD_4$-symmetry. 
\item[Singular solutions:] a solution $(x,u,v,y)$ of   \eqref{eq:lqeleQ} is said to be {\em singular} with respect to an argument, say $x$, if $Q=0$ holds independently of $x$, \cf \cite{ABSII}.  

The vanishing biquadratic polynomial leads to singular solutions of the bulk equation $Q=0$. For instance, if $(x,u) = (X,U)$  such that $\delta_{v,y}Q = 0$, then  $Q$ becomes reducible and can be factorized as a product of two polynomials
  \begin{equation}
    Q\vert_{\delta_{v,y}Q = 0} =F(X,U,v;\alpha,\beta) G(X,U,y;\alpha,\beta)\,. 
  \end{equation}
  In this case,  $G(X,U,y;\alpha,\beta) = 0$ provides a singular solution of $Q = 0$ with respect to $v$. Similarly,   $F(X,U,v;\alpha,\beta) = 0$ provides a solution singular with respect to $y$.

\item[Tetrahedron property:] for a given bulk equation $Q = 0$, it follows from the $3$D consistency property that one can consistently assign values to all vertices on a cube  using, for instance, $x, x_1, x_2, x_3$ as well as the parameters $\alpha_1, \a_2, \alpha_3$ as initial data (see Figure \ref{fig:31tetra}). By tetrahedron property, we mean that the  values on the vertices connected by the diagonals of the faces of the cube (which form a tetrahedron)  satisfy certain bulk equation of  Q-type\footnote{Apparently, the diagonal biquadratic polynomials of all bulk equations coincide with the  edge biquadratic polynomials of certain Q-type equation. Since a bulk equation can be identified with its edge biquadratic polynomials, the equation $Q^\intercal  = 0$ relating the values on the tetrahedron (connected by the black or white vertices on the cube in Figure \ref{fig:31tetra}) is a Q-type equation.}. We use $Q^\intercal = 0$ to denote such equation that is called the {\bf tetrahedron equation} of $Q=0$. 
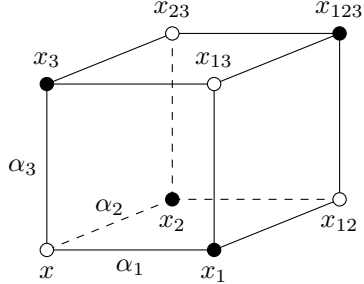
\begin{figure}[h]
	\centering
	\begin{tikzpicture}[scale=.55, decoration={markings,mark=at position 0.55 with {\arrow{latex}}}]
	\tikzstyle{nod1}= [circle, inner sep=0pt, fill=white, minimum size=5pt, draw]
	\tikzstyle{nod}= [circle, inner sep=0pt, fill=black, minimum size=5pt, draw]
	\def\lx{3}%
	\def\ly{1.22}%
	\def\lz{ (sqrt(\x*\x+\y*\y))}%
	\def\l{4}%
	\def\d{4}%
	\coordinate (u00) at (0,0);
	\coordinate (u10) at (\l,0);
	\coordinate (u01) at (\lx,\ly);
	\coordinate (u11) at (\l+\lx,\ly);
	\coordinate (v00) at (0,\d);
	\coordinate (v10) at (\l,\d);
	\coordinate (v01) at (\lx,\d+\ly);
	\coordinate (v11) at (\l+\lx,\d+\ly);
	\draw[-]  (u10)  
	-- (u11) ;
	\draw [dashed]  (u01)  -- (u11);
	\draw [dashed]  (u00)--node [above]{$\alpha_2$}(u01) ;
	\draw[-] (v00) --  (v10) -- (v11) -- (v01) -- (v00);
	\draw[-] (u11) --  (v11);
	\draw[-] (u00) -- node [left]{$\alpha_3$}(v00);
	\draw[-] (u10) -- (v10);
	\draw[dashed] (u01) -- (v01);
	\coordinate (u011) at (1.5*\lx,1.5* \ly);
	
	\draw[-] (u00) 
	-- node [below]{$\alpha_1$} (u10);
	\node[nod] (v00) at (0,\d) [label=above: $x_3$] {};
	\node[nod1] (v10) at (\l,\d) [label=above: $x_{13}$] {};
	\node[nod1] (v01) at (\lx,\d+\ly) [label=above: $x_{23}$] {};
	\node[nod] (v11) at (\l+\lx,\d+\ly) [label=above: $x_{123}$] {};
	\node[nod1] (u00) at (0,0) [label=below: $x$] {};
	\node[nod] (u10) at (\l,0) [label=below: $x_1$] {};
	\node[nod] (u01) at  (\lx,\ly) [label=below: $x_2$] {};
	\node[nod1] (u11) at (\l+\lx,\ly) [label=below: $x_{12}$] {};
      \end{tikzpicture}
	\caption{Tetrahedron property: with initial data $x, x_1, x_2,x_3$ and parameters $\alpha_1, \alpha_2,\alpha_3$,  all values on the cube can be consistently assigned.  Then, the set of values   $x_1, x_2, x_3, x_{123}$ and  $x, x_{12}, x_{13}, x_{23}$ satisfy certain Q-type bulk equation.} \label{fig:31tetra}
\end{figure}

For examples, let $Q$ be  H1, then $x_1, x_2, x_3, x_{123}$ that are the black dots in Figure \ref{fig:31tetra} satisfy
\begin{subequations}\label{QIC}
\begin{equation}\label{eq:QIC}
  Q^\intercal(x_{123},x_1,x_2,x_3;\alpha_3 -\alpha_2,\alpha_3 -\alpha_1  ) = 0\,,
\end{equation}
where $Q^\intercal$ is the Q1($0$) polynomial.  Alternatively, the white dots $x, x_{12}, x_{13}, x_{23}$   satisfy
\begin{equation}\label{eq:QIC1}
  Q^\intercal(x,x_{12},x_{13},x_{23};\alpha_1 -\alpha_2,\alpha_1 -\alpha_3) = 0\,.
\end{equation}
\end{subequations}
In the case of Q-type equations,   $Q^\intercal$ coincides with $Q$ itself. 
\item [Symmetry of Q-type equations:] recall a M\"obius transformation $\fm: \CC\PP^1\to \CC\PP^1$, 
  \begin{equation}\label{eq:mb1}
   \fm( x)=  \frac{m_1 x + m_2}{m_3 x +m_4} 
   \,, \quad m_1 m_4  - m_2  m_3 \neq 0\,. 
  \end{equation}
A M\"obius transformation   can be decomposed as a sequence of elementary transformations such as the translation $x \mapsto x +c$, the scaling transformation $x \mapsto cx $ and the inversion $x \mapsto 1/x$.
  By a symmetry of a bulk equation $Q =0$,  we mean that, under simultaneous   M\"obius transformations on all arguments of $Q$, one has
  \begin{equation}\label{eq:qfm}
    Q(\fm(x),\fm(u),\fm(v),\fm(y);\alpha,\beta) =0\,, 
  \end{equation}    
  when $Q(x,u,v,y;\alpha,\beta) =0$ holds.  Note that this notion of symmetry coincides with the Lie-point symmetry  \cite{papageorgiou2006yang, RH1, XNE}. 

Symmetries of  Q-type equations are listed in Table \ref{tb:qtype}. They are useful devises for the derivation of integrable boundary equations.  The Q1($ 0$) equation has the generic M\"obius transformation as its symmetry, while Q2 only has the trivial symmetry, \ie $x\mapsto x$.    \begin{table}[h]
    \centering
    \begin{tabular}{c|c|c | c |c |c}
      \hline
Q1($ 0$)& Q1($\delta\neq 0$) & Q2 &  Q3($0$) & Q3($\delta \neq0$)& Q4 \\ 
      \hline
      $cx$, $ x+c$, $1/x$ &$\pm x$, $x+c$ & $x$ & $cx$, $1/x$ & $\pm x$ &$ \pm x$, $\pm 1/(kx) $ \\
\end{tabular}
\caption{Symmetries of the bulk equations. For Q4, $k$ denotes  the elliptic modulus. }
\label{tb:qtype}
\end{table}  

Moreover, each Q-type bulk polynomial also satisfies
\begin{equation}\label{eq:qfpara}
Q(x,u,v,y;\alpha,\beta) = -   Q(x,u,v,y;-\alpha,-\beta)\,.
\end{equation}

\item[Equivalence class:] each bulk polynomial $Q$ in its canonical form  is a representative of an equivalence class  $[Q]$  with respect to simultaneous  M\"obius transformations on all arguments of $Q$. We take the M\"obius transformations \eqref{eq:mb1},  the action amounts to 
  \begin{equation}\label{eq:qfm}
 Q_\fm  := \Lambda(x) \Lambda(u) \Lambda(v) \Lambda(y)       Q(\fm(x),\fm(u),\fm(v),\fm(y);\alpha,\beta) \,, 
\end{equation}
where $\Lambda(\ast):= m_3 \ast+m_4$ is understood. Then,  
$    \{ Q \in [Q]: Q \sim  Q_\fm   \}$. 
A remarkable consequence of the ABS classification is that $Q_\fm = 0$ is also $3$D-consistent. The related properties of  $Q_\fm$ can be  derived from those of  $Q$. 
\end{description}

In the rest of paper, we will closely work with the canonical forms of $Q$ in view of solving the boundary consistency. 
Having the canonical forms of bulk equations, the above properties can be directly checked by computations. For completeness, we provide the biquadratic polynomials as well as the tetrahedron equation  $Q^\intercal = 0$  for all bulk equations in Appendix \ref{sec:app1}. Their importance in deriving integrable boundary equations are becoming  apparent.

\subsection{Boundary polynomials and boundary equations}
\label{sec:bpbe}We proceed to formalizing the notions of boundary polynomials and boundary equations. Similar to the way of defining bulk equations, we  use multivariate polynomials to characterize boundary equations.  With some extra properties, these polynomials provide a natural way to define boundary equations considered in this paper. 
\begin{definition}\label{def:q} A multivariate polynomial $q$ depending on three arguments  and two parameters in the form (see Figure \ref{fig:tri})    \begin{equation}
      \label{eq:qxuy}
    q(x,y,z;\alpha,\beta) 
    \end{equation}
is called a {\bf boundary polynomial}, if $1$) it is irreducible in $x,y,z$ and affine-linear with respect to $x$ and $z$, $2$)  it is $\ZZ_2$-symmetric meaning that there exists a {\bf parity parameter} $\gamma = \pm 1$ such that
    \begin{equation}
      \label{eq:7qg}
      q(x,y,z;\alpha,\beta)  = \gamma\, q(z,y,x;\beta,\alpha)\,.     \end{equation}
We call $q =0$ the associated {\bf boundary equation}.   
  \end{definition}

Let us comment on some consequences of the above definition. 
\begin{enumerate}
\item Consider a boundary equation in the form (see Figure \ref{fig:tri})
  \begin{equation}
    q(x, y, z; \alpha, \beta) =0\,. 
  \end{equation}
It is understood that the first  argument $x$ and the third argument $z$ in $q$  represent the values  at the boundary vertices, and the parameters $\alpha,\beta$ are the lattice parameters attached to the edges connecting the value $y$ to the boundary values $x$ and $z$ respectively.

 \item 
 The $\ZZ_2$-symmetry puts the two lattice directions associated with the parameters $\alpha$ and $\beta$ at the same footing. This is in line with the  $\cD_4$-symmetry of the  bulk polynomials. In particular, a new ingredient $\gamma$ called   parity,  enters into the definition of boundary polynomials. A boundary polynomial has parity as its extra characteristic.    Note that the notion of parity for boundary conditions was also mentioned in \cite{ZC2} in the context of integrable PDEs.

\item The  affine-linearity as well as the $\ZZ_2$-symmetry  allows us to express $q$ as
   \begin{equation}\label{eq:q1234}
     q(x,y,z;\alpha,\beta) =  r_1(y;\alpha,\beta)xz  + r_2(y;\alpha,\beta)x + r_3(y;\alpha,\beta)z+ r_4(y;\alpha,\beta)\,,
   \end{equation}
   where $r_j(y;\alpha,\beta)$, $j =1, \dots, 4$, are polynomials in $y$ satisfying
   \begin{equation}
     r_{1\backslash 4}(y;\alpha,\beta) = \gamma\, r_{1\backslash  4}(y;\beta, \alpha)\,,\quad      r_{2\backslash 3}(y;\alpha,\beta) = \gamma\, r_{3\backslash 2}(y;\beta, \alpha)\,.
   \end{equation}
 We do not exclude that the case that some or all $r_j$ are independent of $y$. 
 \item  A boundary polynomial is required to be irreducible. This excludes the case that the quantity (we take $q$ as shown in \eqref{eq:q1234} by omitting the inputs)
   \begin{equation}
     \delta_{x,z}q :=  q\, q_{xz}-q_x\,q_z =r_1 r_4 - r_2r_3  \,, 
   \end{equation}
   vanishes identically for generic input $y$.  
The irreducibility of $q$ also excludes the case that  $q$ is decomposed as  $q = f(y)q^*$  with   $q^*$ being a boundary polynomial.

\end{enumerate}
\begin{rmk}\label{rmk:parity}
Taking account of the parity, a boundary polynomial $q$ in the form \eqref{eq:q1234} can always be reduced to one of the following two forms with opposing parities
  \begin{subequations}\label{eq:gpgm1}
  \begin{align}
   \gamma=1\,,~~ q(x,y,z;\alpha,\beta) &=  r_1(y;\alpha,\beta)xz  + r_2(y;\alpha,\beta)x + r_2(y;\beta,\alpha)z+ r_4(y;\alpha,\beta)\,, \\
   \gamma=-1\,,~~      q(x,y,z;\alpha,\beta)& =  r_2(y;\alpha,\beta)x - r_2(y;\beta,\alpha)z\,,
   \end{align} 
  \end{subequations}
 where $r_{1\backslash 4}(y;\alpha,\beta) = \, r_{1\backslash  4}(y;\beta, \alpha)$ is understood. However, there is still a remaining freedom for the parity. For a given $q$, an equivalence class can be defined using $q\sim  q^\dagger$, if $q^\dagger = g(\alpha, \beta) q$  for some function $g(\alpha, \beta) $ depending on the parameters only. Clearly, an anti-symmetric $g(\alpha, \beta) $, \ie $g(\alpha,\beta) = -g(\beta,\alpha)$,  leads to $q^\dagger$ with opposing parity to $q$. We will see examples later where the parity is fixed in order to have  an appropriate form for $q$. 
\end{rmk}
\begin{rmk}
A boundary equation in the form \eqref{eq:q1234} can be used to define an initial value problem on a single triangle by using, for instance, the values $x,y$ as well as the parameters  $\alpha,\beta$ as initial values to uniquely express   $z$.  This can be generalized  to the study of Cauchy problems for generic quad-graph systems with boundary similar to how this is done for quad-graph systems without boundary \cite{AV, VdK2}.  Note that some well-posed initial-boundary value problems on quad-graphs with a boundary/boundaries have already been considered in \cite{CCZ, CVZ1}.  
\end{rmk}

As explained in Section \ref{sec:prri}, let a bulk equation $Q = 0$ be given, the key step to derive a boundary zero curvature condition for an integrable boundary equation $ q = 0$ is based on the existence of  a {\em dual} boundary equation $p = 0$ as a result of the compatibility of $Q = 0$ and $ q = 0$.   Here, we provide some simple consequences by taking a bulk polynomial $Q$ and a boundary polynomial $q$ defined on the same quadrilateral.

Let $Q$ and $q$ be in the forms
\begin{subequations}\label{eq:Qqc}
\begin{align}\label{eq:QST}
     Q(x,u,v,y;\a,\b)= &S(x,u,v;\alpha,\beta)+T(x,u,v;\alpha,\beta)y \,,  \\  \label{eq:qst1}q(x,u,y;\alpha,\beta) =&s(x,u;\alpha,\beta)+t(x,u;\alpha,\beta)y\,, 
   \end{align}  
\end{subequations}
where $S, T, s, t$ are polynomials in their respective arguments. By irreducibility of $Q$, we require that  $\alpha\neq \beta$. Moreover, we use the following notion of degeneracy for   $q = 0$. 
\begin{definition}\label{def:nondeg}Let $Q$ and $q$  be defined on the same quadrilateral as given in \eqref{eq:Qqc}. Then, the boundary  equation  $q = 0$ is said to be {\bf nondegenerate}, if it does not lead to any singular solution to $Q = 0$.
\end{definition}
By assuming $q=0$ to be  nondegenerate, we exclude  the following two cases: 
\begin{itemize}
\item let $q =  0 $ be a boundary equation that leads to  $\delta_{u,v}Q =0$, then  $Q$ becomes reducible, and $Q = 0$ has singular solutions with respect to either $u$ or $v$. This could happen for H-type  bulk polynomials, and for Q1($\delta$), Q3(0) and A-type bulk polynomials, where  $ \delta_{u,v}Q$ can be factorized as a product of two boundary polynomials (see Appendix \ref{sec:app1} for details). 
\item one can express  $Q$ in \eqref{eq:QST}  as $Q =  S(u,x,y;\alpha,\beta) + T(u,x,y;\alpha,\beta) v$  thanks to the symmetry \eqref{eq:QQs}. Let $q  $ be a linear combination of $S(u,x,y;\alpha,\beta)$ and $T(u,x,y;\alpha,\beta)$, namely, 
  \begin{equation}\label{eq:qc0c1}
    q(x,u,y;\alpha,\beta) = c_1 S(u,x,y;\alpha,\beta) +c_0 T(u,x,y;\alpha,\beta)\,. 
  \end{equation} Due to the symmetry  \eqref{eq:QQs1}, both $S(u,x,y;\alpha,\beta)$ and $ T(u,x,y;\alpha,\beta)$ are affine-linear with respect to $x,y$ and $\ZZ_2$-symmetric by interchanging $x,\alpha$ and $ y, \beta$. Such $q$  is  a boundary polynomial  with parity $\gamma = -1$. This form  is excluded, since  $q = 0$, together with $Q=0$,  implies that either $S(u,x,y;\alpha,\beta) =0 $ and $T(u,x,y;\alpha,\beta)= 0$ should hold simultaneously, or a particular value $v = c_0/c_1$ is a solution of $Q = 0$. Both cases lead to singular solutions to $Q=0$.  \end{itemize}
\begin{rmk}\label{rmk11wd}
In this section, we do not consider degenerate boundary equations discussed in the above two  cases, since they do not lead to any dual boundary equations (see next subsection for details). However, some of them  satisfy the boundary consistency, and contribute to the classification of integrable boundary equations in the  degenerate case (see Section \ref{sec:degbc}). 
 \end{rmk}
By eliminating $y$ in \eqref{eq:Qqc}, the resulting polynomial, denoted by $H$, can be expressed as
   \begin{equation}\label{eq:HH1}
H:=   Q \,q_y - Q_y\, q   =S\,t-T\,s\,. 
   \end{equation}
   It depends on $x,u,v$ as well as on the parameters $\alpha, \beta$.  $ H$ will be  the central object of interest.

   Since the polynomials $S, T$ are of degree $1$ in their respective arguments, and $s, t$ are of degree $1$ in $x$, one has $\deg_x H \leq 2$ and $\deg_v H \leq 1$, and  $H$ can be generically expressed as
     \begin{equation}\label{eq:HH2}
      H = H_0 v+H_1\,,
    \end{equation}
where $H_0$ and $H_1$ are polynomials possibly  depending on $x,u$. 
Under the assumption that $q = 0$ is nondegenerate, we can show $ H$ is affine-linear with respect to $v$, and the equation $H = 0$ yields $v =
-H_1 /H_0$ , which is, in general, a rational expression of $x, u$. Note that due to the $\cD_4$-symmetry of $Q$ and $\ZZ_2$-symmetry of $q$, one can derive similar results for  \eqref{eq:Qqc} by eliminating $x$.

\begin{lemma}\label{lem:31}
  Consider the polynomial $H$ defined in \eqref{eq:HH1}. Then, $H_0$ and $H_1$ in \eqref{eq:HH2} are nonvanishing polynomials for generic $x,u$, and $H_0$ is not proportional to $ H_1$. Moreover, eliminating $v$  in $Q$ and $H$  yields  
    \begin{equation}\label{eq:QHv}
  Q \,H_v - Q_v\, H = -\left(\delta_{v,y}Q\right)  q(x,u,y;a,b)\,,
\end{equation}
where $\delta_{y,v}Q$ is the biquadratic polynomial defined in \eqref{eq:dvy}
\end{lemma}
\prf
 Taking  $Q$ in the form (by omitting the inputs)
\begin{equation}\label{eq:Qr1234}
  Q = R_1 vy +R_2 v  +R_3 y +R_4\,,  
\end{equation}
where $R_j$, $j=1,\dots, 4$, are polynomials of $x,u$, $H_0, H_1$  in  \eqref{eq:HH2} can be expressed as 
\begin{equation}
 H_0 := (R_2 t- R_1 s)\,,\quad H_1: =( R_4 t -R_3 s )\,. 
\end{equation}
Let  $H=0$ as a result of $Q =0$ and $q =0$. For generic  $v$, if $H_{0\backslash 1} = 0$, then $H_{1\backslash 0}=0$. This implies that  
$    R_2 t - R_{1} s =0$ and $   R_3 s-  R_4 t =0$ should  hold  simultaneously  for generic  $x,u$, which means $\delta_{v,y}Q = R_1R_4 - R_2 R_3  =0$. This violates our assumption for $q$.

For non vanishing $H_0$, $H_1$. If $\delta_{v,y}Q = R_1R_4 - R_2 R_3  \neq 0$, assume there is a particular value $v = c_0/c_1$ such that $H = 0$, then $H_0$ is proportional to $ H_1$, namely, 
$
  c_0 H_0+c_1 H_1 =0$. 
This implies, for instance, $H_0=-c_1 h$ and $H_1=c_0 h$ for $h$ being a non vanishing function depending on $x,u$. One could set $h = R_1R_4 - R_2 R_3$, this allows us to express $s,t $ as
\begin{equation}
  s = c_0 R_2 + c_1 R_4\,, \quad t = c_0 R_1 + c_1 R_3\,, 
\end{equation}
and has $q$ in the form
\begin{equation}
  q = c_0 (R_2 + R_1 y) + c_1( R_4 + R_3 y)\,,
\end{equation}
which is precisely \eqref{eq:qc0c1}, since $S, T$ in \eqref{eq:qc0c1} can be read from \eqref{eq:Qr1234} as $S = R_4 + R_3 y $ and  $T =R_2 + R_1 y$. This case is also excluded.    

Knowing that $H$ is affine-linear with respect to $v$, one has
\begin{equation}
    Q \,H_v - Q_v\, H =   Q \,(Q_v q_y-Q_{vy} q) - Q_v\, (Q \,q_y - Q_y\, q )                   = - (Q Q_{vy} - Q_v Q_y)q\,,
\end{equation}
which corresponds to  \eqref{eq:QHv}.  
\finprf

\subsection{Factorization of $Q$ and duality of boundary equations}
\label{sec:factQ}

Consider the polynomials $Q$ and $q$ given in \eqref{eq:Qqc}. Assume $q =0$ to be nondegenerate.  The existence of a dual boundary equation implies that one can extract a boundary polynomial in the sense of Definition \ref{def:q}  from the polynomial $H$. This amounts to the following factorized form of $H$. 
  \begin{definition}\label{def:factor} Consider the polynomial $H$ defined in \eqref{eq:HH1} as a result of eliminating $y$ in \eqref{eq:Qqc}. If $H$ can be factorized as
  \begin{equation}\label{eq:dualq}
    H : = Q \,q_y - Q_y\, q = \chi(x,u;\a,\b)p(u,x,v;\a,\b)\,,
  \end{equation}
where $p$ is also a boundary polynomial and $\chi$ is certain polynomial depending on $x,u$,  then we say that the boundary equation $q=0$ {\bf factorizes} the bulk equation $Q=0$, and  $p=0$ is the   {\bf dual boundary equation} of $q=0$. 
\end{definition}
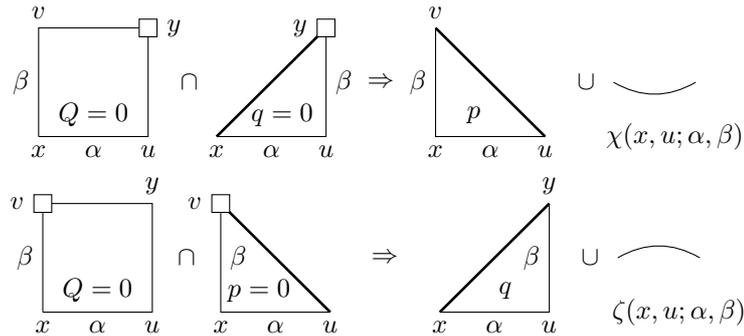
\begin{figure}[htb]
  \begin{center}
  \begin{tikzpicture}[scale=1.8]
    \def\l{.8}%
    \def\d{1.5}%
    \def\le{.3}
    \def\b{.5}
    \tikzstyle{nod}= [rectangle,  fill=white, minimum size=2pt, draw]

    \coordinate (w00) at (0,0);
    \coordinate (w10) at (\l,0);
    \coordinate (w01) at (0,\l);
    \coordinate (w11) at (\l,\l);
    \draw[-] (w00)node [below]{ $x$}  --  node [below]{ $\a$} (w10)node[below]{ $u$} -- (w11) -- (w01) node[above]{ $v$}-- node [left]{ $\beta$}  (w00) ;
    \node (1) at (.5*\l,.2*\l) {$ Q=0  $};
    \coordinate (s00) at (\l+\b,0);
    \coordinate (s10) at (\l+\b+\l,0);
    \coordinate (s11) at (\l+\b+\l,\l);
    \draw[-] (s00)node [below]{ $x$}  --  node [below]{ $\a$} (s10)node[below]{ $u$} --node[right]{ $\beta$} (s11);
    \draw[line width=1pt] (s11)-- (s00);
    \node (1) at (\l+.6*\b,.5*\l) {$ \cap  $};
    \node (1) at (\l+\b+.6*\l,.2*\l) {$ q=0  $};
     \node[nod](w11) at (\l,\l)  [label=right: $y$] {};
     \node[nod](s11) at (\l+\b+\l,\l) [label=left: $y$] {};
    \node (1) at (2* \l+\b+.5*\b+.5*\le,.5*\l) {$ \Rightarrow  $};

    \coordinate (t00) at (\le+2*\l+2*\b,0);
    \coordinate (t10) at (\le+2*\l+2*\b+\l,0);
    \coordinate (t01) at (\le+2*\l+2*\b,\l);
    \draw[-] (t10)node [below]{ $u$}  --  node [below]{ $\alpha$} (t00)node[below]{ $x$} -- node [left]{ $\beta$}(t01)node[above]{ $v$};
    \draw[line width=1pt] (t10)-- (t01);
    \node (1) at (\le+2*\l+2*\b+.35*\l,.2*\l) {$p  $};
    \coordinate (xx1) at (\le+3*\l+2*\b+\b,\l/2);
    \coordinate (xx2) at (\le+3*\l+2*\b+\l+\le,\l/2);
    \draw (xx1) to[out=-30,in=-150](xx2);
    \node (1) at (\le+2*\l+2*\b+\l+  .6*\b,.5*\l) {$ \cup  $};
    \node (1) at (\le+3*\l+2*\b+\l+\le/2,0) {$ \chi(x,u;\alpha,\beta)  $};
  \end{tikzpicture}

  \begin{tikzpicture}[scale=1.8]
    \def\l{.8}%
    \def\d{1.5}%
    \def\le{.3}
    \def\b{.5}

    \tikzstyle{nod}= [rectangle,  fill=white, draw]
    \coordinate (w00) at (0,0);
    \coordinate (w10) at (\l,0);
    \coordinate (w01) at (0,\l);
    \coordinate (w11) at (\l,\l);
    \draw[-] (w00)node [below]{ $x$}  --  node [below]{ $\a$} (w10)node[below]{ $u$} -- (w11) node[above]{ $y$}-- (w01) -- node [left]{ $\beta$}  (w00) ;
    \node (1) at (.5*\l,.2*\l) {$ Q=0  $};
    \coordinate (s00) at (\l+\b,0);
    \coordinate (s10) at (\l+\b+\l,0);
    \coordinate (s01) at (\l+\b,\l);
    \draw[-]   (s01) --node [right]{ $\beta$} (s00)node [below]{ $x$}  --  node [below]{ $\a$} (s10)node[below]{ $u$};
    \draw[line width=1pt] (s01)-- (s10);
    \node (1) at (\l+.5*\b,.5*\l) {$ \cap  $};
    \node (1) at (\l+\b+.35*\l,.2*\l) {$p=0  $};
    \node[nod](w01)  at (0,\l)  [label=left: $v$] {};
    \node[nod](s01)  at (\l+\b,\l)[label=left: $v$] {};
    \node (1) at (2* \l+\b+.5*\b+.5*\le,.5*\l) {$ \Rightarrow  $};

    \coordinate (t00) at (\le+2*\l+2*\b,0);
    \coordinate (t10) at (\le+2*\l+2*\b+\l,0);
    \coordinate (t11) at (\le+2*\l+2*\b+\l,\l);
    \draw[-] (t00)node [below]{ $x$}  --  node [below]{ $\a$} (t10)node[below]{ $u$} -- node [left]{ $\beta$}(t11)node[above]{ $y$};
    \draw[line width=1pt] (t00)-- (t11);
    \node (1) at (\le+2*\l+2*\b+.6*\l,.2*\l) {$q  $};
    \coordinate (xx1) at (\le+3*\l+2*\b+\b,\l/2);
    \coordinate (xx2) at (\le+3*\l+2*\b+\l+\le,\l/2);
    \draw (xx1) to[out=30,in=150](xx2);
    \node (1) at (\le+2*\l+2*\b+\l+  .6*\b,.5*\l) {$ \cup  $};
    \node (1) at (\le+3*\l+2*\b+\l+\le/2,0) {$ \zeta(x,u;\alpha,\beta)  $};
  \end{tikzpicture}
    \end{center}
\caption{\label{fig:factH} Factorization of $Q$: the above figure shows the compatibility of $Q=0$ and $q =0$ by eliminating $y$ results in the product of $p$ and $\chi$; the figure below shows the dual process that is the compatibility of $Q=0$ and $q =0$ by eliminating $v$ yielding the product of $q$ and $\zeta$ .}
\end{figure}

A significant consequence of the factorized form  \eqref{eq:dualq} is that the resulting boundary polynomial $p$ plays exactly the same role as $q$. For simplicity,  we use $\Gamma$ to denote the biquadratic polynomial $\delta_{v,y}Q$, namely,
\begin{equation}\label{eq:Gdq}
  \Gamma(x,u;\alpha,\beta) := \delta_{v,y}Q(x,u,v,y;\alpha,\beta)\,.
\end{equation}

\begin{lemma}\label{lem:6}
  Assume the polynomial $H$ defined in \eqref{eq:HH1} admits the factorized form  \eqref{eq:dualq}. Then, 
\begin{equation}\label{eq:dualp}
  Q \,p_v - Q_v\, p = \zeta(x,u;\alpha,\beta) q(x,u,y;\alpha,\beta)\,,
\end{equation}
where  $ \zeta$ is certain polynomial  depending on $x,u$. Moreover, $\chi$ and $\zeta$ are connected to    $\Gamma$  as
\begin{equation}\label{eq:czd}
\chi(x,u;\alpha,\beta)\, \zeta(x,u;\alpha,\beta)=-\Gamma(x,u;\alpha,\beta)\,. 
\end{equation}
\end{lemma}
\prf
 It follows from the factorized form \eqref{eq:dualq} that  \eqref{eq:QHv} can be written as
\begin{equation}
  Q \,H_v - Q_v\, H =  \chi   (Q \,p_v - Q_v\, p)  = - \Gamma \,q\,. 
\end{equation}
The polynomials $\chi$ and $\Gamma$ are independent of $y$, while  $Q \,p_v - Q_v\, p$ is affine-linear with respect to $y$ as a result of Lemma \eqref{lem:31}. Then,  $Q \,p_v - Q_v\, p$ must be proportional to $ q$, which amounts to  \eqref{eq:dualp}. The last formula \eqref{eq:czd} is a consequence of \eqref{eq:QHv} and \eqref{eq:dualp}. 
\finprf

 The boundary equation $p = 0$  also factorizes  the bulk equation $Q=0$ as shown in \eqref{eq:dualp}. This factorization process is illustrated in  Figure \ref{fig:factH}.  In this sense, we say that  a bulk polynomial $Q$ ({\em resp}. a bulk equation $Q = 0$) admits a pair of boundary polynomials $p$ and $q$ ({\em resp}. a pair of boundary  equations $q =0$ and $p =0$) dual to each other.
 
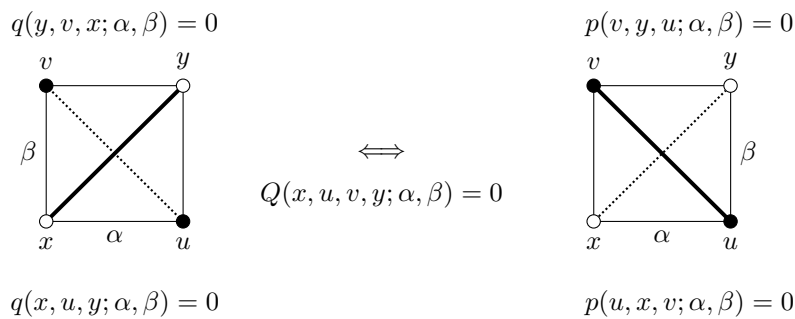
\begin{figure}[htb]
	\begin{center}
		\begin{tikzpicture}[scale=1.8]
		\def\l{1}%
		\def\d{1.5}%
		\def\le{.1}
		\def\b{.3}
		\tikzstyle{nod1}= [circle, inner sep=0pt,fill=white,  minimum size=5pt, draw]
		\tikzstyle{nod}= [circle, inner sep=0pt, fill=black, minimum size=5pt, draw]
		\coordinate (w00) at (0,0);
		\coordinate (w10) at (\l,0);
		\coordinate (w01) at (0,\l);
		\coordinate (w11) at (\l,\l);
		\coordinate (u00) at (\l+\d,0);
		\coordinate (u10) at (\l+\l+\d,0);
		\coordinate (u01) at (\l+\d,\l);
		\coordinate (u11) at (\l+\l+\d,\l);
		\coordinate (v00) at (\l+\d+\d,0);
		\coordinate (v10) at (\l+\d+\d+\l,0);
		\coordinate (v01) at (\l+\d+\d,\l);
		\coordinate (v11) at (\l+\d+\d+\l,\l);

		
		\draw[-] (w00) --  node [below]{ $\alpha$} (w10) -- (w11) -- (w01) -- node [left]{ $\beta$}  (w00) ;
                \draw[densely dotted, thick] (w10) --  (w01);
		\draw[line width=1.5pt] (w00) --  (w11);
		\draw[line width=1.5pt] (v10) --  (v01);
		\node[nod1](w00) at (0,0)  [label=below: $x$] {};
		\node[nod](w10) at (\l,0) [label=below: $u$] {};
		\node[nod](w01) at (0,\l) [label=above: $v$ ] {};
		\node[nod1](w11) at (\l,\l) [label=above: $y$ ] {};

\node (q1) at (.5*\l,-2*\b) {$q(x,u,y;\alpha, \beta)=0$};
\node (q12) at (.5*\l,1.5*\b+\l) {$q(y,v,x;\alpha,\beta)=0$};

		\draw[-] (v10) --  node [below]{ $\alpha$} (v00) -- (v01) -- (v11) -- node [right]{ $\beta$} (v10) ;
		\node[nod] (v10) at (\l+\d+\d+\l,0)[label=below:$u$ ] {};
		\node[nod1] (v00) at (\l+\d+\d,0) [label=below:$x$] {};
		\node[nod1] (v11) at (\l+\d+\d+\l,\l) [label=above:$y$] {};
		\node[nod] (v01) at (\d+\d+\l,\l) [label=above: $v$ ] {};
		\node (q2) at (\l+\d+\d+.7*\l,-2*\b) {$p(u,x,v;\alpha, \beta)=0$};
	\node (q22) at (\l+\d+\d+.7*\l,\l+1.5*\b) {$p(v,y,u;\alpha, \beta)=0$};
        \draw[densely dotted, thick] (v00) --  (v11);
	
		\node (1) at (0.7*\l+\l+.5*\d,.5*\l) {$ \Longleftrightarrow  $};

		\node (2) at (0.7*\l+\l+.5*\d,.2*\l) {$Q(x,u,v,y;\alpha,\beta)=0$};

		
		\end{tikzpicture}
	\end{center}
	\caption{\label{fig:factor2} Factorization of $Q$ along the two diagonals (thick lines).  
	}
\end{figure}

Another important consequence of   \eqref{eq:dualq} is that the set of equations $Q=0$, $p =0$ and $ q =0$  form a compatible system on a single quadrilateral with the values on  the opposite
triangles of the quadrilateral governed by the same boundary equation (see Figure \ref{fig:factor2}).    Consider, for instance, $x,u$ as well as the parameters $\alpha, \beta$ as initial values on a quadrilateral, which are subject to the equations $Q(x,u,v,y;\alpha,\beta) = 0$ and $q(x,u,y;\alpha,\beta) = 0$. Due to the $\cD_4$-symmetry of $Q$ and $\ZZ_2$-symmetry of $q$, one can obtain the dual boundary equation $p =0$  on the two opposite 
triangles, \ie  $ p(v,y,u;\alpha,\beta) =0$ and  $p(u,x,v;\alpha,\beta) =0$,  sharing the same boundary that is  the diagonal  connected by the values $u,v$.   Similarly,  the $\cD_4$-symmetry of $Q$ and $\ZZ_2$-symmetry of $p$ implies that one can obtain the equation $q$ on the two opposite triangles sharing the same  boundary connected by the values $x,y$. 

There is also a duality between the factorized forms \eqref{eq:dualq} and \eqref{eq:dualp}, and the polynomials $\chi$ and $\zeta$ dual to each other are connected by the formula \eqref{eq:czd}. This allows us to determine the degrees of the middle argument of $q$ and $p$. 
When  $Q$ is a Q-type or A-type bulk polynomial, $\Gamma$ is of bidegree $(2,2)$ in $x,u$. It follows from  \eqref{eq:czd} that $\Gamma$ can be factorized as a product of $\chi$ and $\zeta$, therefore $\chi$ can be either of bidegree $(0,0)$ in $x,u$ ($\zeta$ of bidegree $(2,2)$ in $x,u$), or of bidegree $(1,1)$ in $x,u$ ($\zeta$ of bidegree $(1,1)$ in $x,u$). The former case is dual to the case where  $\chi$ is of bidegree $(2,2)$ ($\zeta$  of bidegree $(0,0)$), and the latter case only happens when $Q$ is Q1($\delta$), Q3($\delta =0$), or A-type polynomials where $\Gamma$ can be factorized as a product of two polynomials of bidegree $(1,1)$. 
\begin{lemma}\label{lem:decd}  Let  $Q$ be a Q-type or A-type bulk polynomial admitting a pair of dual boundary polynomials  $q,p$ as given in \eqref{eq:dualq} and \eqref{eq:dualp}. Let $c, d$  denote respectively the degree of the middle argument of $q, p$, and $ e, e'$ denote respectively the degree of $\chi, \zeta$ in one of its arguments, namely, \begin{equation}\label{eq:degree1}
   c = \deg_u q,\quad d =\deg_x p\,,\quad e = \deg_x\chi =\deg_u\chi \,,\quad e' = \deg_x\zeta =\deg_u\zeta\,.  
\end{equation}
Then,  
\begin{equation}
  \label{eq:cd2}
c = e\,,\quad d =e'\,,\quad c+d =e+e'    = 2\,. 
\end{equation}
\end{lemma}
\prf
 One has the explicit degrees of the polynomials
\begin{equation}
Q(\overset{1}{x},\overset{1}{u},\overset{1}{v}, \overset{1}{y}  )\,,  \quad q(\overset{1}{x},\overset{c}{u},\overset{1}{y}  )\,,\quad p(\overset{1}{u},\overset{d}{x},\overset{1}{v}  )\,, \quad    \chi(\overset{e}{x},\overset{e}{u} )\,,     \quad \zeta(\overset{e'}{x},\overset{e'}{u} )\,,  \quad \Gamma(\overset{2}{x},\overset{2}{u} )\,,
\end{equation}
where   $e+e'=2$ due to  \eqref{eq:czd}. Counting the degrees of \eqref{eq:dualq} and \eqref{eq:dualp} in $x,u$ leads to 
\begin{equation}
  1+e\leq 1+c\,,\quad e+d \leq 2\,,  \quad  1+e'\leq 1+d\,,\quad e'+c\leq 2\,. 
\end{equation}
Arranging the above inequalities leads to the results \eqref{eq:cd2}. 
\finprf

\begin{rmk}
 Similar to  Lemma \ref{lem:decd}, for H-type polynomials, one has
  \begin{equation}\label{eq:cdl2}
    c+d \leq 2\,,
  \end{equation}
where $c, d$ denote respectively the degree of the middle argument of $q, p$. The proof is less straightforward. This is because H-type polynomials are degenerate  in the ABS classification in the sense that $\Gamma$ in the canonical form is of bidegree $(0,0)$ (for H1) or $(1,1)$ (for H2, H3), \cf \cite{ABSII}. In these cases, some zeros of $\Gamma$ are located at $\infty$ for one or both of its arguments,   and there are some extra freedoms  for determining $c,d$ (see comments below). Derivations of $q,p$ for H-type polynomials require the action of simultaneous M\"obius transformations  on the factorized forms \eqref{eq:dualq} and \eqref{eq:dualp}. The proof of \eqref{eq:cdl2} requires some explicit computations that are provided in Appendix \ref{app:hp}.\end{rmk}

Let us comment on the action of simultaneous M\"obius transformations on the factorized forms \eqref{eq:dualq} and \eqref{eq:dualp}. Take the M\"obius transformation  \eqref{eq:mb1}. Similar to \eqref{eq:qfm}, one has  \begin{subequations}\label{eq:dualm11}
\begin{align}
  q_\fm :=\Lambda(x)\Lambda^c(u)\Lambda(y)\, q(\fm(x),\fm(u),\fm(y);\alpha,\beta)\,, \quad &\chi_\fm  := \Lambda^e(x)\Lambda^e(u)\, \chi(\fm(x),\fm(u);\alpha,\beta)\,,\\
  p_\fm :=\Lambda(u)\Lambda^d(x)\Lambda(v)\, p(\fm(u),\fm(x),\fm(v);\alpha,\beta)\,, \quad & \zeta_\fm := \Lambda^{e'}(x)\Lambda^{e'}(u)\, \zeta(\fm(x),\fm(u);\alpha,\beta)\,, 
\end{align}
\end{subequations}
where the degrees of the factors $\Lambda $ follow from \eqref{eq:degree1}. Clearly,  $q_\fm, p_\fm $ are boundary polynomials in the sense of Definition \ref{def:q}, if so are $q, p$. For $Q_\fm $ defined in \eqref{eq:qfm}, one also has
\begin{equation}\label{eq:gmm}
      \delta_{v,y}Q_\fm = \Delta_\fm^2 \Gamma_\fm \,,\quad \Gamma_\fm: = \Lambda^{2}(x)\Lambda^{2}(u)\, \Gamma(\fm(x),\fm(u);\alpha,\beta)\,,
\end{equation}
where $ \Delta_\fm  = m_1m_4 - m_2m_3$.  Moreover, it follows from   \eqref{eq:dualq} and \eqref{eq:dualp} that \begin{subequations}
\label{eq:dualm}\begin{align}
Q_\fm q_{\fm,y}-Q_{\fm,y}q_\fm = &\Delta_\fm \Lambda^{2-e-d}(x)\Lambda^{c-e}(u)\chi_\fm p_\fm\,,\\
Q_\fm p_{\fm,v}-Q_{\fm,v}p_\fm = &\Delta_\fm  \Lambda^{2-e'-c}(u)\Lambda^{d-e'}(x)\zeta_\fm q_\fm\,, 
\end{align}  
\end{subequations}
and from \eqref{eq:czd} that 
\begin{equation}\label{eq:eeee}
   \Lambda^{2-e-e'}(x)\Lambda^{2-e-e'}(u)\chi_\fm\zeta_\fm =- \Gamma_\fm\,. 
\end{equation}
If $Q$ is a  Q-type or A-type bulk polynomial, $e+e'=2$, $c=e$, $d = e'$,  the degrees of the factorized forms \eqref{eq:dualm} are completely fixed for a given $\chi$ (or $\zeta$). However, for H-type polynomials, one has $e+e' \leq 1$, the degrees of the factorized forms \eqref{eq:dualm} could have extra freedoms. For instance, for H1, if $e = e' = 0$,  both $c,d$ could take values in   $0,1,2$, and it might happen that $c=d=2$. However, this contradicts \eqref{eq:cdl2} and is indeed excluded following explicit computations (see Appendix \ref{app:hp}). 
\subsection{Classification of boundary polynomials dual to each other }
\label{sec:dbp}
For a  bulk polynomial $Q$ in the ABS classification, we provide two systematic approaches to deriving boundary polynomials that lead to the factorized form  \eqref{eq:dualq}. The boundary polynomials obtained following these two approaches are respectively called {\em M\"obius case} (having $\cM_\pm$ cases and dual $\cM_\pm$ cases as subcases) and {\em singular case}.  As an illustration, explicit computations are provided for the Q1($\delta$) polynomial. We conclude this section by arguing that these two cases exhaust all possible pairs of dual boundary polynomials for the ABS classification. 

Recall the forms of $Q, q, p, \chi, \zeta, \Gamma$ presented in the previous subsection 
\begin{subequations}\label{eq:caqpq}
\begin{align}
  Q :=Q(x,u,v,y;\alpha,\beta)\,,\quad q:=&q(x,u,y;\alpha,\beta)\,,\quad p :=p(u,x,v;\alpha,\beta)\,,\\
\chi :=\chi(x,u;\alpha,\beta)\,,\quad \zeta:=&\zeta(x,u;\alpha,\beta)\,,\quad \Gamma :=\Gamma(x,u;\alpha,\beta)\,.
\end{align}  
\end{subequations}
\begin{description}
\item[M\"obius case:] let $q$ be a boundary polynomial in one of the following two forms called respectively $\cM_+$ and $\cM_-$ cases according to the parity  \begin{subequations}\label{eq:mq}
  \begin{align}
  \cM_+ :\quad\, ~~\gamma = 1\,, \quad   q & =  r_1 x y + r_2 x  + r_2 y+ r_4\,, \label{eq:mq1}\\
  \cM_-:\quad  \gamma = -1\,, \quad   q & =   x  -  y\,,   \label{eq:mqq2}
  \end{align}    
\end{subequations}
where  $r_j$, $j = 1,2, 4$,  in the   $ \cM_+$   case is independent of $u$ but possibly depends on the lattice parameters $\alpha,\beta$ (in the parameter-dependent case, $r_j$ is a symmetric function, \cf Remark \ref{rmk:parity}), and    $r_1r_4 -r_2^2 \neq 0$.   Now, $q$ has $x $ and $y$ connected by a ``constant'' M\"obius transformation independent of $u$. By eliminating $y$ in $Q$ and $q$, one automatically has the factorized form \eqref{eq:dualq}:  let $Q =  S + T y$ as given in \eqref{eq:QST}, one has
  \begin{subequations}
    \label{eq:mqqq}
    \begin{align}
 \text{dual } \cM_+:  \quad H & = r_1 S\, x+ r_2( S -T x)- r_4 T  = p\,, \label{eq:mp1}\\
 \text{dual } \cM_-:\quad     H & =  -S - T x  = \kappa\, p\,,    \label{eq:mpp2}
  \end{align}    
\end{subequations}
 where the factor $\kappa  $ in the dual $\cM_-$ case is introduced as a function of $\alpha,\beta$ satisfying $\kappa(\alpha,\beta) = -\kappa(\beta, \alpha)$. 
 The second equality in both equations serves as a definition for  $p$.  Since the polynomials $S, T$ are affine-linear with respect to $u,v$, and $\ZZ_2$-symmetric with parity $-1$ by interchanging $u,\alpha$ and $v,\beta$ (due to the symmetry \eqref{eq:QQs1}),  $p$ must be affine-linear with respect to $u,v$, and $\ZZ_2$-symmetric. Moreover, one can show that $p$ is  irreducible. 
Therefore, $p$ is a boundary polynomial in the sense of Definition \ref{def:q}.

For the   $ \cM_+$ and dual    $ \cM_+$  cases,  $q$ and $p$ are, in general, three-parameter family of polynomials having $r_1, r_2, r_4$ as parameters.  For the dual $ \cM_-$ case, one has $p: = (-S - T x)/\kappa$. Since $\kappa$ is an anti-symmetric function, $q$ has parity $\gamma = 1$, \cf Remark \ref{rmk:parity}. Related properties of the boundary polynomials $p, q$ in the $\cM_\pm$ and dual $ \cM_\pm$ cases are listed  in Table \ref{tb:mocase}.      \begin{table}[h]
    \centering
    \begin{tabular}{c|c|c }
      \hline
& $\cM_+$ and dual $\cM_+$ & $\cM_-$ and dual $\cM_-$ \\ 
      \hline
      $q$ &   $\gamma = 1$, $c=0$, $\chi = 1$ \eqref{eq:mq1} &  $\gamma = -1$, $c=0$, $\chi = \kappa$  \eqref{eq:mqq2}\\
      \hline
     $p$ &  $\gamma = -1$, $d\leq 2$, $\zeta = -\Gamma $ \eqref{eq:mp1} & $\gamma =  1$, $d\leq 2$, $\zeta = -\Gamma/\kappa$ \eqref{eq:mpp2}\\
      \hline
    \end{tabular}
\caption{M\"obius case:  pairs of dual boundary polynomials $q,p$ constructed using a ``constant'' M\"obius transformation. The degrees of the middle argument are denoted by $c,d$.}
\label{tb:mocase}
\end{table}

\item[Singular case:] this case only applies to Q1($\delta$), Q3($0$), A-type and H-type bulk polynomials. Since H-type polynomials are treated in details in Appendix \ref{app:hp}, we only take $Q$ to  be Q1($\delta$), Q3($0$), or A-type polynomials. In this case,     $\Gamma$ is of bidegree $(2,2)$ but can be factorized as a product of two polynomials of bidegree $(1,1)$. 
Let $\chi$ be one of the polynomials, due to  \eqref{eq:czd}, if $\chi =0$, then $\Gamma =0$ and $H $ in \eqref{eq:dualq} vanishes identically for generic $v$. This implies  $Q$ and $q$ possess a common factor as  \begin{equation}
Q\vert_{\chi=0} = F(x,U(x),v;\alpha,\beta)G(x,U(x),y;\alpha,\beta)\,, \quad   q\vert_{\chi =0} \propto G(x,U(x),y;\alpha,\beta)\,,   
  \end{equation}
  where $(x,u) = (x,U(x))$ such that $\chi =0$. Since $\chi$ is of bidegree $(1,1)$ in $x,u$, $U(x)$ is of degree $1$ in $x$. Due to Lemma \ref{lem:decd}, $q$ is of degree $1$ in its middle argument, and can be  generically expressed as  \begin{equation}\label{eq:genq1}
    q = ( r_{10}  + r_{11} u) x y + ( r_{20}  + r_{21} u) x + ( r_{30}  + r_{31} u) y + ( r_{40}  + r_{41} u)\,,  
  \end{equation}
with $r_{ij}:=r_{ij}(\alpha,\beta)$ to be determined. Such $q$ can be reduced to 
  \begin{equation}\label{eq:genq2}
      q\vert_{\chi = 0} =  (c_0x + c_1) G(x,U(x),y;\alpha,\beta)\,,
  \end{equation}
  where $c_0, c_1$ are assumed to be some constants. By comparing the coefficients of \eqref{eq:genq1} evaluated at  $(x,u) = (x,U(x))$ with those of  \eqref{eq:genq2}, one sets up six equations  allowing us to express six out of eight parameters in \eqref{eq:genq1}  in terms of $c_0, c_1$. The remaining freedoms can be fixed by taking account of the $\ZZ_2$-symmetries of  $q,p$.

As a result, one can always obtain a pair of two-parameter family of boundary polynomials dual to each other in the forms
  \begin{equation}
    \label{eq:qpsc}
    q = c_0 \fq_0 +c_1 \fq_1\,,\quad     p = c_0 \fp_0 +c_1 \fp_1\,,
  \end{equation}
  where $\fq_j$ is dual to $\fp_j$, $j = 1,2$, and $c_0, c_1$ do not  vanish simultaneously. The case $c_0=0$ or $c_1 =0$ represents a subcase. 
\item[Q1($\delta$) as an example:] the bulk polynomial $Q$ reads 
  \begin{equation}
    \alpha (x-v)(u-y)- \beta (x-u)(v-y) + \delta^2 \alpha \beta (\alpha-\beta)\,,  
  \end{equation}with
    \begin{equation} \Gamma = \delta_{v,y}Q = -\b (\a - \b  )(x-u+ \delta\alpha )(x-u- \delta\alpha ) \,,  \end{equation}
which is a product of two polynomials of bidegree $(1, 1)$.  The $\cM_\pm$ and  dual  $\cM_\pm$  cases can be  directly read from \eqref{eq:mq} and \eqref{eq:mqqq}. For $\cM_-$ case, one sets $\chi : = \kappa =\a-\b  $ which yields 
  \begin{equation}
p = u v- x(u +v)  + x^2-\delta\alpha\beta \,, \quad       \zeta  =\b ((x - u)^2 -\delta^2 \alpha^2)\,,
  \end{equation}
for $\delta\neq 0$.  The ${\cal M}_-$ case is excluded for Q1($0$), since $x -y$ leads to $\delta_{u,v}Q=0$.  

For the singular case, let $\{\alpha, \beta\}\to \{\alpha^2, \beta^2\}$  to avoid square roots in the parameters. Without loss of generality, take $\chi \propto x-u+ \delta\alpha^2  $, then $\zeta  \propto  x - u - \delta \a^2$.  Following \eqref{eq:genq1} and \eqref{eq:genq2}, one gets explicitly
$  r_{11} = 0$, $r_{21} = -c_0$, $ r_{31} = c_0 - r_{10}$, $  r_{20}= -c_1 - r_{41} +   \delta \b^2 c_0$, $r_{30}= c_1 - \delta \alpha^2( c_0  -r_{10})$ and $ r_{40}=  \delta c_1(\b^2 -\a^2) - \a^2 r_{41}\delta$.

It remains to fix $r_{10}, r_{41}$. Due to the $\ZZ_2$-symmetries of $q, p$, one can assume
\begin{equation}
  q(x,u,y;\alpha,\beta) = \gamma(\alpha,\beta)q(y,u,x;\beta,\alpha)\,,\quad  p(u,x,v;\alpha,\beta) = \eta(\alpha,\beta)p(v,x,u;\beta,\alpha)\,,
\end{equation}
with $\gamma(\alpha,\beta)\gamma(\beta,\alpha) = \eta(\alpha,\beta)\eta(\beta,\alpha) = 1$. This further leads to $\gamma(\alpha,\beta) = \pm \b/\a$, $ \eta(\alpha,\beta) = -\b^2 / \a^2$, $ r_{10} = c_0(\a\pm \b)/\a$ and $ r_{41} = -c_1(\a\pm \b)/\alpha\,. 
$
Without loss of generality, one can absorb the sign  $\pm $ into $\beta$, since $Q$ is unchanged under $\beta\to -\beta $. Thus, the forms of $q,p$ are completely fixed. Lastly, taking $q \to \a q $ which normalizes $\gamma(\alpha,\beta)$ to $\gamma =  1$, one can express $q,p$ in the forms of \eqref{eq:qpsc}. Precisely, 
\begin{subequations}\label{bq:q11}
\begin{align}
  \fq_0=& (\a + \b)x y -   (\a u - \delta \a \b^2)x - (\b u- \delta \a^2 \b)y  \,,\\
  \fq_1= & \b x +\a y+ (\a + \b) (\delta\a\b-u  ) \,,\\
  \fp_0= & -(\a-\b) u v + \a x u-\b x v+ \delta\a\b (\a - \b) (\delta\a \b+x  )\,,\\
  \fp_1= & \b u -\a v +(\a -\b) (\delta\a \b +x  )\,,
\end{align}
\end{subequations}
 where $q$ has parity $1$ and $p$ has parity $-1$.  The duality also holds under $\delta \to -\delta$.\item[Classification of boundary polynomials dual to each other:] 
it turns out that the two approaches presented here allow us to list all possible dual boundary polynomials $q,p$ for a given $Q$ in the ABS classification. This relies on exhausting $\chi$ (or $\zeta$). The statement can be clearly justified for Q-type and A-type polynomials, since the degrees of $q,p$ can be fixed thanks to Lemma \ref{lem:decd}. The complete list of $q,p$ for  Q-type and A-type polynomials is provided in Appendix \ref{app:cbp}. The situation for H-type polynomials is less straightforward, and explicit computations are provided in Appendix \ref{app:hp}. 
  \begin{itemize}
  \item {\bf Q-type polynomials:} it follows Lemma \ref{lem:decd} that for  a given  $\chi$ (or $\zeta$), the degrees of the  middle argument of $q,p$ are fixed. Then, it suffices to list all possible cases for $\chi$. Due to \eqref{eq:czd}, one has two possibilities:
    \begin{enumerate}
    \item  M\"obius case: $\chi$ is of bidegree $(0, 0)$ in $x, u$ and $\zeta$ is of bidegree $(2, 2)$ in $x, u$. Then, $\deg_u q = 0$, $\deg_x p =2$, the $\deg_u q = 0$ polynomials are exhausted by the $\cM_\pm$ cases.  The case $\chi$ is of bidegree $(2, 2)$ is dual.
   \item singular case: if $Q$ is Q1($\delta$) or Q3($0$), it is possible that both $\chi$ and $\zeta$ are of bidegree $(1,1)$ in $x, u$ since $\Gamma$ can be factorized as a product of two polynomials of bidegree $(1,1)$. This case can also be exhausted in the singular case approach.      
    \end{enumerate}
      \item {\bf A-type polynomials:} the arguments for Q-type polynomials hold here, since A1($\delta$), A2 can be  obtained respectively from Q1($\delta$) and Q3($0$) (see Appendix \ref{app:Aty}). 
      \item {\bf H-type polynomials:}  for H1 in its canonical form, $\Gamma$ is of bidegree $(0, 0)$ in $x, u$. Under the simultaneous M\"obius transformations in sense of \eqref{eq:qfm} and \eqref{eq:dualm11}, one transforms $ \Gamma$ to $\Gamma_\fm$  which is of bidegree $(2, 2)$ in $x,u$. Then,   all possible $q,p$ can be derived following the singular
        case approach by exhausting all possible $\chi$ (or $\zeta$) following \eqref{eq:eeee}. 
        By explicit computations, it turns out that all possible $q,p$ of H1 belong to the  $\cM_+$ and dual  $\cM_+$   cases.
        
For  H2, H3($\delta$), $\Gamma$ is of bidegree $(1, 1)$ in $x, u$.     Similarly to H1,   under suitable simultaneous M\"obius transformations, all possible $q,p$ can be obtained following the singular
        case approach by exhausting all possible $\chi$ (or $\zeta$).  Besides the M\"obius case, it turns out that both H2 and H3 possess a pair of boundary polynomials in the forms \eqref{eq:qpsc} that can only be obtained from the singular case.    
  \end{itemize}

\end{description}

Let us summarize the above results.
\begin{theorem}\label{theo:pq}
  The boundary polynomials $q,p$ dual to each other satisfying the factorization property \eqref{eq:dualq}  for $Q$ in the ABS classification can be exhausted into the following three categories. 
  \begin{enumerate}
  \item  $\cM_+$ and dual $\cM_+$: this applies to all bulk equations. In general,  $q,p$ are  three-parameter families of boundary polynomials in the forms
    \begin{equation}\label{eq:MMMp}
           \text{$\cM_+$:}\quad q= r_{1}x y+ r_{2} x+ r_{2} y+ r_{4} \,,\quad 
 \text{dual $\cM_+$:} \quad    p = r_1\mathbi{p}_1 + r_2\mathbi{p}_2 +r_4  \mathbi{p}_4\,.     
\end{equation}
where $r_j$, $j = 1,2,4$, is a symmetric function of $\alpha,\beta$, and $r_1r_4 -r_2^2 \neq  0$. 
\item   $\cM_-$ and dual $\cM_-$: this  applies to all bulk equations except for Q1($0$), A1($0$), H1  due to the nondegeneracy requirement.  $q,p$ are in the forms
  \begin{equation}\label{eq:MMMm}
\cM_ -:\quad       q= x-y\,,\quad  \text{dual $\cM_-$:} \quad        p=   \mathbi{p}\,.
\end{equation}
\item  Singular case: for Q3($0$), Q1($\delta$), A2, A1($\delta$), H3($\delta \neq  0$) and H2, exactly one pair of $q,p$, not obtainable as a subcase of \eqref{eq:MMMp}, exists. The pair $q,p$ are, in general, two-parameter families of boundary polynomials in the forms
  \begin{equation}\label{eq:MMMs}
q = c_0 \fq_0 +c_1 \fq_1  \,,\quad p = c_0 \fp_0 +c_1 \fp_1\,,
\end{equation}
where $\fq_j$ is dual to $\fp_j$, $j = 1,2$, and $c_0, c_1$ are constants that do not vanish simultaneously. 
\end{enumerate}
\end{theorem}
\begin{rmk}\label{rmk:speq}
There are still some subtleties here. For H2 and H3($\delta\neq 0$) as  degenerate polynomials in the ABS classification, it happens that the pair $\fq_1, \fp_1$ in the singular case coincides with the $\cM_-$ and dual $\cM_-$ cases (see Appendix \ref{app:hp}). For Q3($0$), Q1($\delta$),  $q,p$ in the singular case were obtained following certain changes of the lattice parameters (see Appendix \ref{app:cbp}) to avoid square roots in the expressions, and certain $\fq_j$  either coincides with, or is equivalent to (in the sense of equivalence given in Remark \ref{rmk:parity}) $\fp_j$, $j = 0,1$:  for Q1$(0)$, $\fq_{0/1}$ coincides with $\fp_{0/1}$  under irrelevant change $\beta \to -\beta$, so that $q$ coincides with $p$ in the singular case;  for Q1$(\delta\neq 0)$, $\fq_{1}$ coincides with $\fp_{1}$ under $\beta \to -\beta$; for Q3$(0)$, $\fq_{0/1}$ is equivalent to $\fp_{0/1}$ under irrelevant change $\beta \to -\beta$, but  $q, p$  are different from each other.
\end{rmk}

\section{Factorization of consistency on rhombic dodecahedron and classification of integrable boundary equations}
\label{sec:fcrd}Based on the factorization of $Q =0$ by pairs of boundary equations $p=0$ and $q=0$ dual to each other, we  provide a criterion to select admissible boundary equations that are boundary consistent with $Q=0$. This relies on the factorization of a consistent system around a rhombic dodecahedron into two equivalent halves. In particular, the construction allows us to identify the involution relation $\sigma$ needed in Definition \ref{def:bc1} for nondegenerate boundary consistency. 
The degenerate boundary equations $q=0$ (see Definition \ref{def:nondeg}) are also considered. Some of them contribute to degenerate integrable boundary equations.
\subsection{Boundary consistency as a factorized rhombic dodecahedron}
\begin{figure}[th!]
	\centering
	\begin{tikzpicture}[scale=.55, decoration={markings,mark=at position 0.55 with {\arrow{latex}}}]
	\tikzstyle{nod1}= [circle, inner sep=0pt, fill=white, minimum size=5pt, draw]
	\tikzstyle{nod}= [circle, inner sep=0pt, fill=black, minimum size=5pt, draw]
        \def\a{.7}
        \def\lx{3*\a }%
	\def\ly{1.5*\a}%
	\def\lz{ (sqrt(\x*\x+\y*\y))*\a}%
	\def\l{4*\a}%
	\def\d{4*\a}%
	\coordinate (u00) at (0,0);
	\coordinate (u10) at (\l,0);
	\coordinate (u01) at (\lx,\ly);
	\coordinate (u11) at (\l+\lx,\ly);
	\coordinate (v00) at (0,\d);
	\coordinate (v10) at (\l,\d);
	\coordinate (v01) at (\lx,\d+\ly);
	\coordinate (v11) at (\l+\lx,\d+\ly);
	\draw[-]  (u10)  
	-- (u11) ;
	\draw [dashed]  (u01)  -- (u11);
	\draw [dashed]  (u00)--(u01) ;
	\draw[-] (v00) --  (v10) -- (v11) -- (v01) -- (v00);
	\draw[-] (u11) --  (v11);
	\draw[-] (u00) -- (v00);
	\draw[-] (u10) -- (v10);
	\draw[dashed] (u01) -- (v01);
	\coordinate (u011) at (1.5*\lx,1.5* \ly);
	\draw[-] (u00) 
	-- (u10);
	\node[nod] (v00) at (0,\d) [label=left: $x_3$] {};
	\node[nod1] (v10) at (\l,\d) [label=right: $x_{13}$] {};
	\node[nod1] (v01) at (\lx,\d+\ly) [label=left: $x_{23}$] {};
	\node[nod] (v11) at (\l+\lx,\d+\ly) [label=right: $x_{123}$] {};
	\node[nod1] (u00) at (0,0) [label=left: $x$] {};
	\node[nod] (u10) at (\l,0) [label=right: $x_1$] {};
	\node[nod] (u01) at  (\lx,\ly) [label=below: $x_2$] {};
	\node[nod1] (u11) at (\l+\lx,\ly) [label=right: $x_{12}$] {};

        \def\b{1.3}
	\def\lxx{4*\b}%
	\def\lyy{2*\b}%
	\def\lzz{ (sqrt(\x*\x+\y*\y))}%
	\def\ll{6*\b}%
	\def\dd{6*\b}%
        \def\dca{-3.5}
	\coordinate (uu00) at (0+\dca,0+\dca);
	\coordinate (uu10) at (\ll+\dca,0+\dca);
	\coordinate (uu01) at (\lxx+\dca,\lyy+\dca);
	\coordinate (uu11) at (\ll+\lxx+\dca,\lyy+\dca);
	\coordinate (vv00) at (0+\dca,\dd+\dca);
	\coordinate (vv10) at (\ll+\dca,\dd+\dca);
	\coordinate (vv01) at (\lxx+\dca,\dd+\lyy+\dca);
	\coordinate (vv11) at (\ll+\lxx+\dca,\dd+\lyy+\dca);
	\draw[-]  (uu10)  
	-- (uu11) ;
	\draw [dashed]  (uu01)  -- (uu11);
	\draw [dashed]  (uu00)--  node [below]{$\alpha_2$}(uu01) ;
	\draw[-] (vv00) --   (vv10) -- (vv11) -- (vv01) -- (vv00);
	\draw[-] (uu11) --  (vv11);
	\draw[-] (uu00) -- node [left]{$\alpha_3$}(vv00);
	\draw[-] (uu10) -- (vv10);
	\draw[dashed] (uu01) -- (vv01);
	\coordinate (u011) at (1.5*\lx,1.5* \ly);
	\draw[-] (uu00) 
	-- node [below]{$\alpha_1$}(uu10);
	\node[nod1] (vv00) at  (0+\dca,\dd+\dca) [label=above: $x_{34}$] {};
	\node[nod] (vv10) at (\ll+\dca,\dd+\dca) [label=above: $x_{134}$] {};
	\node[nod] (vv01) at (\lxx+\dca,\dd+\lyy+\dca) [label=above: $x_{234}$] {};
	\node[nod1] (vv11) at(\ll+\lxx+\dca,\dd+\lyy+\dca) [label=above: $x_{1234}$] {};
	\node[nod] (uu00) at  (0+\dca,0+\dca) [label=below: $x_4$] {};
	\node[nod1] (uu10) at  (\ll+\dca,0+\dca) [label=below: $x_{14}$] {};
	\node[nod1] (uu01) at (\lxx+\dca,\lyy+\dca) [label=below: $x_{24}$] {};
	\node[nod] (uu11) at(\ll+\lxx+\dca,\lyy+\dca) [label=below: $x_{124}$] {};

	\draw[densely dotted] (u01) -- (uu01);
        \draw[densely dotted] (u00) --node [above]{$\alpha_4$} (uu00);
        \draw[densely dotted] (u10) -- (uu10);
        \draw[densely dotted] (u11) -- (uu11);
	\draw[densely dotted] (v01) -- (vv01);
        \draw[densely dotted] (v00) -- (vv00);
        \draw[densely dotted] (v10) -- (vv10);
        \draw[densely dotted] (v11) -- (vv11);

      \end{tikzpicture}
	\caption{Consistency around a hypercube
} \label{fig:4d}
\end{figure}
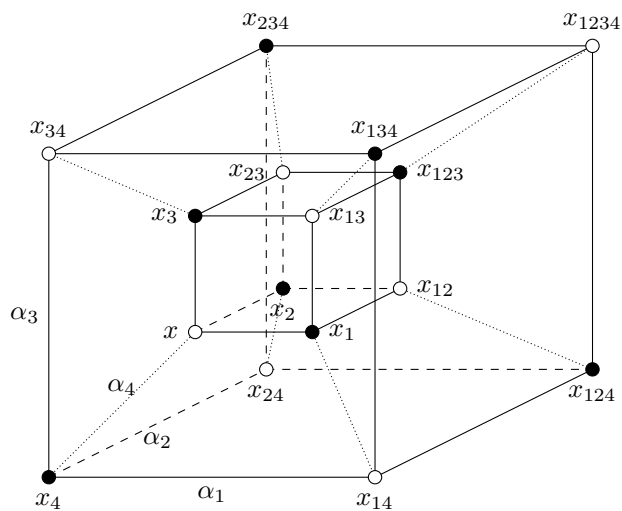
Any bulk equation $Q=0$ in the ABS classification is consistent on a hypercube that is a four-dimensional cube (see Figure \ref{fig:4d}). This implies that $Q=0$ is consistent around a  rhombic dodecahedron.

\begin{figure}[th]
  \centering
  \begin{subfigure}[b]{0.4\textwidth}
    \begin{tikzpicture}[scale=2.1]
    \tikzstyle{nod1}= [circle, inner sep=0pt, minimum size=5pt, draw]
    \tikzstyle{nod}= [circle, inner sep=0pt, fill=black, minimum size=5pt, draw]
    \tikzstyle{vertex}=[circle,minimum size=20pt,inner sep=0pt]
    \tikzstyle{selected vertex} = [vertex, fill=red!24]
    \tikzstyle{selected edge} = [draw,line width=2pt,-]
    \tikzstyle{edge} = [draw, thin,-,black] 
    \tikzstyle{ddedge} = [draw, densely dotted,-,black] 
    \tikzstyle{dedge} = [draw, dashed,-,black] 
    \tikzstyle{eedge} = [draw, thick,-,black] 

    \pgfmathsetmacro \ll {6}
    \pgfmathsetmacro \rl {7}
    \pgfmathsetmacro \h {2}
    \pgfmathsetmacro \an {28}
    \pgfmathsetmacro \bn {20}
    \pgfmathsetmacro \cn {15}
    \pgfmathsetmacro \ra {.15}

    \pgfmathsetmacro \llx {{\ra*\ll*cos(\an)}}
    \pgfmathsetmacro \lly {{\ra*\ll*sin(\an)}}
    \pgfmathsetmacro \rlx {{\ra*\rl*cos(\bn)}}
    \pgfmathsetmacro \rly {{\ra*\rl*sin(\bn)}}
    \pgfmathsetmacro \dh {{\h*tan(\cn)}}
    \node[nod1] (v) at (0,0) [label=below:$x$] {};
    \node[nod] (v1) at (-\llx,\lly) [label=left:$x_1$] {};
    \node[nod] (v4) at (\rlx,\rly) [label=right:$x_4$] {};
    \node[nod1] (v14) at (\rlx-\llx,\rly+\lly) [label=below:$x_{14}$] {};
    \draw[dedge] (v1) -- (v14) -- (v4) ;
    \draw[edge] (v1) --node [below]{$\alpha_1$} (v) --  node [below]{$\alpha_4$}(v4);
    \node[nod1] (w) at (0,\h) [label=above:$x_{23}$] {};
    \node[nod] (w1) at (-\llx,\lly+\h) [label=left:$x_{123}$] {};
    \node[nod] (w4) at (\rlx,\rly+\h) [label=right:$x_{234}$] {};
    \node[nod1] (w14) at (\rlx-\llx,\rly+\lly+\h) [label=above:$x_{1234}$] {};
    \draw[edge] (w1) -- (w14) -- (w4) ;
    \draw[edge] (w1) -- (w) -- (w4);
    \node[nod] (u1) at (-\dh,\h/2) [label=below:$x_2$] {};
    \node[nod] (u2) at (\dh,\h/2) [label=below:$x_3$] {};
    \draw[edge] (v) --  node [left]{$\alpha_2$}(u1) -- (w) -- (u2)  -- node [right]{$\alpha_3$}(v) ;
    \node[nod1] (u3) at (\dh+\rlx,\h/2+\rly) [label=left:$x_{34}$] {};
    \node[nod] (u4) at (\dh+ \rlx-\llx,\rly+\lly+\h/2) [label=right:$x_{134}$] {};
    \draw[edge] (u2) -- (u3);
    \draw[dedge] (u3) -- (u4);
    \node[nod1] (u5) at (-\dh -\llx,\h/2+\lly) [label=right:$x_{12}$] {};
    \node[nod] (u6) at (-\dh+\rlx-\llx,\rly+\lly+\h/2) [label= left :$x_{124}$] {};
    \draw[edge] (u1) -- (u5);
    \draw[dedge] (u5) -- (u6);
    \draw[edge] (v1) -- (u5) --(w1);
    \draw[edge] (v4) -- (u3) --(w4);
    \draw[dedge] (v14) -- (u6) -- (w14) -- (u4) --(v14);
    \node[nod1] (u0) at (-\dh+\rlx,\h/2+\rly)  [label=right:$x_{24}$] {};

    \draw[ddedge] (u0) -- (u1);
    \draw[ddedge] (u0) -- (u6);
    \draw[ddedge] (u0) -- (w4);
    \draw[ddedge] (u0) -- (v4);

  \end{tikzpicture}
\end{subfigure}\hspace{.3cm}
  \begin{subfigure}[b]{0.5\textwidth}
  \begin{tikzpicture}[scale=2.1]
    \tikzstyle{nod1}= [circle, inner sep=0pt, minimum size=5pt, draw]
    \tikzstyle{nod}= [circle, inner sep=0pt, fill=black, minimum size=5pt, draw]
    \tikzstyle{vertex}=[circle,minimum size=20pt,inner sep=0pt]
    \tikzstyle{selected vertex} = [vertex, fill=red!24]
    \tikzstyle{selected edge} = [draw,line width=2pt,-]
    \tikzstyle{edge} = [draw, thin,-,black] 
    \tikzstyle{ddedge} = [draw, densely dotted,-,black] 
    \tikzstyle{dedge} = [draw, dashed,-,black] 
    \tikzstyle{eedge} = [draw, thick,-,black] 

    \pgfmathsetmacro \ll {6}
    \pgfmathsetmacro \rl {7}
    \pgfmathsetmacro \h {2}
    \pgfmathsetmacro \an {28}
    \pgfmathsetmacro \bn {20}
    \pgfmathsetmacro \cn {15}
    \pgfmathsetmacro \ra {.15}

    \pgfmathsetmacro \llx {{\ra*\ll*cos(\an)}}
    \pgfmathsetmacro \lly {{\ra*\ll*sin(\an)}}
    \pgfmathsetmacro \rlx {{\ra*\rl*cos(\bn)}}
    \pgfmathsetmacro \rly {{\ra*\rl*sin(\bn)}}
    \pgfmathsetmacro \dh {{\h*tan(\cn)}}
    \node[nod1] (v) at (0,0) [label=below:$x$] {};
    \node[nod] (v1) at (-\llx,\lly) [label=left:$x_1$] {};
    \node[nod] (v4) at (\rlx,\rly) [label=right:$x_4$] {};
    \node[nod1] (v14) at (\rlx-\llx,\rly+\lly) [label=below:$x_{14}$] {};
    \draw[dedge] (v1) -- (v14) -- (v4) ;
    \draw[edge] (v1) --  node [below]{$\alpha_1$}(v) -- node [below]{$\alpha_4$} (v4);
    \node[nod1] (w) at (0,\h) [label=above:$x_{23}$] {};
    \node[nod] (w1) at (-\llx,\lly+\h) [label=left:$x_{123}$] {};
    \node[nod] (w4) at (\rlx,\rly+\h) [label=right:$x_{234}$] {};
    \node[nod1] (w14) at (\rlx-\llx,\rly+\lly+\h) [label=above:$x_{1234}$] {};
    \draw[edge] (w1) -- (w14) -- (w4) ;
    \draw[edge] (w1) -- (w) -- (w4);
    \node[nod] (u1) at (-\dh,\h/2) [label=below:$x_2$] {};
    \node[nod] (u2) at (\dh,\h/2) [label=below:$x_3$] {};
    \draw[edge] (v) --  node [left]{$\alpha_2$}(u1) -- (w) -- (u2)  -- node [right]{$\alpha_3$}(v) ;
    \node[nod1] (u3) at (\dh+\rlx,\h/2+\rly) [label=left:$x_{34}$] {};
    \node[nod] (u4) at (\dh+ \rlx-\llx,\rly+\lly+\h/2) [label=right:$x_{134}$] {};
    \draw[edge] (u2) -- (u3);
    \draw[dedge] (u3) -- (u4);
    \node[nod1] (u5) at (-\dh -\llx,\h/2+\lly) [label=right:$x_{12}$] {};
    \node[nod] (u6) at (-\dh+\rlx-\llx,\rly+\lly+\h/2) [label= left:$x_{124}$] {};
    \draw[edge] (u1) -- (u5);
    \draw[dedge] (u5) -- (u6);
    \draw[edge] (v1) -- (u5) --(w1);
    \draw[edge] (v4) -- (u3) --(w4);
    \draw[dedge] (v14) -- (u6) -- (w14) -- (u4) --(v14);
    \node[nod1] (0u) at (\dh-\llx,\h/2+\lly) [label= right:$x_{13}$]     {};

    \draw[ddedge] (0u) -- (u2);
    \draw[ddedge] (0u) -- (u4);
    \draw[ddedge] (0u) -- (w1);
    \draw[ddedge] (0u) -- (v1);
  \end{tikzpicture}

\end{subfigure}
  \caption{The rhombic dodecahedron can be constructed by four cubes in two ways}
    \label{fig:rd1}

\end{figure}
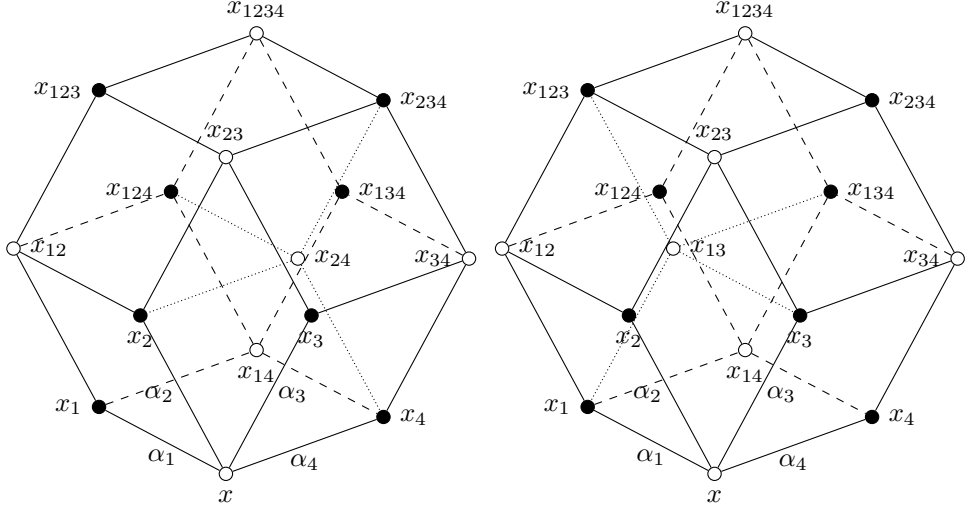

\begin{theorem}\label{theo:rd4c}
Let a $Q$ in the ABS classification be given. Consider an initial value problem for $Q=0$ on a rhombic-dodecahedron with initial values $x, x_1,x_2,x_3, x_4$ as well as the parameters $\alpha_1, \alpha_2, \alpha_3, \alpha_4$ (see Figure \ref{fig:rd1}). Then, $Q=0$ is consistent around a rhombic-dodecahedron which means the four ways to compute  $x_{1234}$ give the same result. 
\end{theorem}
\prf
   A  rhombic-dodecahedron can be constructed from two possible ``shadow projections'' of a hypercube cube (see Figure \ref{fig:rd1}). Then, the consistency for $Q=0$ around a rhombic-dodecahedron follows from the consistency around a  hypercube. 
\finprf

Note that the consistency around a rhombic dodecahedron for $Q=0$  was also used in \cite{JNY} as an important devise to reduce the bulk equation to certain discrete Painlev\'e equations. In our context, the consistency around a rhombic dodecahedron is needed since the boundary consistency is derived by factorizing the consistent system into two equivalent ones around the two halves of the rhombic dodecahedron (see Figure  \ref{fig:facrd}). In this process,   the notion of factorization of $Q$ provided in Section \ref{sec:factQ} is the key ingredient. 

\begin{theorem}\label{th:frb}
Let a $Q$ in the ABS classification be given. Let $q = 0$ and $p=0$ be a  pair of boundary equations dual to each other of $Q=0$. Consider an initial value problem for $Q=0$ on a rhombic-dodecahedron (see Figure  \ref{fig:facrd}) with   $x, x_1,x_2$ and $\alpha_1, \alpha_2, \alpha_3, \alpha_4$ as the initial data. Moreover, assume the values  $x_3, x_4$ are obtained using one of the boundary equations, say $p = 0$, by solving
\begin{subequations}\label{eq:ppim}
  \begin{align}
  p(x_1, x, x_4;\alpha_1,\alpha_4) = &0\,,\label{eq:px14}\\   p(x_2, x, x_{3};\alpha_2,\alpha_3) = &0\,.     \label{eq:px23}
  \end{align}
\end{subequations}
We call $p =0$ the {\bf  companion boundary equation} for the boundary consistency. If the values of  $x_{124}, x_{14}, x_{134}$ satisfy again the boundary equation $p=0$, namely, 
\begin{equation}\label{eq:pxit}
p(x_{124}, x_{14}, x_{134};\alpha_2,\alpha_3) = 0\,,
\end{equation}
which is accompanied by certain constraints on the lattice parameters, then the boundary equation $q=0$  dual to $p=0$ is boundary consistent with $Q =0$.  

\end{theorem}

\begin{figure}[ht]
  \centering
  \begin{tikzpicture}[scale=2.5]
    \tikzstyle{nod1}= [circle, inner sep=0pt, minimum size=5pt, draw]
    \tikzstyle{nod}= [circle, inner sep=0pt, fill=black, minimum size=5pt, draw]
    \tikzstyle{vertex}=[circle,minimum size=20pt,inner sep=0pt]
    \tikzstyle{selected vertex} = [vertex, fill=red!24]
    \tikzstyle{selected edge} = [draw,line width=2pt,-]
    \tikzstyle{edge} = [draw, thin,-,black] 
    \tikzstyle{ddedge} = [draw, densely dotted,-,black] 
    \tikzstyle{dedge} = [draw, dashed,-,black] 
    \tikzstyle{eedge} = [draw, thick,-,black] 

    \pgfmathsetmacro \ll {6}
    \pgfmathsetmacro \rl {7}
    \pgfmathsetmacro \h {2}
    \pgfmathsetmacro \an {28}
    \pgfmathsetmacro \bn {20}
    \pgfmathsetmacro \cn {15}
    \pgfmathsetmacro \ra {.15}

    \pgfmathsetmacro \llx {{\ra*\ll*cos(\an)}}
    \pgfmathsetmacro \lly {{\ra*\ll*sin(\an)}}
    \pgfmathsetmacro \rlx {{\ra*\rl*cos(\bn)}}
    \pgfmathsetmacro \rly {{\ra*\rl*sin(\bn)}}
    \pgfmathsetmacro \dh {{\h*tan(\cn)}}
    \node[nod1] (v) at (0,0) [label=below:$x$] {};
    \node[nod] (v1) at (-\llx,\lly) [label=left:$x_1$] {};
    \node[nod] (v4) at (\rlx,\rly) [label=right:$x_4$] {};
    \node[nod1] (v14) at (\rlx-\llx,\rly+\lly) [label=below:$x_{14}$] {};
    \draw[dedge] (v1) -- (v14) -- (v4) ;
    \draw[edge] (v1) -- node[below]{$\alpha_1$} (v) --node[below]{$\alpha_4$} (v4);
    \node[nod1] (w) at (0,\h) [label=above:$x_{23}$] {};
    \node[nod] (w1) at (-\llx,\lly+\h) [label=left:$x_{123}$] {};
    \node[nod] (w4) at (\rlx,\rly+\h) [label=right:$x_{234}$] {};
    \node[nod1] (w14) at (\rlx-\llx,\rly+\lly+\h) [label=above:$x_{1234}$] {};
    \draw[edge] (w1) -- (w14) -- (w4) ;
    \draw[edge] (w1) -- (w) -- (w4);
    \node[nod] (u1) at (-\dh,\h/2) [label=above:$x_{2}$] {};
    \node[nod] (u2) at (\dh,\h/2) [label=above:$x_3$] {};
    \draw[densely dotted, thick] (u1) -- (u2);
    \draw[thick] (v) -- (w);
    \draw[edge] (v) -- node[left]{$\alpha_2$}  (u1) -- (w) -- (u2)  --node[right]{$\alpha_3$}(v) ;
    \node[nod1] (u3) at (\dh+\rlx,\h/2+\rly) [label=left:$x_{34}$] {};
    \node[nod] (u4) at (\dh+ \rlx-\llx,\rly+\lly+\h/2) [label=right:$x_{134}$] {};
    \draw[edge] (u2) -- (u3);
    \draw[dedge] (u3) -- (u4);
    \node[nod1] (u5) at (-\dh -\llx,\h/2+\lly) [label=right:$x_{12}$] {};
    \node[nod] (u6) at (-\dh+\rlx-\llx,\rly+\lly+\h/2) [label= left :$x_{124}$] {};
    \draw[edge] (u1) -- (u5);
    \draw[dedge] (u5) -- (u6);
    \draw[edge] (v1) -- (u5) --(w1);
    \draw[edge] (v4) -- (u3) --(w4);
    \draw[dedge] (v14) -- (u6) -- (w14) -- (u4) --(v14);
    \draw [dotted, fill=gray, opacity =.1] (0,0) --(0,\h) -- (\rlx-\llx,\rly+\lly+\h)                  --  (\rlx-\llx,\rly+\lly) --cycle;
    \draw[densely dotted, thick] (v1) -- (v4);
    \draw[thick] (v) -- (v14);
    \draw[densely dotted, thick] (u4) -- (u6);
    \draw[thick] (w14) -- (v14);

        \draw[densely dotted, thick] (w1) -- (w4);
    \draw[thick] (w) -- (w14);

  \end{tikzpicture}
  \caption{Factorization of a rhombic dodecahedron into two halves: the factorization of the ``front'' equation involving $x,x_2,x_3,x_{23}$ ({\em resp}. of the ``bottom'' equation involving $x,x_1,x_4,x_{14}$) is transformed to the ``back'' equation involving $x_{14},x_{124},x_{134},x_{1234}$ ({\em resp}. to the ``top'' equation  involving $x_{23},x_{123},x_{234},x_{1234}$). }
  \label{fig:facrd}
\end{figure}

\prf
  The sets of values $x,x_1,x_4,x_{14}$ and $x,x_2,x_3,x_{23}$ satisfy the bulk equations $Q =0$
\begin{subequations}
  \begin{align}
\label{eqQQ}  Q(x,x_1,x_4,x_{14};\alpha_1,\alpha_4) = &0\,,\\   Q(x,x_2,x_3,x_{23};\alpha_2,\alpha_3) = &0\,.     
  \end{align}
\end{subequations} Since the set of equations $Q = 0$, $p=0$ and $q=0$ form a compatible system on a single quadrilateral (see Figure \ref{fig:factor2}), by imposing \eqref{eq:ppim},   $x, x_1, x_{14}$ and $x, x_2, x_{23}$ obey the boundary equations $q =0$
\begin{subequations}
  \begin{align}
  q(x,x_1,x_{14};\alpha_1,\alpha_4) = &0\,, \label{eq:qx14}\\   q(x,x_2,x_{23};\alpha_2,\alpha_3) =&0\,.    
  \end{align}
\end{subequations}

If the condition  \eqref{eq:pxit} is satisfied, using symmetry arguments by interchanging the indices $1,2$ and $3,4$, one can show $x_{123}, x_{23}, x_{234}$ also satisfy $p=0$, namely, \begin{equation}\label{eq:pxit2}
p(x_{123}, x_{23}, x_{234};\alpha_1,\alpha_4) = 0  \,. 
\end{equation}

Based on the duality between $q=0$ and $p=0$, the values on the rhombic dodecahedron can be uniquely determined with the initial data $x, x_1, x_2$ and $\alpha_1,\alpha_2, \alpha_3, \alpha_4 $ using $Q =0$ and $q = 0$, since the ``front'', ``back'', ``top'' and ``bottom'' equations of the  rhombic dodecahedron (see Figure \ref{fig:facrd}) admit the factorization property as shown in Figure \ref{fig:factor2} thanks to the equations \eqref{eq:ppim}, \eqref{eq:pxit} and \eqref{eq:pxit2}. This implies the sets of values  $x_{14}, x_{124}, x_{1234}$ and $x_{23}, x_{123}, x_{1234}$ also satisfy $q =0$, namely,
\begin{subequations}
  \begin{align}
  q(x_{14},x_{124},x_{1234};\alpha_2,\alpha_3) = &0\,, \\   q(x_{23},x_{123},x_{1234};\alpha_1,\alpha_4) =& 0\,.    
  \end{align}
\end{subequations}

Finally, the consistency around the rhombic dodecahedron for $Q=0$ implies the consistency around a half of the rhombic dodecahedron for $Q = 0$ and $q =0$ modulo certain constraints on the lattice parameters.
\finprf

We have just shown that  \eqref{eq:pxit} provides a sufficient condition that enables us to select integrable ones among the list of  dual boundary equations of a given $Q=0$.
In the following, the  companion boundary equation for the boundary consistency is denoted by  $p =0$, so that its dual boundary equation $q=0$ is  the ``potentially'' integrable one satisfying the boundary consistency. 

\begin{rmk}
One could interpret the condition  \eqref{eq:pxit} as a {\em B\"acklund transformation} of \eqref{eq:px23}. By the tetrahedron property, the sets of values $x_{124}, x_{1}, x_{2}, x_{4}$ and $x_{134}, x_{1}, x_{3}, x_{4} $ satisfy the tetrahedron equations
\begin{equation}
  \label{eq;bakl}
  Q^\intercal (x_{124}, x_{4}, x_{2}, x_{1};\alpha_{12}, \alpha_{14 })=0\,,\quad   Q^\intercal (x_{134}, x_{4}, x_{3}, x_{1};\alpha_{13}, \alpha_{14})=0\,, 
\end{equation}where $\alpha_{ij} = \alpha_i-\alpha_j$ and  $Q^\intercal =0 $ is given in \eqref{eq:QIC}. Then, $x_{124}, x_{134}$ can be expressed as
\begin{subequations}
\label{eq:backll}\begin{equation}
  x_{124} =   \mathbi{g}^\intercal_{(x_{1}, x_{4};\alpha_{14})}( \alpha_{12} ) \,[ x_{2}]  \,,  \quad x_{134} =  \mathbi{g}^\intercal_{(x_{1}, x_{4};\alpha_{14})}(\alpha_{13} ) \, [x_{3}]\,, 
\end{equation}  
where $\mathbi{g}^\intercal$ is the transition matrix associated to $Q^\intercal =0$  (see \eqref{eq:3mllm1}).   Using the boundary equations \eqref{eq:qx14} and \eqref{eq:px14}, one can express $x_{14}$ and $x_4$  as  
\begin{equation}
x_{14} = \mathbi{h}_{x_1}(\alpha_1,\alpha_4)\, [x]\,, \quad x_{4} = \mathbi{k}_{x}(\alpha_1,\alpha_4)\, [x_1]\,, 
\end{equation}
\end{subequations}
where $\mathbi{h}, \mathbi{k}$ are the some boundary matrices similar to \eqref{eq:4k}. 
If the condition   \eqref{eq:pxit} holds, then
\begin{equation}\label{eq:pxitex}
  p(  \mathbi{g}^\intercal_{(x_{1}, x_{4}; \alpha_{14})}(\alpha_{12} ) [x_{2}],  \mathbi{h}_{x_1}(\alpha_1,\alpha_4)\,[x],  \mathbi{g}^\intercal_{(x_{1}, x_{4};\alpha_{14})}( \alpha_{13} ) [x_{3}];\alpha_2,\alpha_3) = 0\,,
\end{equation}
provided that  $x_4 =\mathbi{k}_{x}(\alpha_1,\alpha_4)\, [x_1]$,  which is a  B\"acklund transformation of \eqref{eq:px23}. 
In other words, using the set of transformations \eqref{eq:backll}, one transforms the factorization property of $Q$  from the ``front'' equation to the ``back'' equation, and similarly from the ``bottom'' equation to the ``top'' equation.
\end{rmk}

The condition \eqref{eq:pxit}, or alternatively \eqref{eq:pxitex}, amounts to certain constraints on the parameters $\alpha_1, \alpha_2, \alpha_3, \alpha_4$. This constraint gives rise to the involution relation $\sigma$ as the consistency condition needed in Definition \ref{def:bc1} among the parameters. 

\subsection{Deriving integrable boundary equations and classification}
\label{sec:cibe}


To derive integrable boundary equations, one could readily check the condition \eqref{eq:pxit}, or alternatively \eqref{eq:pxitex}, by ``brute force'' using the list of dual boundary equations $q=0$ and $ p=0$ derived in Section \ref{sec:2fact}. This will done for $q,p$ in the singular case.  
For the boundary equations derived from the M\"obius case, \ie $q,p$ in the $\cM_\pm$ and dual  $\cM_\pm$ cases, we provide an efficient method for determining whether they are integrable.
The full list of integrable boundary equations obtained following  Theorem \ref{th:frb} is given in Table \ref{table:Mcas1} and \ref{ta:spg}.  

\begin{description}
\item[M\"obius case I: $q$ in the $\cM_\pm$ case, $p$ in the dual $\cM_\pm$ case.] Assuming the sets of values $x,x_1,x_{14}$ and $x, x_2, x_{23}$ satisfying $q=0$ (see Figure \ref{fig:rd1}) with $q$ being in the $\cM_\pm$ cases \eqref{eq:mq}. We make a further restriction that the parameters $r_1, r_2, r_4$ appearing in  the $\cM_+$ case are independent of $\alpha,\b$. Then,  $x_{14}, x_{23}$ are M\"obius transformations of $x$ as
  \begin{equation}\label{eq:defm}
          \cM_+: ~ x_{14} = x_{23} = \fm^+_{\fb}(x)  = -\frac{r_2 x +r_4}{r_1 x +r_2}\,,  \quad
      \cM_-: ~  x_{14} = x_{23} =\fm^-_{\fb}(x)  = x\,. 
  \end{equation}
  It follows from the tetrahedron property \eqref{eq:QIC1} that \begin{subequations}
    \begin{align}
     Q^\intercal(x,x_{12},x_{23},x_{13};\alpha_{21} ,\alpha_{23})
      = &  -  Q^\intercal(x,x_{12},\fm^\pm_\fb(x),x_{13};\alpha_{12} ,\alpha_{32}) = 0\,,\label{eq:QIC1111} \\
                    Q^\intercal(x_{1234},x_{12},x_{14},x_{13};\alpha_{34},\alpha_{32})  =  &Q^\intercal(x_{1234},x_{12},\fm^\pm_\fb(x),x_{13};\alpha_{34},\alpha_{32}) = 0 \,, 
    \end{align}
\end{subequations}
where $\alpha_{ij} =\alpha_i -\alpha_j$, and the second equality in \eqref{eq:QIC1111} is a result of \eqref{eq:qfpara}.  Comparing the above two equations, one has $x_{1234} = x$, if  $\alpha_{12} = \alpha_{34}$ holds. This leads to 
\begin{equation}\label{eq:s1234}
  \alpha_3 -\alpha_4 = \alpha_1 -\alpha_2\quad \Rightarrow \quad  \alpha_1 +\alpha_4  = \alpha_2 +\alpha_3 = \mu\,, 
\end{equation}
where $\mu$ is an arbitrary constant. This allows us to set 
\begin{equation}\label{eq:sam}
 \sigma(\alpha) = -\alpha +\mu\,, 
\end{equation}
as an involution relation on the parameters, so that $\alpha_4 = \sigma(\alpha_1)$ and  $\alpha_3 = \sigma(\alpha_2)$. Then,  $x_{1234}$ is connected to $x_{14}$ as $x_{1234} = \fm^\pm_\fb(x_{14})$, and the condition \eqref{eq:pxit} holds.

\item[M\"obius case II: $q$ in the dual $\cM_\pm$ case,   $p$  in the $\cM_\pm$ case.] The companion boundary equation $p =0$ is in the $\cM_\pm$ cases. Let  $r_1, r_2, r_4$ be independent of $\alpha,\b$. Then,  (see Figure \ref{fig:rd1})
  \begin{equation}\label{eq:mbpm}
x_{4/1} = \fm^\pm_\fb(x_{1/4})\,,\quad x_{3/2} = \fm^\pm_\fb(x_{2/3})\,.
\end{equation}
It also follows from the tetrahedron property that
\begin{equation}
Q^\intercal(x_{124},x_4,x_2,x_1,\alpha_{12},\alpha_{14})= 0\,,\quad Q^\intercal(x_{134},x_1,x_3,x_4,\a_{34},\a_{14})=0\,.   \label{eq:qi2}
\end{equation}
Assume $\fm^\pm_\fb$ is a M\"obius transformation making $Q^\intercal =0$ invariant. Then, 
\begin{align} \label{eq:mtqq}
  0=&  \, Q^\intercal(x_{124},x_4,x_2,x_1,\alpha_{12},\alpha_{14})  \propto   \, Q^\intercal(\fm^\pm_\fb (x_{124}), {\fm}^\pm_\fb(x_4), {\fm}^\pm_\fb(x_2), {\fm}^\pm_\fb(x_1);\alpha_{12},\alpha_{14})\nonumber\\
  = & \, Q^\intercal({\fm}^\pm_\fb (x_{124}),x_1,x_3,x_4;\alpha_{12},\alpha_{14}) \,,
\end{align}
Comparing this with the second equation in \eqref{eq:qi2},   the condition \eqref{eq:pxit} holds as
\begin{equation}
{\fm}^\pm_\fb (x_{124}) =x_{134}\,, 
\end{equation}
provided that  $\alpha_{12} = \alpha_{34}$, which imposes a $\sigma$ in the form of \eqref{eq:sam}. 

\end{description}
\begin{table}[htb]
    \centering
    \begin{tabular}{c|c|c}
       $ Q$& $q= r_1\mathbi{q}_1 + r_2\mathbi{q}_2+r_4\mathbi{q}_4$  & $q = \mathbi{q}$\\
        \hline
        Q1($0$)  & $r_1 \mathbi{q}_1+r_2 \mathbi{q}_2+r_4 \mathbi{q}_4$ & N/A \\
        Q1($\delta \neq 0$)    & $r_2 \mathbi{q}_2+r_4 \mathbi{q}_4$ & $\mathbi{q}$ \\
        Q2              & N/A & $\mathbi{q}$ \\
        Q3($0$)  & $r_1 \mathbi{q}_1+r_4 \mathbi{q}_4\,, \mathbi{q}_2 $ & $\mathbi{q}$ \\
        Q3($\delta \neq 0$)    & $\mathbi{q}_2$ & $\mathbi{q}$ \\
      Q4              & $\mathbi{q}_2\,,k \mathbi{q}_1\pm\mathbi{q}_4$ & $\mathbi{q}$ \\
        H1              & $r_1 \mathbi{q}_1+r_2 \mathbi{q}_2+r_4 \mathbi{q}_4$ & N/A \\
        H2              & $r_2 \mathbi{q}_2+r_4 \mathbi{q}_4$ & $\mathbi{q}$ \\
        H3($\delta $)    & $r_1 \mathbi{q}_1+r_4 \mathbi{q}_4\,, \mathbi{q}_2 $ & $\mathbi{q}$
    \end{tabular}
    \caption{M\"obius case II: with  $\sigma(\alpha)=-\a+\mu$, $q= 0$ is integrable. We take $q= r_1\mathbi{q}_1 + r_2\mathbi{q}_2+r_4\mathbi{q}_4$ for $q$  in the dual $\cM_+$ case, and $q= \mathbi{q}$ for $q$  in the dual $\cM_-$ case. Their explicit forms are given in Appendix \ref{app:cbp} and \ref{app:hp}  (represented by $p$). For Q4, $k$ denotes  the elliptic modulus.}. 
    \label{table:Mcas1}
\end{table}

Collecting the above arguments, the integrable boundary equations among the boundary equations dual to each other in the $\cM_\pm$ and dual $\cM_\pm$ cases can be stated as follows. 

\begin{theorem}\label{theo:class}
  For a given $Q$ in the ABS classification, let $q,p$ be a pair of dual boundary polynomials in the $\cM_\pm$ and dual $\cM_\pm$ cases as classified in Theorem \ref{theo:pq}. Assume $r_1,r_2,r_4$ in the $\cM_+$ case are independent of the lattice parameters $\alpha,\beta$ and obey $r_1r_4-r_2^2 \neq  0$.   Then,  $q=0$ is boundary consistent with $Q=0$ in the sense of Definition \ref{def:bc1} with $\sigma(\alpha) = -\alpha +\mu$, where $\mu$ is an arbitrary parameter, in the following two cases
  \begin{enumerate}
  \item M\"obius case I: $q$ is  in the $\cM_\pm$ case;
  \item M\"obius case II: $p$ is in the $\cM_\pm$ case such that $\fm^\pm_\fb$ is a symmetry of the tetrahedron equation $Q^\intercal=0$. 
  \end{enumerate}
\end{theorem}

The proof follows from the  arguments given above.  For $q = 0$ being a nondegenerate boundary equation either in the   $\cM_\pm$ case, or in the dual $\cM_-$ case (the companion equation $p=0$ is in the $\cM_-$ case which is a trivial symmetry for any $Q^\intercal=0$), the boundary consistency is always satisfied provided that  $\sigma(\alpha) = -\alpha +\mu$. For $q = 0$ in the dual $\cM_+$ case, the requirement that $\fm^+_\fb$ is a symmetry for $Q^\intercal =0$ (which is a Q-type bulk equation) imposes certain restriction on the parameters  $r_1,r_2,r_4$. Comparing to the symmetries of the Q-type bulk equations shown in Table \ref{tb:qtype}, this restriction can be easily obtained.
The complete list of integrable $q =0$ with $q$ being in the  dual $\cM_\pm$ case (then, $p$ is in the $\cM_\pm$ case with   $p=0$ being the companion equation) for Q-type and H-type bulk polynomials  is given in Table \ref{table:Mcas1}. The A2 ({\em resp}. A1($\delta$)) equation has the same results as   Q3($0$) ({\em resp}. Q1($\delta$)).

\begin{table}[htb]
    \centering
    \begin{tabular}{c|c|c}
      $  Q$& \eqref{eq:MMMs} &  $\sigma$    \\
        \hline
        Q1($0$)  & $c_0 \fq_0+c_1 \fq_1$ & $\sigma(\alpha)=\mu/\a$ \\
        Q1($\delta\neq  0$)    & $\fq_1$ & $\sigma(\alpha)=\mu/\a$ \\
        Q3($0$)  & $\fq_0$ & $\sigma(\alpha)=\sinh^{-1}(\mu/\sinh(\a))$ \\
               Q3($0$)         & $\fq_1$ & $\sigma(\alpha)=\tanh^{-1}(\mu/\tanh(\a))$ \\
        H3($\delta\neq  0$)    & $\fq_0\,, \fp_0$ & $\sigma(\alpha)=\log(\sqrt{-e^{2 \a}+ \mu})$
    \end{tabular}
    \caption{Singular case: integrable boundary equations for $q, p$ as a pair of dual boundary polynomials in the forms \eqref{eq:MMMs}. The parameter $\mu $ can be taken freely ($\mu \neq 0$ expect for     H3($\delta\neq  0$)). }
\label{ta:spg}  \end{table}

Now turn to $q,p$ as a pair of dual boundary polynomials in the singular case given in Theorem \ref{theo:pq}. The condition \eqref{eq:pxit}, or alternatively \eqref{eq:pxitex}, can be checked by direct computations respectively for $\fq_j$ and $\fp_j$, $j =0,1$.  The integrable ones give rise to certain constraint on the parameters that can be cast into an involution relation  $\sigma$  similar to \eqref{eq:s1234}. For instance, for  Q1($0$),  the condition \eqref{eq:pxit} implies that both $\fq_0 =0$ and $\fq_1 =0$  are  integrable provided that
\begin{equation}
\alpha_1 \alpha_4  = \alpha_2 \alpha_3     \quad \Rightarrow \quad \alpha_{3/4} = \sigma(\alpha_{2/1}) = \frac{\mu}{\alpha_{2/1}}\,,  
\end{equation}
with $\mu \neq  0$ being an arbitrary constant. It turns out that $q = 0$ with $q =c_0\fq_0 +c_1\fq_1 $ as a linear combination of  $\fq_0 $ and $\fq_1$ is also integrable with $\sigma(\alpha) = \mu/\alpha$. 

The complete list of integrable boundary equations for Q3($0$), Q1($\delta$), H3($\delta\neq 0$) is provided in Table \ref{ta:spg}. It turns out that there is also a free parameter $\mu$ in $\sigma$ in all cases. For Q3($0$) and H3($\delta\neq 0$), the branches of the parameters should, {\em a priori},  be fixed in order to have $\sigma$ as an involution. H2 and H3($\delta\neq 0$) have their pairs $\fq_1 =0, \fp_1 =0$  satisfying the boundary consistency with $\sigma(\alpha)  = -\alpha +\mu$. However, they coincide with the $\cM_-$ and dual $\cM_-$ cases. For Q1($0$), Q1($\delta\neq 0$) and Q3($0$), the integrable boundary equations coincide with their dual boundary equations (see Remark \ref{rmk:speq}), it suffices to list one of them. 

Following Theorem \ref{th:frb}, the results listed in Table \ref{table:Mcas1} and \ref{ta:spg}, together with the boundary equations in the M\"obius case I, exhaust all integrable boundary equations in the sense of Definition \ref{def:bc1}, for Q-type and H-type equations in ABS classification. The results of A1($\delta$) (or A2) follow trivially from those of Q1($\delta$) (or Q3($0$)), and are omitted.  Therefore, for bulk equations in the ABS classification, we claim a {\bf classification of integrable boundary equations}  based on the factorization approaches: first the quad-graph equations are factorized  into  pairs of dual boundary equations, and then the boundary consistency is obtained by factorizing the consistency around a rhombic dodecahedron into two equivalent halves. 
Comparing to  known results obtained in \cite{CCZ}, equations listed in Table \ref{table:Mcas1} and \ref{ta:spg} are, in general new, and have the partial list of integrable boundary equations  in  \cite{CCZ} as  subcases.  The involution relation $\sigma$,  a result of the boundary consistency,  enters systematically into the definition integrable boundary equation. Namely, for given $\sigma$ and $q$, the integrable boundary equation is defined as
\begin{equation}
 q(x,u,y;\alpha,\sigma (\alpha)) =0 \,, \quad \text{or } q(x,u,y;\alpha) =0 \,. 
\end{equation}
To each integrable $q=0$, one can associate a discrete boundary zero curvature condition as given in \eqref{eq:3mllm} where the boundary matrix  is obtained using the companion  (dual) equation $p=0$.   

\begin{rmk}Let us also comment on the action of simultaneous M\"obius transformations on the boundary consistency. Recall the equivalence class $[Q]$ having $3$D-consistent $Q_\fm =0$. As a result of Theorem \ref{theo:rd4c},  $Q_\fm$ is also consistent around a rhombic dodecahedron.  Similarly, it follows from Theorem \ref{th:frb} that the boundary consistency is  preserved under simultaneous M\"obius transformations in the sense of \eqref{eq:qfm} and \eqref{eq:dualm11}  on both the bulk equation $Q = 0$ and its integrable boundary equation $q =0$ . Therefore, with a given $\sigma$,  $Q_\fm=0$ has $q_\fm = 0$ as its integrable boundary equation. 
\end{rmk}

\begin{table}[htb]
    \centering
    \begin{tabular}{c|c|c}
       $ Q$& $q= r_1\mathbi{q}_1 + r_2\mathbi{q}_2+r_4\mathbi{q}_4$  & $q = r_1 x y+r_2(x+y)+r_4$\\
        \hline
        Q1($\delta$)    & $\mathbi{q}_2$ & $x+y$ \\
           Q3($0$)  & $\mathbi{q}_1+\mathbi{q}_4\,, \mathbi{q}_2$ & $x y+1\,, x+y$ \\
        Q3($\delta \neq 0$)    & $\mathbi{q}_2$ & $x+y$ \\
      	Q4              & $k \mathbi{q}_1+1\,, \mathbi{q}_2$ & $k x y+1\,, x+y$ \\
        H1              & $\mathbi{q}_1+c \mathbi{q}_4\,, \mathbi{q}_2$ & $x y+ c\,, x+y$  \\
        H2              & $\mathbi{q}_2 + c \mathbi{q}_4$ & $x+y+c$ \\
        H3($0$)    & $\mathbi{q}_1+\mathbi{q}_4\,, \mathbi{q}_2 $ & $x y +1\,, x+y$\\
        H3($\delta \neq 0$) & $\mathbi{q}_1+ c \mathbi{q}_4\,, \mathbi{q}_2 $ & $x  y +c \,, x+y$
    \end{tabular}
	\caption{Canonical forms of integrable $q=0$ in the $\cM_+$  and dual $\cM_+$ cases. For H2, $c$ is an arbitrary parameter; for H1 and H3($\delta\neq 0$), $c\neq  0$. For Q4,  $k$ denotes  the elliptic modulus.}
\label{ta:ca}\end{table}

\begin{rmk}One can further reduce the integrable boundary equations $q=0$ appearing  the  $\cM_+$ and dual $\cM_+$ cases to their {\em canonical forms}  by factoring out the action of simultaneous M\"obius transformations that are also symmetries of $Q=0$. For instance, for  Q1($0$), the generic M\"obius transformation is a symmetry. One can reduce the integrable boundary equation $q =  r_1 x y+r_2(x+y)+r_4$ in the $\cM_+$ case to $q = x+y$ that is a parameter-free boundary equation without changing the form of the bulk equation $Q = 0$. This is also accompanied by reducing the parameters appearing in the dual $\cM_+$ case.  For a bulk-equation in its canonical form, its integrable boundary equations with least number of parameters are referred to as canonical forms (they are not unique). The canonical forms are listed in Table \ref{ta:ca}. For the derivation of canonical forms of H-type equations, their symmetries are needed (they are $\pm x$, $ x+c$ for H1; $x$ for H2;  $c x$, $ 1/x$ for H3($\delta=0$); $\pm x$ for H3($\delta \neq 0$)). Moreover, for   Q1($0$), the canonical form of the integrable boundary equation $q =0$ in the singular case given in Table \ref{ta:spg} can be either $\fq_0 = 0$ or $\fq_1 =0$ (they can be transformed from one to the other using $\fm (x) = 1/x$).
\end{rmk}

\subsection{Degenerate integrable boundary equations}\label{sec:degbc}
The above derivations exclude the degenerate boundary equations, since they do not lead to any dual boundary equations through \eqref{eq:HH1}. Here, we provide a criterion to select integrable degenerate boundary equations. It turns out that the boundary consistency in this case can be understood as certain ``degeneration'' of the consistent system around the rhombic dodecahedron. In particular, the boundary consistency does not require any involution relation $\sigma$, this is in contrast to the nondegenerate boundary consistency given in Definition \ref{def:bc1}.

There are two possibilities for a $q =0$ to be degenerate (see discussions right below Definition \ref{def:nondeg}). They are respectively called degenerate case I and II. 
\begin{description}

\item[Degenerate case I:] let $q =  0 $ be a boundary equation leading to  $\delta_{u,v}Q =0$. This applies to the H-type equations,  Q1($\delta$), Q3($0$) and A-type equations. 
  Take Q1($\delta$), Q3($0$) as examples. Arguments for H-type equations and A-type equations follow similarly.
  
  Let $Q$ to be Q1($\delta$) or Q3($0$), one has
   \begin{equation}
    \text{Q1($\delta$)}:~~ q =x-y\pm(\a-\b)\delta\,, \quad \text{Q3($0$)}:~~ q= e^{\a}x -e^{\b} y\,,  \text{~or~} q =e^{\b}x -e^{\a}y\,.
  \end{equation}
It can be checked by direct computations that $q=0$ is consistent with the associated $Q=0$ around  a half of a rhombic dodecahedron without any restriction on the lattice parameters. This can be understood as a ``degeneration'' of the consistency around a rhombic dodecahedron: following Figure \ref{fig:rd1}, by imposing the above $q$ to the ``bottom'' ({\em resp}.  ``front'') bulk equation $Q=0$ with $    q(x,x_1,x_{14};\alpha_1, \alpha_4 )=0$ ({\em resp}.   $    q(x,x_2,x_{23};\alpha_2,\alpha_3 )=0$),  then the ``bottom'' ({\em resp}. ``front'') equation can have a singular solution with respect to $x_1$  ({\em resp}.  to $x_2$), which allows us to assign generic value to $x_1$  ({\em resp}.  to $x_2$). The tetrahedron equation $Q^\intercal(x,x_{12}, x_{14},x_{24};\alpha_{12},\alpha_{14}) =0$, where $\alpha_{ij}=\alpha_i-\alpha_j$, 
also has a singular solution with respect to $x_{12}$, and the values $x_{14}, x_{23}, x_{24}$ are fixed and depend on $x$ only. Moreover, the tetrahedron equation  $Q^\intercal(x_{1234},x_{12}, x_{23},x_{24};\alpha_{43},\alpha_{41} ) = 0$ with $x_{23}, x_{14}$ obtained above is also singular with respect to $x_{12}$, and the values $x_{1234},x_{23}$ can be connected using $q = 0$. The values  $x_{1234},x_{14}$ can also be connected using  $q=0$, since $Q^\intercal(x_{1234},x_{14}, x_{24},x_{34};\alpha_{32},\alpha_{31} ) = 0$ is singular with respect to $x_{34}$. The boundary consistency around a half of the rhombic dodecahedron follows from the consistency around the whole rhombic dodecahedron. 

    \begin{table}[htb]
        \centering
    \begin{tabular}{c|c|c | c |c |c}
      \hline
        Q1($0$)& Q1($\delta\neq 0$) & Q2 &  Q3($0$) & Q3($\delta \neq0$)& Q4 \\
      \hline
       $ (c_0, c_1)\neq (0,0)$ & $ c_1 = 0$ & $ c_1 = 0$ & $c_0 =0$ or $c_1=0$ & $c_1 = 0$ & N/A  \\
\end{tabular}
        \caption{Possibilities of $c_0, c_1$ not vanishing simultaneously to make a Q-type bulk equation \eqref{eq:cccbulk} holds for generic $y$.}
\label{ta:geny}    \end{table}

\item[Degenerate case II:]
  take $q$ in the form \eqref{eq:qc0c1}. In this case, the consistency between $Q=0$ and $q=0$ around a half of the rhombic dodecahedron can also be understood as certain ``degeneration'' of the consistency around the whole rhombic dodecahedron.

  Following Figure \ref{fig:rd1}, one imposes $q= 0$ to the ``bottom''  and  ``front'' equations, and uses  $x, x_1, x_2$ as well as $ \a_1, \a_2, \a_3, \a_4$ as initial data. This leads to   $x_4= x_3=c_0/c_1$ (see discussions after \eqref{eq:qc0c1}), and the ``bottom''  and  ``front'' bulk equations reduce to $q =0$, and are independent of $x_4$ and $x_3$ respectively. If the ``top'' and ``back'' equations also hold independently of $x_{234}$ and $x_{134}$ respectively with $x_{234}=x_{134}=c_0/c_1$, then they reduce to $q =0$, the consistency around the left half of the rhombic dodecahedron can be fulfilled without any restriction on the lattice parameters.

Having $x_4= x_3=c_0/c_1$, the condition  $x_{234}=x_{134}=c_0/c_1$ can be checked using the tetrahedron equations
  \begin{equation}
	Q^\intercal(x_{134},x_1,x_4,x_3; \alpha_{43}, \alpha_{41})=0\,,\quad 		Q^\intercal(x_{234},x_2,x_3,x_4;\alpha_{42},\alpha_{43})=0\,.    
  \end{equation}
  In other words, one needs the first equation holds with $x_4=x_3=x_{134}=c_0/c_1$ for generic input $x_1$, and similarly to the second equation with $x_4=x_3=x_{234}=c_0/c_1$ for generic input $x_2$.  This amounts to examining if the  tetrahedron equation 
  \begin{equation}\label{eq:cccbulk}
Q^\intercal(c_0/c_1,c_0/c_1,c_0/c_1,y;\alpha,\beta)=0\,,
  \end{equation}
as a  Q-type bulk equation, holds for generic $y$. This is related to properties of the Q-type bulk equations, and possible values of  $c_0, c_1$ are listed in Table \ref{ta:geny}.

\end{description}
Let us summarize the above results.
\begin{table}[ht]
    \centering
    \begin{tabular}{c|c|c}
       $ Q$& degenerate case I & degenerate case II ($q$ in the form \eqref{eq:qc0c1})   \\
        \hline
        Q1($0$), H1  & $x-y$ & $c_1 S+ c_0 T$   \\
        Q1($\delta$), H2    & $x-y\pm(\a-\b)\delta$ & $T$  \\
        Q2             & N/A   & $T$ \\
        Q3($0$), H3($\delta$)   & $e^{\a}x -e^{\b} y,e^{\b}x -e^{\a} y$ & $S,T$ \\
        Q3($\delta$)   & N/A   & $T$ 
    \end{tabular}
    \caption{Degenerate integrable boundary equations: the boundary consistency holds without any restriction on the parameters. For H2 in the  degenerate case I, $ \delta =1$. Degenerate case II is derived using Table \ref{ta:geny}, and explicit forms of $S,T$ are given in Table \ref{tal:st1}.    }\label{tal:st}
\end{table}

\begin{table}[ht]
    \centering
    \begin{tabular}{c|c|c}
        $Q$ & $S$ & $T$  \\
        \hline
      Q1($0$)  & $\alpha x(u-y)-\beta y (x-u)$ 	& $\a( y-u)-\beta(x-u)u$   \\
      Q1($\delta\neq 0$)  & not needed ({\it n.n.})	& $\a( y-u)-\beta(x-u)u$    \\
        Q2             & {\it n.n.}		   & $\a y-\b x +(\a-\b)(\a \b- u)$ \\
        Q3($0$)   & $\sinh(\a) u x -\sinh(\b)u y-\sinh(\a-\b) x y $          & $\sinh(\a) y-\sinh(\b)x  - u \sinh(\a-\b)$ \\
        Q3($\delta\neq 0$)   &{\it n.n.}  & $\sinh(\a) y-\sinh(\b)x  - u \sinh(\a-\b)$ \\
        H1  & $ u (x-y)-\a+\b$                    & $y-x$ \\
        H2 & {\it n.n.}                      & $y- x -\a +\b$  \\
        H3($\delta$)    & $(e^\a  x- e^\b y)u+ \delta(e^{2 \a}-e^{2 \b}) $        		& $ e^{\a} y-e^{\b} x $ 
    \end{tabular}
    \caption{Explicit forms of $S,T$  needed in Table \ref{tal:st}.}
\label{tal:st1}\end{table}

\begin{theorem}\label{theo:deg}
Let $q=0$ be a degenerate boundary equation, and  $Q=0$ a bulk equation in the ABS classification. We say $q=0$ is boundary consistent with $Q=0$, if the initial value problem on the half rhombic dodecahedron, \cf left picture in Figure \ref{fig:bccy}, is well-posed.  The list of such $q=0$ belonging to degenerate case I and II as discussed above is given in Table \ref{tal:st}. 
\end{theorem}

\begin{rmk}
In \cite{AJ1, AJ2}, the so-called ``singular boundary'' reductions were considered for Q-type bulk equations in the ABS classification, where the initial data of an  initial value problem  on a $\ZZ^2$-lattice were taken as singular solutions to $Q=0$. Some of their results are in connection to the integrable degenerate boundary equations of case II. For instance, for Q1($0$), one has $q =0$ with $q$ in the form \eqref{eq:qc0c1} which provides a singular solution to Q1($0$). Then, an initial-boundary value problem on a $\ZZ^2$-lattice with a boundary with $q =0$ being the boundary conditions corresponds to an initial value problem with a ``singular boundary'' considered in       \cite{AJ1, AJ2}. This is also the cases for other Q-type equations except for  Q4, where the ``singular boundary'' solutions provided in  \cite{AJ1, AJ2} can not be written as a boundary equation formulated in this paper, since Q4 does not have any integrable degenerate boundary equations.     
\end{rmk}
\section{Integrable boundary equations for   H1$^\epsilon$}\label{sec:cd4}
In this section, we consider integrable boundary equations for the H1$^\epsilon$ equation which belongs to the H$^\epsilon$-type (also called H$^4$-type in \cite{RB1}) equations as an extension of H-type equations in ABS classification \cite{ABS3}. The derivation of integrable $q=0$ is based on the successive factorization approaches presented in this paper. Here, we do not intend to provide the complete list of integrable $q=0$, since it requires a substantial amount of computations but following the exactly same techniques as presented above.

\begin{figure}[h!]
\begin{center}
\begin{tikzpicture}[scale=.9]
\draw[very thin]
    (4,4)--(4,1)
    (5,0)--(7,0)
    (4,1)--(2,1)
    (1,1)--(1,-1)--(2,-1)--(2,0)
    (2,1)--(2,2)--(5,2)--(5,3)--(3,3)--(3,0)
    (2,0)--(0,0)
    (5,0)--(5,-2)
    (4,-2)--(3,-2)--(3,-3)--(4,-3)--(4,-2)
    (4,-1)--(6,-1)--(6,1)--(7,1)--(7,0)
    (4,-4)--(4,-3)--(5,-3)
    (5,-2)--(6,-2)--(6,-1)--(7,-1)--(7,0)--(8,0);
\draw[thick, dashed]
	(1,1)--(2,1)--(2,0)--(3,0)
	(4,-1)--(4,-2)--(5,-2)--(5,-3);
\draw[very thick]
	(0,0)--(4,4)
	(4,-4)--(8,0);
\draw[dotted, thin]
    (2,-2)--(3,-2)--(3,0)--(5,0)--(5,2)--(6,2)--(6,1)--(4,1)--(4,-1)--(2,-1)--(2,-2);
\fill
    \foreach \x in {(2,1),(3,0),(4,-1),(5,-2),(1,0),(2,-1),(3,-2),(4,-3),(4,3),(3,2),(5,2),(4,1),(6,-1),(5,0),
    		(7,0),(6,1)	}
            {\x circle (0.08cm)};
\fill[color=white]
    \foreach \x in { (1,1),(2,0),(4,-2),(5,-3),(0,0),(1,-1),(3,-3),(4,-4),(4,4),(3,3),(5,3),(2,2),(4,2),(3,1),(6,-2),
    		(5,-1),(7,-1),(6,0),(8,0),(7,1)}
            {\x circle (0.08cm)};
\draw
    \foreach \x in {(1,1),(2,1),(2,0),(3,0),(4,-1),(4,-2),(5,-2),(5,-3),(0,0),(1,0),(1,-1),(2,-1),(3,-2),(3,-3),(4,-3),(4,-4),(4,4),(3,3),(4,3),
    (5,3),(2,2),(3,2),(4,2),(5,2),(3,1),(4,1),(6,-2),(5,-1),(6,-1),(7,-1),(5,0),
    (6,0),(7,0),(8,0),(6,1),(7,1)}
            {\x circle (0.08cm)};
\draw   (1,1) node[above left] {\footnotesize$x_1$}
        (2,1) node[above left] {\footnotesize$x_2$}
        (2,0) node[above left] {\footnotesize$x_3$}
        (3,0) node[above left] {\footnotesize$x_4$}
        (4,-1) node[above left] {\footnotesize$x_{2n-2}$}
        (4,-2) node[above left] {\footnotesize$x_{2n-1}$}
        (5,-2) node[above left] {\footnotesize$x_{2n}$}
        (5,-3) node[above left] {\footnotesize$x_{2n+1}$};
\end{tikzpicture}
\begin{tikzpicture}[scale=.9]
\draw[very thin]
    (4,4)--(4,1)
    (5,0)--(7,0)
    (4,1)--(2,1)
    (1,1)--(1,-1)--(2,-1)--(2,0)
    (2,1)--(2,2)--(5,2)--(5,3)--(5,3)--(3,3)--(3,0)
    (2,0)--(0,0)
    (5,0)--(5,-2)
    (4,-2)--(3,-2)--(3,-3)--(4,-3)--(4,-2)
    (4,-1)--(6,-1)--(6,1)--(7,1)--(7,0);
\draw[thick, dashed]
	(1,1)--(2,1)--(2,0)--(3,0)
	(4,-1)--(4,-2)--(5,-2);
\draw[very thick]
	(0,0)--(4,4)
	(4,-3)--(7,0);
\draw[dotted,thin]
    (2,-2)--(3,-2)--(3,0)--(5,0)--(5,2)--(6,2)--(6,1)--(4,1)--(4,-1)--(2,-1)--(2,-2);
\fill
    \foreach \x in {(2,1),(3,0),(4,-1),(5,-2),(1,0),(2,-1),(3,-2),(4,-3),(4,3),(3,2),(5,2),(4,1),(6,-1),(5,0),
    		(7,0),(6,1)	}
            {\x circle (0.08cm)};
\fill[color=white]
    \foreach \x in { (1,1),(2,0),(4,-2),(0,0),(1,-1),(3,-3),(4,4),(3,3),(5,3),(2,2),(4,2),(3,1),
    		(5,-1),(6,0),(7,1)}
            {\x circle (0.08cm)};
\draw
    \foreach \x in {(1,1),(2,1),(2,0),(3,0),(4,-1),(4,-2),(5,-2),(0,0),(1,0),(1,-1),(2,-1),(3,-2),(3,-3),(4,-3),(4,4),(3,3),(4,3),
    (5,3),(2,2),(3,2),(4,2),(5,2),(3,1),(4,1),(5,-1),(6,-1),(5,0),
    (6,0),(7,0),(6,1),(7,1)}
            {\x circle (0.08cm)};            
\draw   (1,1) node[above left] {\footnotesize$x_1$}
        (2,1) node[above left] {\footnotesize$x_2$}
        (2,0) node[above left] {\footnotesize$x_3$}
        (3,0) node[above left] {\footnotesize$x_4$}
        (4,-1) node[above left] {\footnotesize$x_{2n-2}$}
        (4,-2) node[above left] {\footnotesize$x_{2n-1}$}
        (5,-2) node[above left] {\footnotesize$x_{2n}$};       
\end{tikzpicture}
\end{center}
\caption{Well-posed quad-graph systems on trips with two parallel boundaries with odd (left) and even (right) initial data (denoted by dashed lines and the parameters are omitted). The odd case has the same patterns of triangles as boundaries, while the even case has boundaries with different patterns.  }
	\label{fig:dynamicwithboundary}\end{figure}
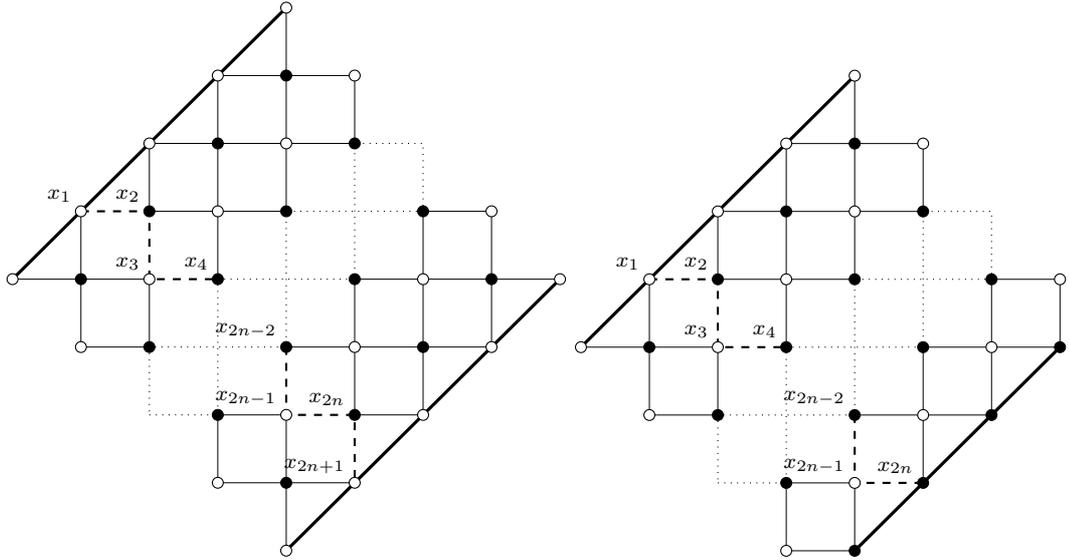

One particularity of  the H1$^\epsilon$ and other H$^\epsilon$-type equations is that their bulk polynomials are rhombic-symmetric only, which implies that one should also take the patterns of the quadrilateral  (orientation of the white and black vertices) into account. Quad-graph systems with boundary for the H$^\epsilon$-type equations could have two types of boundary conditions depending on the patterns of the triangles assigned to the boundary, and the boundary consistency conditions should also be adapted to the patterns of the quadrilaterals and triangles. Here, we provide examples of integrable boundary equations for H1$^\epsilon$, and also give quad-graph systems on a strip with two boundaries that are of different pattern types (see Figure \ref{fig:dynamicwithboundary}).

       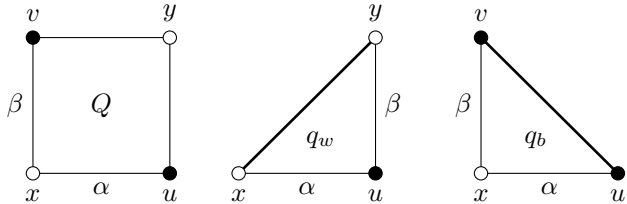
\begin{figure}[h]
    \centering
      \begin{tikzpicture}[scale=1.8]
    \def\d{1}%
    \def\le{0.2}%
    \def\r{0.06}
    \def\ld{1.5}%
    \tikzstyle{nod1}= [circle, inner sep=0pt, fill=white, minimum size=5pt, draw]
    \tikzstyle{nod}= [circle, inner sep=0pt, fill=black, minimum size=5pt, draw]

    \coordinate (u00) at (0,0);
    \coordinate (u10) at (\d,0);
    \coordinate (u01) at (0,\d);
    \coordinate (u11) at (\d,\d);
    \draw[-] (u00)   -- node [below]{ $\a$} (u10)  -- (u11) -- (u01) -- node [left]{ $\b$}(u00);

    \node[nod1] (u00) at (0,0) [label=below: $x$] {};
    \node[nod] (u10) at (\d,0) [label=below: $u$] {};
    \node[nod] (u01) at (0,\d) [label=above:$v$] {};
    \node[nod1] (u11) at (\d,\d) [label=above:$y$] {};
    \node at (0.5*\d,0.5*\d) {$Q$};
\end{tikzpicture}
\quad
  \begin{tikzpicture}[scale=1.8]
    \def\d{1}%
    \def\le{0.2}%
    \def\r{0.06}
    \def\ld{1.5}%
    \tikzstyle{nod1}= [circle, inner sep=0pt, fill=white, minimum size=5pt, draw]
    \tikzstyle{nod}= [circle, inner sep=0pt, fill=black, minimum size=5pt, draw]

    \coordinate (u00) at (0,0);
    \coordinate (u10) at (\d,0);
    \coordinate (u01) at (0,\d);
    \coordinate (u11) at (\d,\d);
    \draw[-] (u00)   -- node [below]{ $\a$} (u10);
    \draw[-] (u11)   -- node [right]{ $\b$} (u10);
    \draw[line width=1pt] (u00)-- (u11);

    \node[nod1] (u00) at (0,0) [label=below: $x$] {};
    \node[nod] (u10) at (\d,0) [label=below: $u$] {};
    \node[nod1] (u11) at (\d,\d) [label=above:$y$] {};
    \node  at (0.6*\d,0.25*\d) {$q_w$};
\end{tikzpicture}
    \quad
  \begin{tikzpicture}[scale=1.8]
    \def\d{1}%
    \def\le{0.2}%
    \def\r{0.06}
    \def\ld{1.5}%
    \tikzstyle{nod1}= [circle, inner sep=0pt, fill=white, minimum size=5pt, draw]
    \tikzstyle{nod}= [circle, inner sep=0pt, fill=black, minimum size=5pt, draw]

    \coordinate (u00) at (0,0);
    \coordinate (u10) at (\d,0);
    \coordinate (u01) at (0,\d);
    \coordinate (u11) at (\d,\d);
    \draw[-] (u00)   -- node [below]{ $\a$} (u10);
    \draw[-] (u01) -- node [left]{ $\b$}(u00);
    \draw[line width=1pt] (u10)-- (u01);

    \node[nod1] (u00) at (0,0) [label=below: $x$] {};
    \node[nod] (u10) at (\d,0) [label=below: $u$] {};
    \node[nod] (u01) at (0,\d) [label=above:$v$] {};
    \node  at (0.4*\d,0.25*\d) {$q_b$};
\end{tikzpicture}
	\caption{A rhombic-symmetric $Q$ has two types of boundary equations depending on the patterns of the underlying triangles. They are denoted by $q_w$ whose boundary vertices are white dots, and $q_b$ whose boundary vertices are black dots. }
	\label{fig:rhombicboundary}
\end{figure}

The bulk polynomial $Q$ of  H1$^\epsilon$  reads
\begin{equation}
Q:=Q(x,u,v,y;\alpha,\beta) =  (x-y)(u-v)+(\b-\a)(1+\epsilon u v)\,, 
\end{equation}
which is affine-linear with respect to each of its arguments. When $\epsilon =0$, it coincides with H1. In contrast to H1, or any  $Q$ in the ABS classification, it is  rhombic-symmetric
    \begin{equation}\label{eq:rs}
    Q(x,u,v,y,\a,\b)=-Q(x,v,u,y,\b,\a)\,,\quad    Q(x,u,v,y,\a,\b)=-Q(y,u,v,x,\b,\a)\,, 
    \end{equation}
meaning that it is irrelevant to interchange the black and white vertices (one gets a different bulk equation by doing so). We set the convention following Figure \ref{fig:rhombicboundary}: the first and fourth arguments of $Q$ are assigned to the white vertices, while the second and third arguments are assigned to the black vertices. The biquadratic polynomials of H1$^\epsilon$ are
  \begin{equation}
\Gamma:=    \delta_{v,y}Q  =  -(\alpha-\beta)(1+\epsilon u^2) \,, \quad
    \delta_{u,v}Q  =  (x-y)^2+\epsilon (\a-\b)^2 \,, \quad
    \delta_{x,y}Q  =  (u-v)^2 \,. 
  \end{equation}
  Due to the symmetry \eqref{eq:rs}, the biquadratic polynomials attached to the edges of the underlying quadrilateral take similar forms, but the two biquadratic polynomials attached to the two crossing diagonals are of different forms.

  The H1$^\epsilon$ equation is   3$D$-consistent following the assignments of white and black vertices as shown in Figure \ref{fig:31tetra}. It also processes the tetrahedron properties with two different tetrahedron equations  $Q^{\intercal}_w = 0$  and  $Q^{\intercal}_b = 0$ that are satisfied respectively by the values on the white and  black vertices of the cube. Following Figure \ref{fig:31tetra}, one has 
        \begin{equation}
Q^{\intercal}_w(x,x_{12},x_{13},x_{23};\alpha_{12},\alpha_{13}) = 0\,, \quad             Q^{\intercal}_b(x_{123},x_1,x_2,x_3;\alpha_{32},\alpha_{31}  ) = 0\,,  
        \end{equation}
where $\alpha_{ij} = \alpha_i -\alpha_j$, and $Q^{\intercal}_w =0$ is Q1($\sqrt{-\epsilon}$) and  $Q^{\intercal}_b = 0$ is Q1($0$). 

 Now, we turn to the factorization of H1$^\epsilon$ to derive its pairs of dual boundary equations. We follow the definition of boundary equations given in section \ref{sec:bpbe}.  In addition to consider the boundary polynomial \eqref{eq:qst1} whose boundary vertices are white dots, one could also take a boundary polynomial whose boundary vertices are black dots. These correspond to two patterns of triangles  (see Figure \ref{fig:rhombicboundary}),  and the associated boundary polynomials are respectively denoted by $q_w$ and $q_b$ as
 \begin{equation}
   q_w(x,u,y;\alpha,\beta) \,,\quad  q_b  (u,x,v;\alpha,\beta)\,. 
 \end{equation}
Both $q_w$ and $q_b$  should have dual boundary polynomials, and they are not necessarily dual to each other. Let us illustrate this property by taking $q_w$ and $q_b$ in the $\cM_+$ case. 
\begin{figure}[thb]
  \centering
  \begin{subfigure}[b]{0.4\textwidth}
\begin{tikzpicture}[scale=1.1]
    \tikzstyle{nod1}= [circle, inner sep=0pt, minimum size=5pt, draw]
    \tikzstyle{nod}= [circle, inner sep=0pt, fill=black, minimum size=5pt, draw]
    \tikzstyle{vertex}=[circle,minimum size=20pt,inner sep=0pt]
    \tikzstyle{selected vertex} = [vertex, fill=red!24]
    \tikzstyle{selected edge} = [draw,line width=1.5pt,-]
    \tikzstyle{edge} = [draw, thick,-,black] 
    \tikzstyle{ddedge} = [draw, densely dotted,-,black] 
    \tikzstyle{dedge} = [draw, dashed,-,black] 
    \tikzstyle{eedge} = [draw, line width=1.5pt,-,black] 
    \pgfmathsetmacro \lv {1.1}
    \pgfmathsetmacro \lh {0.8}
    \pgfmathsetmacro \ld {2}
    \pgfmathsetmacro \an {38}

    \pgfmathsetmacro \llx {{\ld*tan(\an)}}
    \pgfmathsetmacro \lly {{\ld/sin(\an)}}
    \node (em) at (0,0) [label=right:${}$] {};
    \node[nod1] (z) at (0+\lh,\lv) [label=below:$z$] {};
    \node[nod1] (x) at (2*\ld+\lh,0+\lv) [label=below:$x$] {};
    \node[nod1] (bz) at (0+\lh,2*\ld+\lv) [label=above:$t$] {};
    \node[nod1] (bx) at (2*\ld+\lh,2*\ld+\lv) [label=above:$r$] {};
    \draw[eedge] (z) -- node[below]{\footnotesize $\alpha = \sigma(\beta)       $}   (x)  -- node[right]{\footnotesize $\eta = \sigma(\lambda)       $}   (bx) --node[above]{\footnotesize $\beta = \sigma(\alpha)       $}   (bz) --node[left]{\footnotesize $\lambda = \sigma(\eta)       $}   (z);

    \node[nod] (by) at (0+\lh+\ld,\lv+\lly) [label=below right:$s$] {};
    \node[nod] (w) at (2*\ld+\lh-\lly,0+\lv+\ld) [label=left:$w$] {};
    \node[nod] (u) at (0+\lh+\lly,\lv+\ld) [label=right:$u$] {};
    \node[nod] (y) at (0+\lh+\ld,2*\ld+\lv-\lly) [label=above right :${y}$] {};
    \draw[edge] (bz) -- node [below left] {$\beta$} (by)  -- node [below right] {$\alpha$}(bx) -- node [left] {$\eta$} (u) --node [left] {$\lambda$} (x) --node [above right] {$\alpha$} (y) --node [above left] {$\beta$} (z) -- node [right] {$\lambda$} (w) -- node [right] {$\eta$}(bz);
    \node[nod1] (v) at (0+\lh+\ld,\lv+\ld) [label=below right:$v$] {};
    \draw[edge] (w) --  
    (v) --  
    (u);
    \draw[edge] (by) --  
    (v) -- 
    (y);

    \coordinate (Y0) at (2*\ld+\lh,\ld+\lv);
    \coordinate (Y1) at (2*\ld/2+\lh/2+0+\lh/2+\lly/2,2*\ld/2+\lv/2+\lv/2+\ld/2);
    \coordinate (Y2) at ( \lh/2+\ld/2+\lh/2+\ld/2,\lv/2+\lly/2+\lv/2+\ld/2);
    \coordinate (Y3) at ( 2*\ld/2+\lh/2-\lly/2+\lh/2,2*\ld/2+\lv/2+\lv/2+\ld/2);
    \coordinate (Y4) at (0+\lh,2*\ld/2+\lv);
    \coordinate (Y5) at ( 2*\ld/2+\lh/2-\lly/2+\lh/2,\lv/2+\lv/2+\ld/2);
    \coordinate (Y6) at ( \lh/2+\ld/2+\lh/2+\ld/2,2*\ld/2+\lv/2-\lly/2+\lv/2+\ld/2);
    \coordinate (Y7) at  (2*\ld/2+\lh/2+0+\lh/2+\lly/2,\lv/2+\lv/2+\ld/2);

    \coordinate (X0)  at (\ld+\lh,0+\lv);
    \coordinate (X1) at (\lh/2+\ld/2+2*\ld/2+\lh/2, 2*\ld/2+\lv/2-\lly/2+\lv/2);
  \coordinate (X2) at (\lh/2+\ld/2+\lh/2+\lly/2, \lv/2+\ld/2+\lv/2+\ld/2);
  \coordinate (X3) at ( 2*\ld/2+\lh/2+\lh/2+\ld/2,2*\ld/2+\lv/2+\lv/2+\lly/2 );
  \coordinate (X4) at  (0+\lh/2+0+\lh/2+\ld/2,2*\ld/2+\lv/2+\lv/2+\lly/2);
  \coordinate (X41) at (\ld+\lh,0+2*\ld+\lv);
  \coordinate (X5) at  (0+\lh/2+\ld/2+2*\ld/2+\lh/2-\lly/2,\lv/2+\ld/2+0+\lv/2+\ld/2);
  \coordinate (X6) at (\lh/2+\ld/2+\lh/2,\lv/2+2*\ld/2+\lv/2-\lly/2);

  \end{tikzpicture}
  \end{subfigure}
  \qquad \qquad
  \begin{subfigure}[b]{0.4\textwidth}
\begin{tikzpicture}[scale=1.1]
    \tikzstyle{nod1}= [circle, inner sep=0pt, minimum size=5pt, draw]
    \tikzstyle{nod}= [circle, inner sep=0pt, fill=black, minimum size=5pt, draw]
    \tikzstyle{vertex}=[circle,minimum size=20pt,inner sep=0pt]
    \tikzstyle{selected vertex} = [vertex, fill=red!24]
    \tikzstyle{selected edge} = [draw,line width=1.5pt,-]
    \tikzstyle{edge} = [draw, thick,-,black] 
    \tikzstyle{ddedge} = [draw, densely dotted,-,black] 
    \tikzstyle{dedge} = [draw, dashed,-,black] 
    \tikzstyle{eedge} = [draw, line width=1.5pt,-,black] 
    \pgfmathsetmacro \lv {1.1}
    \pgfmathsetmacro \lh {0.8}
    \pgfmathsetmacro \ld {2}
    \pgfmathsetmacro \an {38}

    \pgfmathsetmacro \llx {{\ld*tan(\an)}}
    \pgfmathsetmacro \lly {{\ld/sin(\an)}}
    \node (em) at (0,0) [label=right:${}$] {};
    \node[nod] (z) at (0+\lh,\lv) [label=below:$z$] {};
    \node[nod] (x) at (2*\ld+\lh,0+\lv) [label=below:$x$] {};
    \node[nod] (bz) at (0+\lh,2*\ld+\lv) [label=above:$t$] {};
    \node[nod] (bx) at (2*\ld+\lh,2*\ld+\lv) [label=above:$r$] {};
    \draw[eedge] (z) -- node[below]{\footnotesize $\alpha = \sigma(\beta)       $}   (x)  -- node[right]{\footnotesize $\eta = \sigma(\lambda)       $}   (bx) --node[above]{\footnotesize $\beta = \sigma(\alpha)       $}   (bz) --node[left]{\footnotesize $\lambda = \sigma(\eta)       $}   (z);

    \node[nod1] (by) at (0+\lh+\ld,\lv+\lly) [label=below right:$s$] {};
    \node[nod1] (w) at (2*\ld+\lh-\lly,0+\lv+\ld) [label=left:$w$] {};
    \node[nod1] (u) at (0+\lh+\lly,\lv+\ld) [label=right:$u$] {};
    \node[nod1] (y) at (0+\lh+\ld,2*\ld+\lv-\lly) [label=above right :${y}$] {};
    \draw[edge] (bz) -- node [below left] {$\beta$} (by)  -- node [below right] {$\alpha$}(bx) -- node [left] {$\eta$} (u) --node [left] {$\lambda$} (x) --node [above right] {$\alpha$} (y) --node [above left] {$\beta$} (z) -- node [right] {$\lambda$} (w) -- node [right] {$\eta$}(bz);
    \node[nod] (v) at (0+\lh+\ld,\lv+\ld) [label=below right:$v$] {};
    \draw[edge] (w) --  
    (v) --  
    (u);
    \draw[edge] (by) --  
    (v) -- 
    (y);

    \coordinate (Y0) at (2*\ld+\lh,\ld+\lv);
    \coordinate (Y1) at (2*\ld/2+\lh/2+0+\lh/2+\lly/2,2*\ld/2+\lv/2+\lv/2+\ld/2);
    \coordinate (Y2) at ( \lh/2+\ld/2+\lh/2+\ld/2,\lv/2+\lly/2+\lv/2+\ld/2);
    \coordinate (Y3) at ( 2*\ld/2+\lh/2-\lly/2+\lh/2,2*\ld/2+\lv/2+\lv/2+\ld/2);
    \coordinate (Y4) at (0+\lh,2*\ld/2+\lv);
    \coordinate (Y5) at ( 2*\ld/2+\lh/2-\lly/2+\lh/2,\lv/2+\lv/2+\ld/2);
    \coordinate (Y6) at ( \lh/2+\ld/2+\lh/2+\ld/2,2*\ld/2+\lv/2-\lly/2+\lv/2+\ld/2);
    \coordinate (Y7) at  (2*\ld/2+\lh/2+0+\lh/2+\lly/2,\lv/2+\lv/2+\ld/2);

    \coordinate (X0)  at (\ld+\lh,0+\lv);
    \coordinate (X1) at (\lh/2+\ld/2+2*\ld/2+\lh/2, 2*\ld/2+\lv/2-\lly/2+\lv/2);
  \coordinate (X2) at (\lh/2+\ld/2+\lh/2+\lly/2, \lv/2+\ld/2+\lv/2+\ld/2);
  \coordinate (X3) at ( 2*\ld/2+\lh/2+\lh/2+\ld/2,2*\ld/2+\lv/2+\lv/2+\lly/2 );
  \coordinate (X4) at  (0+\lh/2+0+\lh/2+\ld/2,2*\ld/2+\lv/2+\lv/2+\lly/2);
  \coordinate (X41) at (\ld+\lh,0+2*\ld+\lv);
  \coordinate (X5) at  (0+\lh/2+\ld/2+2*\ld/2+\lh/2-\lly/2,\lv/2+\ld/2+0+\lv/2+\ld/2);
  \coordinate (X6) at (\lh/2+\ld/2+\lh/2,\lv/2+2*\ld/2+\lv/2-\lly/2);

  \end{tikzpicture}
  \end{subfigure}

  \caption{Two configurations of patterns for the boundary consistency conditions.}
\label{fig:tp}\end{figure}
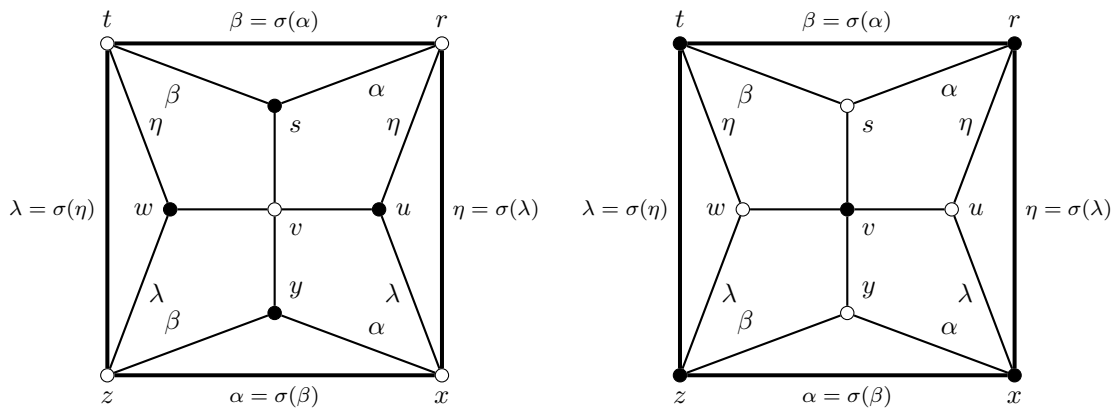

One can express $Q$ as 
            \begin{equation}
                Q=S+Ty=M+Nv\,, 
              \end{equation}
              where $S,T$, and $ M,N$ are polynomials affine-linear with respect to its arguments. They are in the forms
        \begin{subequations}              \begin{align}
                S=(\b+\alpha) (1 + \epsilon u v) + (u - v) x\,, \quad &  T = v-u\,, \\ \quad M =   \b-\a  + u (x - y)\,,  \quad & N =\epsilon(\b-\a)  u - x + y\,. 
              \end{align}
            \end{subequations}
If $\epsilon = 0$, then $S$ ({\em resp}. $N$) coincides with $M$ ({\em resp}. $T$). Let $q_w$ be in the $\cM_+$ case, \ie $q_w  =  r_1 x y + r_2 x  + r_2 y+ r_4$, then it has a  dual boundary polynomial  $p_b$ as
\begin{equation}
           \text{dual } \cM_+ \text{ case}: \quad p_b  =r_1 S\, x+ r_2( S -T x)- r_4 T  \,. 
\end{equation}
Alternatively, let $q_b$ be in  the $\cM_+$ case, \ie $q_b  =  \rho_1 u v + \rho_2 u  + \rho_2 v+ \rho_4$, it has
\begin{equation}
\text{dual } \cM_+ \text{ case}: \quad p_w = \rho_1 M\, u+ \rho_2( M -N u)- \rho_4 N  \,, 
\end{equation}
as its dual boundary polynomial. 
For simplicity, we assume $r_j$ and $\rho_j$, $j = 1,2,4$, are independent of the lattice parameters.       One obtains two pairs of dual boundary polynomials $\{q_w, p_b\}$ and $\{q_b, p_w\}$, and clearly, $p_b$ is different from $p_w$.

The criterion for them to be boundary consistent with H1$^\epsilon$ is explained in Section \ref{sec:cibe}. 
One needs to respectively consider the symmetries of $Q^\intercal_w = 0$ and $Q^\intercal_b =0$ which are different equations here. One also need to interchange the patterns of white and black vertices appearing in the rhombic dodecahedron of Figure \ref{fig:rd1}, which results in two boundary consistency conditions with different patterns of quadrilaterals and triangles as shown in Figure \ref{fig:tp}.  Based on arguments given in M\"obius cases I and II in  Section \ref{sec:cibe}, it is straightforward to conclude that both $q_b = 0$ and $q_w =0$ are integrable with $\sigma(\alpha) = -\alpha +\mu$. With the same involution relation  $\sigma(\alpha) = -\alpha +\mu$,  $p_w =0$ is integrable provided that $\rho_1 =0$ and $\rho_2\neq 0$ since $Q^\intercal_w = 0$ is Q1($\sqrt{-\epsilon}$), and $p_b =0$ is integrable with $r_1r_4-r_2^2\neq 0$  since $Q^\intercal_b = 0$ is Q1($0$).
Interestingly, the integrable boundary equations also depend on the patterns of the underlying triangles.  In Figure \ref{fig:dynamicwithboundary}, we provide two examples of  quad-graph systems on a trip with two parallel boundaries, where the patterns of the boundary equations are of different types.

\section{Conclusion}\label{sec:conc}
In this paper, we provide a classification of boundary equations that are boundary-consistent with quad equations in the ABS classification.

First, we provide the notions of boundary polynomials and boundary equations in Section \ref{sec:bpbe} as the natural objects to characterizes boundary conditions for quad-graph systems with boundary. The classification is based on factorization approaches that are introduced along the paper. The factorization of quad equation is formulated in Section  \ref{sec:factQ}, which, in turn, leads to the duality property of its boundary equations. The exhausted list of nondegenerate boundary equations dual to each other is provided in Section \ref{sec:dbp}. This list is taken in Section \ref{sec:fcrd} to factorize the consistent system around a rhombic dodecahedron into two equivalent halves, which amounts to the boundary consistency and the associated integrable boundary equations. In particular, the involution relation $\sigma$ appearing in the definition of boundary consistency is systematically derived. 
A criterion for degenerate boundary equations that are boundary consistent with a given quad equation is also provided, and the degenerate boundary consistency does not require any involution $\sigma$. We extend our method to derive integrable boundary equations for the H1$^\epsilon$ equation, that is rhombic-symmetric only. In this case, the patterns of boundary equations should be taken into account, and the boundary consistency conditions are also adapted according to the patterns. Examples of quad-graph systems on a strip with different type of patterns are also provided.

The results obtained in this paper lays foundations for what we believe could be some new areas of research in discrete integrable systems. Among many open questions we can think of, we
would mention some that are of importance:  an inverse scattering transform for discrete integrable equations on a ``half-plane'' can be readily developed, notably for some Q-type equations including  Q1($0$) and Q4 equations. This is in analogy to a half-line problem for soliton equations, and the technique could follow the half-line inverse scattering transform recently introduced by one of the authors \cite{ZC2}.
  We would expect that discrete soliton solutions as given in \cite{NAH, HZ11} reflect at the boundary with soliton parameters changing according to $\sigma$. More generic  quad-graph systems with boundary could also be investigated as have been done in  \cite{AV}.
The quad-graph systems without boundary are in conjunction with examples of discrete Riemann surface and discrete complex analysis \cite{M1, BIts1, BS}. Similar aspects for quad-graph systems with boundary are completely open. Moreover, the factorization approach developed in this paper could also be adapted to systems of $3$D consistent equations such as the list of discrete Boussinesq-type equations obtained in \cite{HBSQ}.  Lastly, we would like to mention that classification problems for discrete integrable systems represent a very active research areas in connection to many aspects of mathematical physics. Recent results in various contexts can be found, for instance,  in  \cite{ABS3, JNNN, kels1}.

\appendix
\section{ABS classification}
\label{sec:app1}
The canonical forms of the bulk polynomial $Q:=Q(x,u,v,y;\alpha,\beta)$, the associated biquadratic polynomials, and the tetrahedron equation $Q^\intercal =0$ are listed below. 
\subsection{Q-type bulk polynomials}

\begin{description}
  \item[Q1($\delta$):] the bulk polynomial $Q$ reads 
    \[\alpha (x-v)(u-y)- \beta (x-u)(v-y) + \delta^2 \alpha \beta (\alpha-\beta)\,.\]
 It has Q1($\delta$) as $Q^\intercal =0$. The factorized forms of the biquadratic polynomials are 
    \begin{align*} \delta_{v,y}Q                               & =  -\b (\a - \b  )((x-u)+ \delta\alpha )((x-u)- \delta\alpha ) \,,\\
      \delta_{u,v}Q    & =  \a \b((x-y)+  \delta(\alpha-\beta) )((x-y)- \delta(\alpha-\beta) )\,.\end{align*}
  \item[Q2:] the bulk polynomial $Q$ reads 
    \[           \alpha (x-v)(u-y)- \beta (x-u)(v-y)+\alpha \beta(\alpha-\beta)(x+u+v+y)     - \alpha \beta ( \alpha- \beta)(\alpha^2-\alpha \beta +\beta^2)  \,.      
    \]
  It has Q2 as $Q^\intercal =0$. The  biquadratic polynomials are    
    \begin{align*}
      \delta_{v,y}Q & = -(\alpha -\beta)\beta ((x - u)^2 - 2 (x + u) \alpha^2 + \alpha^4)\,,  \\
      \delta_{u,v}Q & =\alpha \beta ((x - y)^2 - 2 (x + y) (\alpha-\beta)^2 + (\alpha-\beta)^4)   \,.
    \end{align*}

  \item[Q3($\delta$):] the bulk polynomial $Q$ reads 
    \[
 \sinh (\alpha) (x u + v y) -  \sinh (\beta)(x v + u y)- \sinh (\alpha-\beta) (x y +  u v) + \delta \sinh (\alpha-\beta) \sinh(\alpha)\sinh(\beta)
\]  It has Q3($\delta$) as $Q^\intercal =0$. The  biquadratic polynomials are    
\begin{align*}
  \delta_{v,y}Q  = &  - \sinh(\alpha - \beta)\sinh(\beta)( x^2 + u^2  - 
   2  xu \cosh(\alpha)- \delta\sinh^2 (\alpha)) \,, \\
    \delta_{u,v}Q  = &  \sinh(\alpha)\sinh(\beta)( x^2 + y^2  - 
   2 xy \cosh(\alpha-\beta)- \delta\sinh^2 (\alpha-\beta))\,. 
\end{align*}
When $\delta = 0$,  the factorized forms of the biquadratic polynomials are
\begin{align*}
  \delta_{v,y}Q  = &  - \sinh(\alpha - \beta)\sinh(\beta)( u -e^{-\alpha}x )( u - e^\alpha x ) \,, \\
    \delta_{u,v}Q  = &  \sinh(\alpha)\sinh(\beta)( x - e^{-\alpha+\beta}  y )( x -e^{\alpha-\beta}  y )\,. 
\end{align*}
\item[Q4:] the bulk polynomial $Q$ reads 
  \[       \text{sn}(\alpha)(x u+vy) - \text{sn}(\beta)(x v + uy)                 - \text{sn}(\alpha-\beta)(x y +uv) +  \text{sn}(\alpha-\beta)\text{sn}(\alpha)\text{sn}(\beta)(1+ k^2 x u v y )\,,\]
where $ \text{sn}$ is the Jacobi elliptic functions of modulus $k$, $k \neq 0, \pm 1$. It has Q4 as $Q^\intercal=0$, and   \begin{align*}
   \delta_{v,y}Q  = & -\text{sn}(\alpha-\beta)\text{sn}(\beta)(x^2 + u^2- k^2\text{ sn}^2 (\alpha)x^2 u^2 - 2\text{cn} (\alpha) \text{dn} (\alpha) xu - \text{sn}^2 (\alpha) ) \,, \\
    \delta_{x,y}Q  = &   \text{sn}(\alpha)\text{sn}(\beta)(u^2 + v^2- k^2\text{ sn}^2 (\alpha-\beta) u^2v^2 - 2\text{cn} (\alpha-\beta) \text{dn} (\alpha-\beta) uv - \text{sn}^2 (\alpha-\beta) ) \,.
\end{align*}
\end{description}

\subsection{H-type bulk polynomials}
\begin{description}
\item[H1:]the bulk polynomial $Q$ reads
  \[(x-y)(u-v)+\beta-\alpha \,.\]
  It has Q1($0$)  as $Q^\intercal =0$. The  biquadratic polynomials are
\[
  \delta_{v,y}Q  =  -\alpha+\beta \,, \quad    \delta_{u,v}Q  =  (x-y)^2\,.
  \]
  \item[H2:]the bulk polynomial $Q$ reads
  \[(x-y)(u-v)+(\beta-\alpha)(x+u+v+y)+\beta^2-\alpha^2 \,.\]
  It has Q1($1$) as $Q^\intercal =0$. The  biquadratic polynomials are
\[
  \delta_{v,y}Q  =  -2 (\alpha - \beta) (\alpha + u + x) \,, \quad
    \delta_{u,v}Q                    =  (x-y+\alpha-\beta)(x-y-\alpha+\beta)  \,.
  \]
  \item[H3($\delta$):]
  the bulk polynomial $Q$ reads
  \[e^\alpha(x u + v y)-e^\beta(x v +  uy )+\delta(e^{2\alpha} -e^{2\beta}) \,.\]
  It has Q3($0$) as $Q^\intercal =0$. The  biquadratic polynomials are
  \[
  \delta_{v,y}Q  =  (e^{2\alpha}-e^{2\beta})(u x +e^\alpha \delta) \,, \quad
  \delta_{u,v}Q  =  (e^\alpha x - e^\beta y)(e^\beta x- e^\alpha y)\,.
  \]
\end{description}

\subsection{A-type bulk polynomials}
\label{app:Aty}  \begin{description}
  \item[A1($\delta$):]
  the bulk polynomial $Q$ reads
  \[\alpha (x+v)(u+y)- \beta (x+u)(v+y) - \delta^2 \alpha \beta (\alpha-\beta) \,.\]
  It can be obtained from Q1($\delta$) using $u\to -u$, $v\to -v$. It has Q1($\delta$) as $Q^\intercal=0$, and the biquadratic polynomials follow from those of  Q1($\delta$).  
  \item[A2:]
  the bulk polynomial $Q$ reads
  \[\sinh (\alpha)(x v + v y)-\sinh(\b)(x u +  vy )-\sinh(\a-\b)(1+ x u v y) \,.\]
  It can be obtained from Q3($0$) using $u\to 1/u$, $v\to 1/v$. It has Q3($0$) as $Q^\intercal=0$, and  the biquadratic polynomials follow from those of  Q3($0$). 
 \end{description}

 \section{Boundary polynomials for Q- and A-type polynomials}  \label{app:cbp}
 We follow the forms of $Q, q, p, \chi, \zeta, \Gamma$  given in \eqref{eq:caqpq}. 
    \subsection{Q-type polynomials}
  \begin{description}
  \item[Q1($\delta$):] 

    \begin{description}
    \item[$\cM_+$ case:] 
    \begin{align*}
      \fp_1 = & (\a-\b) u vx - x^2(\a u -\b v)  -\delta^2(\a-\b) \a \b  x \,,\\
      \fp_2 =& (\a-\b) u v- (\a +\b)x(u-v) - (\a-\b)(x^2+ \a \b \delta^2)\,,   \\
      \fp_4 =& -\b u +\a v -(\a-\b)x\,.
    \end{align*}
     \item[~~ $\cM_-$ case:]  set  $\kappa =\alpha-\beta$,  then  \[        \mathbi{p}=  u v- x(u+v)+x^2 -\delta^2 \a \b\,. \]
 When $\delta=0$, the  $\cM_-$ boundary polynomial is excluded since  $q=0$ leads to  $Q_{u,v} = 0$.  
\item[~~~Singular case:] let $\{\alpha, \beta\}\to \{\alpha^2, \beta^2\}$, then
  \begin{align*}
  \fq_0=& (\a + \b)x y -   (\a u - \delta \a \b^2)x - (\b u- \delta \a^2 \b)y  \,,\\
  \fq_1= & \b x +\a y+ (\a + \b) (\delta\a\b-u  ) \,,\\
  \fp_0= & -(\a-\b) u v + \a x u-\b x v+ \delta\a\b (\a - \b) (\delta\a \b+x  )\,,\\
  \fp_1= & \b u -\a v +(\a -\b) (\delta\a \b +x  )\,. 
\end{align*}
    \end{description}

  \item[Q2:]
        \begin{description}
    \item[$\cM_+$ case:]     \begin{align*}
      \mathbi{p}_1 =& x(-(\a-\b) u v + \a \b (\a - \b)(u + v) + \a x u-\b x v  - \a \b (\a - \b) (\a^2 - \a \b + \b^2 - x))\, ,  \\
      \mathbi{p}_2 =& -(\a - \b) u v + (\a + \b) x(u-v)+ \a \b(\a - \b)(u+v) -(\a- \b)(\a \b (\a^2 -\a \b +\b^2)- x^2)\,, \\
      \mathbi{p}_4 =& \b u -\a v -(\a - \b) (\a \b - x)\,.
    \end{align*}
\item[$\cM_-$ case:] set  $\kappa =\alpha-\beta$, then
    \[\mathbi{p} = u v - (\a \b + x)(u + v)+ \a \b (\a^2 - \a \b + \b^2) - 2 \a \b x + x^2\,.\]

  \end{description}

 \item[Q3($\delta$):]
\begin{description}
    \item[$\cM_+$ case:]
    \begin{align*}
      \mathbi{p}_1 =&  x (-\sinh(\a-\b) u v +\sinh(\a) x  u- \sinh(\b) x  v  + \delta \sinh(\a)\sinh(\b) \sinh(\a-\b))\,, \\
      \mathbi{p}_2 =& -\sinh(\a-\b) u v+ x(\sinh(\a)+ \sinh(\b))(u-v) + \sinh(\a-\b)(x^2+ \delta\sinh(\a)\sinh(\b))\,,  \\
      \mathbi{p}_4 =& \sinh(\b) u - \sinh(\a) v + x \sinh(\a-\b)\,.
    \end{align*}
    \item[~~ $\cM_-$ case:]  set $\kappa=\sinh(\a-\b)$, then
    \[\mathbi{p}=u v - \text{csch}(\a-\b) (\sinh(\a)-\sinh(\b)) x (u+v)-\delta \sinh(\a)\sinh(\b) + x^2\,.\]

\item[~~ Singular case:] when $\delta=0$, let $\{e^\a,e^\b\}\mapsto\{ \cosh(\a), \cosh(\b)\}$, then
    \begin{align*}
    \fq_0=& (\sinh(\a)+\sinh(\b)) x y -u \sinh(\a) \cosh(\b) x - u \cosh(\a)\sinh(\b)y\,, \\
    \fq_1=& \sinh(\b)x +\sinh(\a) y-u \sinh(\a+\b)\,, \\
    \fp_0=& (\sinh(\a)+\sinh(\b))(-(
    \sinh(\a)-\sinh(\b)) u v +x \sinh(\a) \cosh(\b) u - x \cosh(\a)\sinh(\b)v )\,, \\
    \fp_1=& \sinh(\a+\b)(\sinh(\b)u - \sinh(\a) v + x \sinh(\a-\b))\,.
    \end{align*}
\end{description}

  \item[Q4:]

    \begin{description}
    \item[$\cM_+$ case:]
    \begin{align*}
      \mathbi{p}_1 =& x(-\text{sn}(\a-\b) u v + (\text{sn}(\a)-\text{sn}(\b))x(u-v)+\text{sn}(\a)\text{sn}(\b)\text{sn}(\a-\b))\,, \\
      \mathbi{p}_2 =& -\text{sn}(\a-\b)(1+ k^2 \text{sn}(\a)\text{sn}(\b) x^2) u v+ (\text{sn}(\a)+\text{sn}(\b))x ( u - v)+ \text{sn}(\a-\b)(\text{sn}(\a)\text{sn}(\b)+ x^2)\,,   \\
      \mathbi{p}_4 =& -k^2 \text{sn}(a) \text{sn}(\b)\text{sn}(\a-\b) u v+ \text{sn}(\b) u -\text{sn}(\a) v+ \text{sn}(\a-\b) x\,.
    \end{align*}

  \item[$\cM_-$ case:] set  $\kappa =\text{sn}(\a-\b)$, then
    \begin{align*}
        \mathbi{p}& = -(k^2 \text{sn}(\a)\text{sn}(\b)x^2-1) u v- \text{ns}(\a-\b)(\text{sn}(\a)-\text{sn}(\b))x (u + v) -\text{sn}(\a)\text{sn}(\b)+x^2\,.
    \end{align*}

    \end{description}
  \end{description}

    \subsection{A-type polynomials}
A1($\delta$) (or A2) can be derived from Q1($\delta$) (or Q3($0$)) using $\{u, v\}\to\{ \phi(u), \phi(v)\}$ ($\phi$ is a M\"obius transformation, \cf \ref{app:Aty}).  
Let $Q$ denote Q1($\delta$) (or Q3($0$)), and  $\widetilde{Q}$ denote A1($\delta$) (or A2), then,
\[
    \widetilde{Q}(x,u,v,y;\alpha,\beta)=\Lambda(u) \Lambda(v) Q(x,\phi(u),\phi(v),y;\alpha,\beta)\,, 
\]
with $\Lambda(\ast)$ defined in \eqref{eq:qfm}. Let  $q,p$ be a pair of dual boundary polynomials of $Q$, it follows from some simple considerations that  $\widetilde{q}, \widetilde{p}$ in the forms
\begin{align*}
    \widetilde{q}(x,u,y;\alpha,\beta)&=\Lambda^c(u) q(x,\phi(u),y;\alpha,\beta)\,,\\
    \widetilde{p}(u,x,v;\alpha,\beta)&=\Lambda(u)\Lambda(v) p(\phi(u),x,\phi(v);\alpha,\beta)\,,
\end{align*}
are boundary polynomials dual to each other for $\widetilde{Q}$. Precisely, one gets
\begin{description}
  \item[A1($\delta$):]

    \begin{description}
    \item[$\cM_+$ case:]
    \begin{align*}
      \mathbi{p}_1 = & x(-(\a-\b) u v - x(\a u -\b v)  +\delta^2(\a-\b) \a \b)  \,,\\
      \mathbi{p}_2 =& -(\a-\b) u v - (\a +\b)x(u-v) + (\a-\b)(x^2+ \delta^2 \a \b )\,,   \\
      \mathbi{p}_4 =& -\b u +\a v +(\a-\b)x\,.
    \end{align*}
     \item[~~ $\cM_-$ case:]  set  $\kappa =\alpha-\beta$, then
  \[        \mathbi{p}=  u v+ x(u+v)+x^2 -\delta^2 \a \b \,  . \]
 When $\delta=0$, the  $\cM_-$ boundary polynomial is excluded since  $q=0$ leads to  $Q_{u,v} = 0$.
\item[~~ Singular case:] let $\{\a, \b\} \to \{\alpha^2, \b^2\}$, then
    \begin{align*}
    \fq_0=& (\alpha+\beta) x y+\alpha(u+ \delta \beta^2 )x+ \beta (u+  \delta \alpha^2) y\,, \\
    \fq_1=& \beta x + \alpha y +(\alpha+\beta)(u+  \delta\alpha \beta)\,, \\
    \fp_0=& (\alpha-\beta) u v + \alpha x u -\beta x v -\delta \alpha \beta (\alpha - \beta) (x + \delta \alpha \beta )\,,\\
    \fp_1=& \beta u - \alpha v - (\alpha-\beta)(x + \delta \alpha \beta )\,.
    \end{align*}
    \end{description}

  \item[A2:]
\begin{description}
    \item[$\cM_+$ case:]
    \begin{align*}
      \mathbi{p}_1 =&  x (- \sinh(\b) x  u+ \sinh(\a) x  v - \sinh(\a-\b)   \,, \\
      \mathbi{p}_2 =&  \sinh(\a-\b)x^2 u v- (\sinh(\a)+ \sinh(\b))x(u-v) -\sinh(\a-\b)  \,,  \\
      \mathbi{p}_4 =& \sinh(\a-\b)x u v- \sinh(\a) u + \sinh(\b) v  \,.
    \end{align*}
    \item[$\cM_-$ case:] one set $\kappa=\sinh(\a-\b)$, then
    \[\mathbi{p}=x^2 u v - \text{csch}(\a-\b) (\sinh(\a)-\sinh(\b)) x (u+v)+1  \,.\]

\item[~~ Singular case:] let  $\{e^\a,  e^\b\} \to \{ \cosh(\a),   \cosh(\b)\}$, then
    \begin{align*}
    \fq_0=& (\sinh(\a)+\sinh(\b))u x y - \sinh(\a) \cosh(\b) x -  \cosh(\a)\sinh(\b)y \,, \\
    \fq_1=& \sinh(\b)u x +\sinh(\a) u y- \sinh(\a+\b)\,, \\
    \fp_0=& (\sinh(\a)+\sinh(\b))(  -  \cosh(\a)\sinh(\b) x u+ \sinh(\a) \cosh(\b) x v -(\sinh(\a)-\sinh(\b)) )\,, \\
    \fp_1=& \sinh(\a+\b)( \sinh(\a-\b) x u v - \sinh(\a) u +\sinh(\b)v  )\,.
    \end{align*}
\end{description}

\end{description}

\section{Boundary polynomials for H-type polynomials}\label{app:hp}

We follow the forms of $Q, q, p, \chi, \zeta, \Gamma$  given in \eqref{eq:caqpq}. 
\subsection{Results according to  Theorem \ref{theo:pq}}
 \begin{description}
 \item[H1:]
   \begin{description}
 \item[$\cM_+$ case:]
     \begin{align*}
        \mathbi{p}_1=&  x (x (u-v) -\a + \b))\,,  \\
        \mathbi{p}_2=&  2 x (u-v) - \a + \b\,, \\
        \mathbi{p}_4=&  u - v\,.
     \end{align*}
   \end{description}

\item[H2:]
\begin{description}
    \item[$\cM_+$ case:]
    \begin{align*}
        \mathbi{p}_1=& x(-(\a-\b)(u+v)+ x (u-v) -(\a-\b)(\a+ \b + x))\,,  \\
        \mathbi{p}_2=& -(\a-\b)(u+v)+2 x (u - v) - \a^2 +\b^2\,, \\
        \mathbi{p}_4=& u - v +\a - \b\,.
   \end{align*}
    \item[$\cM_-$ case:] one set $\kappa=\a-\b$, then
    \[
    \mathbi{p}= u + v + \a + \b + 2 x\,.
    \]
    \item[Singular case:]  
    \begin{align*}
        \fq_0 & = 2 x y +u (x + y)+ \b x +\a y\,, \\
        \fq_1 & = x + y + \a + \b + 2 u\,, \\
        \fp_0 & = -(\a-\b)(u+v)+2x (u-v) -(\a-\b)(\a+\b+ 2 x)\,,\\
        \fp_1 & = 2(u - v)\,.
    \end{align*}
Here, $\fp_1,\fq_1$ coincide with the $\cM_-$ and dual $\cM_-$ cases. 
  \end{description}

\item[H3($\delta$):]

\begin{description}
    \item[$\cM_+$ case:]
    \begin{align*}
      \mathbi{p}_1 =& x (e^\a x u - e^\b x v + \delta(e^{2 \a}-e^{2 \b}))\,, \\
      \mathbi{p}_2 =& (e^\a + e^\b) (x (u - v) + \delta(e^\a - e^\b))\,,\\
      \mathbi{p}_4 =& e^\b u - e^\a v\,.
    \end{align*}
    \item[~~~$\cM_-$ case:] one set $\kappa=e^\a-e^\b$ then
    \[
    \mathbi{p}= -x(u+v)-\delta(e^\alpha+e^\beta)\,.
    \]
    \item[~~~Singular case:] when $\delta \neq 0$,
    \begin{align*}
    \fq_0=& 2u x y + \delta(e^\beta x+ e^\alpha y)\,,\\
    \fq_1=& u(x+y)+  \delta(e^\alpha+e^\beta)\,,\\
    \fp_0=& 2 x(e^\alpha u- e^\beta v)+ \delta(e^{2 \alpha}-e^{2 \beta})\,,\\
    \fp_1=& (e^\alpha+ e^\beta)(u-v)\,.
    \end{align*}
    Here,  $\fp_1,\fq_1$ coincide with the $\cM_-$  and dual $\cM_-$ cases. 
\end{description}

\end{description}
\subsection{Proof}

It is stated in Theorem \ref{theo:pq} that H-type bulk polynomials only have the above list of $q,p$. Here, we provide a direct proof.
\begin{description}
\item[H1:]  $\Gamma$ is $\b-\a$ that is of bidegree $(0,0)$ in $x,u$. Transform $Q$ to $Q_\fm $ using $\fm(x)=1/x$, then 
  \begin{equation}
    \label{qm1}
    Q_\fm=(x-y)(u-v)+(\b-\a)x y u v\,, \quad \Gamma_\fm=(\b-\a)x^2 u^2\,.   \end{equation}
 We aim to classify all possible $q,p$ for \eqref{qm1}, which, through the equivalence class with respect to simultaneous M\"obius transformations,  provides all $q,p$ for $Q$ in the canonical form.  For simplicity, the subscript $\fm $ is omitted here. Take a generic boundary polynomial $q$ with $12$ parameters
\begin{equation}
    q=(\sum^2_{j=0} r_{1j}u^j) x y+ (\sum^2_{j=0} r_{2j}u^j) x + (\sum^2_{j=0} r_{3j}u^j) y +\sum^2_{j=0} r_{4j}u^j\,,     
\end{equation}

with $r_{ij}$ to be determined.  The idea is to combine the above forms of $Q$ and $q$ following the singular case approach provided in Section \ref{sec:dbp}.  To classify all possible $q,p$, it suffices to classify all possible $\chi, \zeta$ dual to each other, which,  as a result of \eqref{eq:eeee}, obey
\begin{equation}
\chi \, \zeta  =- (\b-\a)x^2 u^2\,, 
\end{equation}
Due to the duality between $\chi$ and $\zeta$, one only needs to take six possible cases of $\chi$, namely, $\chi \propto x^2u^2 $, $\chi \propto x^2u $, $ \chi \propto x^2 $, $\chi \propto u^2 $, $\chi \propto x u^2 $,  $ \chi \propto xu $. 

Let $\deg_x \chi$ be $2$, which contains the first three cases. Assuming the existence of $p$, by the analysis of the degrees of the middle argument of $q,p$ provided in Section \ref{sec:factQ},  the polynomial $Q q_{y}-Q_{y}q$ defined in \eqref{eq:dualm} should be of  degree $2$ in $x$, which implies that $\deg_x p=0$. This means $p$ belongs to the $\cM_+$ case. This amounts to the only possible $q,p$, and the results are listed above. 

One can show that the remaining three cases are not possible.  For instance, let $\chi \propto u^2$, then  $Q q_{y}-Q_{y}q$ has a factor $x^2$. This leads to restrictions on the parameters of $q$ as $r_{10} = 0$, $r_{40} = 0$, $r_{30} = -r_{20}$, $r_{41} = 0$, $r_{31} = -r_{21}$ and $r_{11} = (\a - \b)r_{20}$.     Since $q, p$ are irreducible, we have $q\vert _{u=0}=r_{20}(x-y) \neq 0$ and $ p\vert_{x=0}=r_{42}(u-v) \neq 0$. This imposes
\begin{equation}
 r_{20}\neq 0\,,\quad  r_{42} \neq 0 \,. 
\end{equation}
On the other hand, due to the $   \ZZ_2$-symmetry of $q,p$, assume 
\begin{equation}
  q(x,u,y;\alpha,\beta) = \gamma(\alpha,\beta)q(y,u,x;\beta,\alpha)\,,\quad  p(u,x,v;\alpha,\beta) = \eta(\alpha,\beta)p(v,x,u;\beta,\alpha)\,,
\end{equation}
with $\gamma(\alpha,\beta)\gamma(\beta,\alpha) = \eta(\alpha,\beta)\eta(\beta,\alpha) = 1$. Together with the above results, this leads to 
\begin{subequations}
  \begin{align}
  r_{20}(\alpha,\beta)=-\gamma(\alpha,\beta) r_{20}(\beta,\alpha)= &-\eta(\alpha,\beta) r_{20} (\beta,\alpha)\,, \\
    r_{42}(\alpha,\beta)=\gamma(\alpha,\beta) r_{42}(\beta,\alpha)=&-\eta(\alpha,\beta) r_{42} (\beta,\alpha)\,, 
  \end{align}
\end{subequations}
 which require that $\gamma(\alpha,\beta)=\eta(\alpha,\beta),\gamma(\alpha,\beta)=-\eta(\alpha,\beta)$. One has $\gamma(\alpha,\b)=\eta(\alpha,\b)=0$, which is impossible. 

\item[H2, H3:] they can be treated similarly, since $\Gamma$ in both cases is of bidegree $(1,1)$ in $x,u$.  Take H2 as an  example. Using transformation $\fm(x)=1/x$, one gets
  \begin{equation}
\label{qm2}
    Q_\fm=(x-y)(u-v)+(\beta-\alpha)(xuv+xvy+xuy+uvy)+(\beta^2-\alpha^2)xuvy\,, 
  \end{equation}
 with
 \begin{equation}
  \Gamma_\fm=-2(\a-\b) x u (x+ u +\a x u)\,. 
 \end{equation}
One aims  to classify all possible $q,p$ for $Q_\fm$. Drop the subscript $\fm$ for simplicity. Due to the duality between $\chi$ and $\zeta$, one needs to take four possibilities for $\chi$, namely, $\chi \propto x u (x+u+\a x u) $, $\chi \propto x(x+u+\a x u) $, $\chi \propto u(x+u+\a x u) $, $ \chi \propto x+u+\a x u $. The first two cases have $\deg_{x} \chi=2$, which implies the dual boundary polynomials belong to the  $\cM_+$ or $\cM_-$ case.  In the case $ \chi \propto x+u+\a x u $, $\chi$ and $\zeta$  are of bidegree $(1,1)$ in $x,u$, and all possible $q,p$ can be exhausted following the singular case approach. Lastly,  the case $\chi \propto u(x+u+\a x u) $ is impossible. This follows the same arguments  presented above for H1.

\end{description}





\end{document}